\pdfoutput=1

\documentclass[12pt,a4paper]{article}

\usepackage{ifthen} 
\newboolean{pdflatex}
\setboolean{pdflatex}{true} 

\newboolean{articletitles}
\setboolean{articletitles}{true} 

\newboolean{uprightparticles}
\setboolean{uprightparticles}{false} 

\def\paperauthors{LHCb collaboration} 
\def\paperasciititle{Study of b-hadron decays to Lambdac+ h- h'- final states}
\def\papertitle{Study of $b$-hadron decays to $\mathit{\Lambda}_{c}^+ h^- h^{\prime -}$ final states}
\def\paperkeywords{{High Energy Physics}, {LHCb}} 
\def\papercopyright{\the\year\ CERN for the benefit of the LHCb collaboration} 
\def\paperlicence{CC BY 4.0 licence}
\def\paperlicenceurl{https://creativecommons.org/licenses/by/4.0/}

\usepackage[top=1in, bottom=1.25in, left=1in, right=1in]{geometry}

\usepackage{microtype}
\usepackage{lineno}  
\usepackage{xspace} 
\usepackage{caption} 

\usepackage{graphicx}  
\usepackage{color}
\usepackage{colortbl}
\graphicspath{{./figs/}} 

\usepackage{amsmath} 
\usepackage{amssymb}
\usepackage{amsfonts}
\usepackage{upgreek} 

\newcommand*\patchAmsMathEnvironmentForLineno[1]{%
\expandafter\let\csname old#1\expandafter\endcsname\csname #1\endcsname
\expandafter\let\csname oldend#1\expandafter\endcsname\csname
end#1\endcsname
 \renewenvironment{#1}%
   {\linenomath\csname old#1\endcsname}%
   {\csname oldend#1\endcsname\endlinenomath}%
}
\newcommand*\patchBothAmsMathEnvironmentsForLineno[1]{%
  \patchAmsMathEnvironmentForLineno{#1}%
  \patchAmsMathEnvironmentForLineno{#1*}%
}
\AtBeginDocument{%
\patchBothAmsMathEnvironmentsForLineno{equation}%
\patchBothAmsMathEnvironmentsForLineno{align}%
\patchBothAmsMathEnvironmentsForLineno{flalign}%
\patchBothAmsMathEnvironmentsForLineno{alignat}%
\patchBothAmsMathEnvironmentsForLineno{gather}%
\patchBothAmsMathEnvironmentsForLineno{multline}%
\patchBothAmsMathEnvironmentsForLineno{eqnarray}%
}

\usepackage{hyperxmp}

\usepackage[pdftex,
            pdfauthor={\paperauthors},
            pdftitle={\paperasciititle},
            pdfkeywords={\paperkeywords},
            pdfcopyright={Copyright (C) \papercopyright},
            pdflicenseurl={\paperlicenceurl}]{hyperref}

\usepackage[bottom,flushmargin,hang,multiple]{footmisc}

\usepackage[all]{hypcap} 

\usepackage{xspace} 
\usepackage{upgreek}

\def\lhcb   {\mbox{LHCb}\xspace}

\def\MagUp {\mbox{\em Mag\kern -0.05em Up}\xspace}

\ifthenelse{\boolean{uprightparticles}}%
{

 \def\Ppi         {\ensuremath{\uppi}\xspace}

 \def\Ppsi        {\ensuremath{\uppsi}\xspace}

 \def\PDelta      {\ensuremath{\Delta}\xspace}                 
 \def\PXi         {\ensuremath{\Xi}\xspace}                 
 \def\PLambda     {\ensuremath{\Lambda}\xspace}                 
 \def\PSigma      {\ensuremath{\Sigma}\xspace}                 
 \def\POmega      {\ensuremath{\Omega}\xspace}                 
 \def\PUpsilon    {\ensuremath{\Upsilon}\xspace}
 \let\oldPi\Pi
 \def\PPi         {\ensuremath{\oldPi}\xspace}

 \def\PB      {\ensuremath{\mathrm{B}}\xspace}                 
                  
 \def\PD      {\ensuremath{\mathrm{D}}\xspace}

 \def\PJ      {\ensuremath{\mathrm{J}}\xspace}                 
 \def\PK      {\ensuremath{\mathrm{K}}\xspace}

 \def\Pb      {\ensuremath{\mathrm{b}}\xspace}                 
 \def\Pc      {\ensuremath{\mathrm{c}}\xspace}

 \def\Pi      {\ensuremath{\mathrm{i}}\xspace}

 \def\Pp      {\ensuremath{\mathrm{p}}\xspace}

 \def\Ps      {\ensuremath{\mathrm{s}}\xspace}

 \def\thebaroffset{0.0em}
}
{

 \def\Ppi         {\ensuremath{\pi}\xspace}

 \def\Ppsi        {\ensuremath{\psi}\xspace}                 
                  
 \mathchardef\PDelta="7101
 \mathchardef\PXi="7104
 \mathchardef\PLambda="7103
 \mathchardef\PSigma="7106
 \mathchardef\POmega="710A
 \mathchardef\PUpsilon="7107
 \mathchardef\PPi="7105
                  
 \def\PB      {\ensuremath{B}\xspace}                 
                  
 \def\PD      {\ensuremath{D}\xspace}

 \def\PJ      {\ensuremath{J}\xspace}                 
 \def\PK      {\ensuremath{K}\xspace}

 \def\Pb      {\ensuremath{b}\xspace}                 
 \def\Pc      {\ensuremath{c}\xspace}

 \def\Pi      {\ensuremath{i}\xspace}

 \def\Pp      {\ensuremath{p}\xspace}

 \def\Ps      {\ensuremath{s}\xspace}

 \def\thebaroffset{0.18em}
}
\newcommand{\offsetoverline}[2][\thebaroffset]{\kern #1\overline{\kern -#1 #2}}%

\makeatletter
\ifcase \@ptsize \relax
  \newcommand{\miniscule}{\@setfontsize\miniscule{4}{5}}
\or
  \newcommand{\miniscule}{\@setfontsize\miniscule{5}{6}}
\or
  \newcommand{\miniscule}{\@setfontsize\miniscule{5}{6}}
\fi
\makeatother

\DeclareRobustCommand{\optbar}[1]{\shortstack{{\miniscule (\rule[.5ex]{1.25em}{.18mm})}
  \\ [-.7ex] $#1$}}

\def\squark    {{\ensuremath{\Ps}}\xspace}

\def\cquark    {{\ensuremath{\Pc}}\xspace}

\def\bquark    {{\ensuremath{\Pb}}\xspace}

\def\pion   {{\ensuremath{\Ppi}}\xspace}

\def\pip    {{\ensuremath{\pion^+}}\xspace}
\def\pim    {{\ensuremath{\pion^-}}\xspace}

\def\kaon    {{\ensuremath{\PK}}\xspace}

\def\KorKbar {\kern \thebaroffset\optbar{\kern -\thebaroffset \PK}{}\xspace}

\def\Km      {{\ensuremath{\kaon^-}}\xspace}

\def\D       {{\ensuremath{\PD}}\xspace}

\def\DorDbar {\kern \thebaroffset\optbar{\kern -\thebaroffset \PD}\xspace}
\def\Dz      {{\ensuremath{\D^0}}\xspace}

\def\Dp      {{\ensuremath{\D^+}}\xspace}
\def\Dm      {{\ensuremath{\D^-}}\xspace}

\def\DpDm    {\ensuremath{\Dp {\kern -0.16em \Dm}}\xspace}

\def\Dstarp  {{\ensuremath{\D^{*+}}}\xspace}

\def\B       {{\ensuremath{\PB}}\xspace}

\def\BorBbar {\kern \thebaroffset\optbar{\kern -\thebaroffset \PB}\xspace}

\def\Bd      {{\ensuremath{\B^0}}\xspace}

\def\BdorBdbar {\kern \thebaroffset\optbar{\kern -\thebaroffset \Bd}\xspace}
\def\Bu      {{\ensuremath{\B^+}}\xspace}
\def\Bub     {{\ensuremath{\B^-}}\xspace}

\def\Bm      {{\ensuremath{\Bub}}\xspace}

\def\Bs      {{\ensuremath{\B^0_\squark}}\xspace}

\def\BsorBsbar {\kern \thebaroffset\optbar{\kern -\thebaroffset \Bs}\xspace}

\def\jpsi     {{\ensuremath{{\PJ\mskip -3mu/\mskip -2mu\Ppsi}}}\xspace}

\def\Y#1S{\ensuremath{\PUpsilon{(#1S)}}\xspace}

\def\proton      {{\ensuremath{\Pp}}\xspace}
\def\antiproton  {{\ensuremath{\overline \proton}}\xspace}

\def\Lz          {{\ensuremath{\PLambda}}\xspace}

\def\LorLbar     {\kern \thebaroffset\optbar{\kern -\thebaroffset \PLambda}\xspace}

\def\Sigmares    {{\ensuremath{\PSigma}}\xspace}

\def\Xires       {{\ensuremath{\PXi}}\xspace}

\def\Omegares    {{\ensuremath{\POmega}}\xspace}

\def\Lc          {{\ensuremath{\Lz^+_\cquark}}\xspace}

\def\Sigmac      {{\ensuremath{\Sigmares_\cquark}}\xspace}
\def\Sigmacpp    {{\ensuremath{\Sigmares_\cquark^{++}}}\xspace}
\def\Sigmacp     {{\ensuremath{\Sigmares_\cquark^+}}\xspace}

\def\Lb           {{\ensuremath{\Lz^0_\bquark}}\xspace}

\def\Xibz         {{\ensuremath{\Xires^0_\bquark}}\xspace}
\def\Xibm         {{\ensuremath{\Xires^-_\bquark}}\xspace}

\def\Omegab       {{\ensuremath{\Omegares^-_\bquark}}\xspace}

\def\to                 {\ensuremath{\rightarrow}\xspace}

\def\CP                {{\ensuremath{C\!P}}\xspace}

\def\AT#1     {\ensuremath{A_{\mathrm{T}}^{#1}}\xspace}           

\def\C#1      {\ensuremath{\mathcal{C}_{#1}}\xspace}                       
\def\Cp#1     {\ensuremath{\mathcal{C}_{#1}^{'}}\xspace}                    
\def\Ceff#1   {\ensuremath{\mathcal{C}_{#1}^{\mathrm{(eff)}}}\xspace}        
\def\Cpeff#1  {\ensuremath{\mathcal{C}_{#1}^{'\mathrm{(eff)}}}\xspace}       
\def\Ope#1    {\ensuremath{\mathcal{O}_{#1}}\xspace}                       
\def\Opep#1   {\ensuremath{\mathcal{O}_{#1}^{'}}\xspace}                    

\newcommand{\nospaceunit}[1]{\ensuremath{\text{#1}}}       
\newcommand{\aunit}[1]{\ensuremath{\text{\,#1}}}       

\newcommand{\tev}{\aunit{Te\kern -0.1em V}\xspace}
\newcommand{\gev}{\aunit{Ge\kern -0.1em V}\xspace}
\newcommand{\mev}{\aunit{Me\kern -0.1em V}\xspace}
\newcommand{\kev}{\aunit{ke\kern -0.1em V}\xspace}
\newcommand{\ev}{\aunit{e\kern -0.1em V}\xspace}
 
\newcommand{\mevc}{\ensuremath{\aunit{Me\kern -0.1em V\!/}c}\xspace}
\newcommand{\gevc}{\ensuremath{\aunit{Ge\kern -0.1em V\!/}c}\xspace}
\newcommand{\mevcc}{\ensuremath{\aunit{Me\kern -0.1em V\!/}c^2}\xspace}
\newcommand{\gevcc}{\ensuremath{\aunit{Ge\kern -0.1em V\!/}c^2}\xspace}

\def\mum  {\ensuremath{\,\upmu\nospaceunit{m}}\xspace}

\def\fb   {\ensuremath{\aunit{fb}}\xspace}
\def\invfb   {\ensuremath{\fb^{-1}}\xspace}

\newcommand{\stat}{\aunit{(stat)}\xspace}
\newcommand{\syst}{\aunit{(syst)}\xspace}

\newcommand{\chisq}{\ensuremath{\chi^2}\xspace}

\newcommand{\chisqip}{\ensuremath{\chi^2_{\text{IP}}}\xspace}

\def\gsim{{~\raise.15em\hbox{$>$}\kern-.85em
          \lower.35em\hbox{$\sim$}~}\xspace}
\def\lsim{{~\raise.15em\hbox{$<$}\kern-.85em
          \lower.35em\hbox{$\sim$}~}\xspace}

\def\pt         {\ensuremath{p_{\mathrm{T}}}\xspace}

\def\ptot       {\ensuremath{p}\xspace}

\def\evtgen     {\mbox{\textsc{EvtGen}}\xspace}

\def\geant      {\mbox{\textsc{Geant4}}\xspace}

\def\photos     {\mbox{\textsc{Photos}}\xspace}

\def\pythia     {\mbox{\textsc{Pythia}}\xspace}

\def\tell1  {TELL1\xspace}
\def\ukl1   {UKL1\xspace}

\newcommand{\ie}{\mbox{\itshape i.e.}\xspace}

\newcommand{\lhcborcid}[1]{\href{https://orcid.org/#1}{\hspace*{0.1em}\raisebox{-0.45ex}{\includegraphics[width=1em]{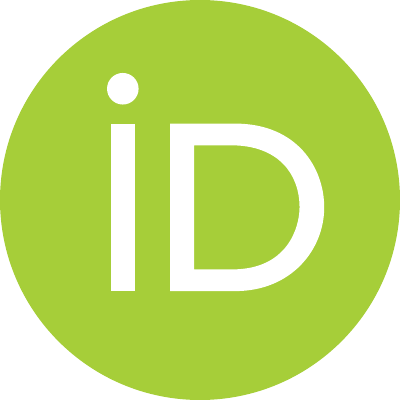}}}}

\newcommand{\runone}{\mbox{Run~1}\xspace}
\newcommand{\runtwo}{\mbox{Run~2}\xspace}

\usepackage{cite} 
\usepackage{mciteplus}

\usepackage{comment}

\begin{document}

\renewcommand{\thefootnote}{\fnsymbol{footnote}}
\setcounter{footnote}{1}


\begin{titlepage}
\pagenumbering{roman}

\vspace*{-1.5cm}
\centerline{\large EUROPEAN ORGANIZATION FOR NUCLEAR RESEARCH (CERN)}
\vspace*{1.5cm}
\noindent
\begin{tabular*}{\linewidth}{lc@{\extracolsep{\fill}}r@{\extracolsep{0pt}}}
\ifthenelse{\boolean{pdflatex}}
{\vspace*{-1.5cm}\mbox{\!\!\!\includegraphics[width=.14\textwidth]{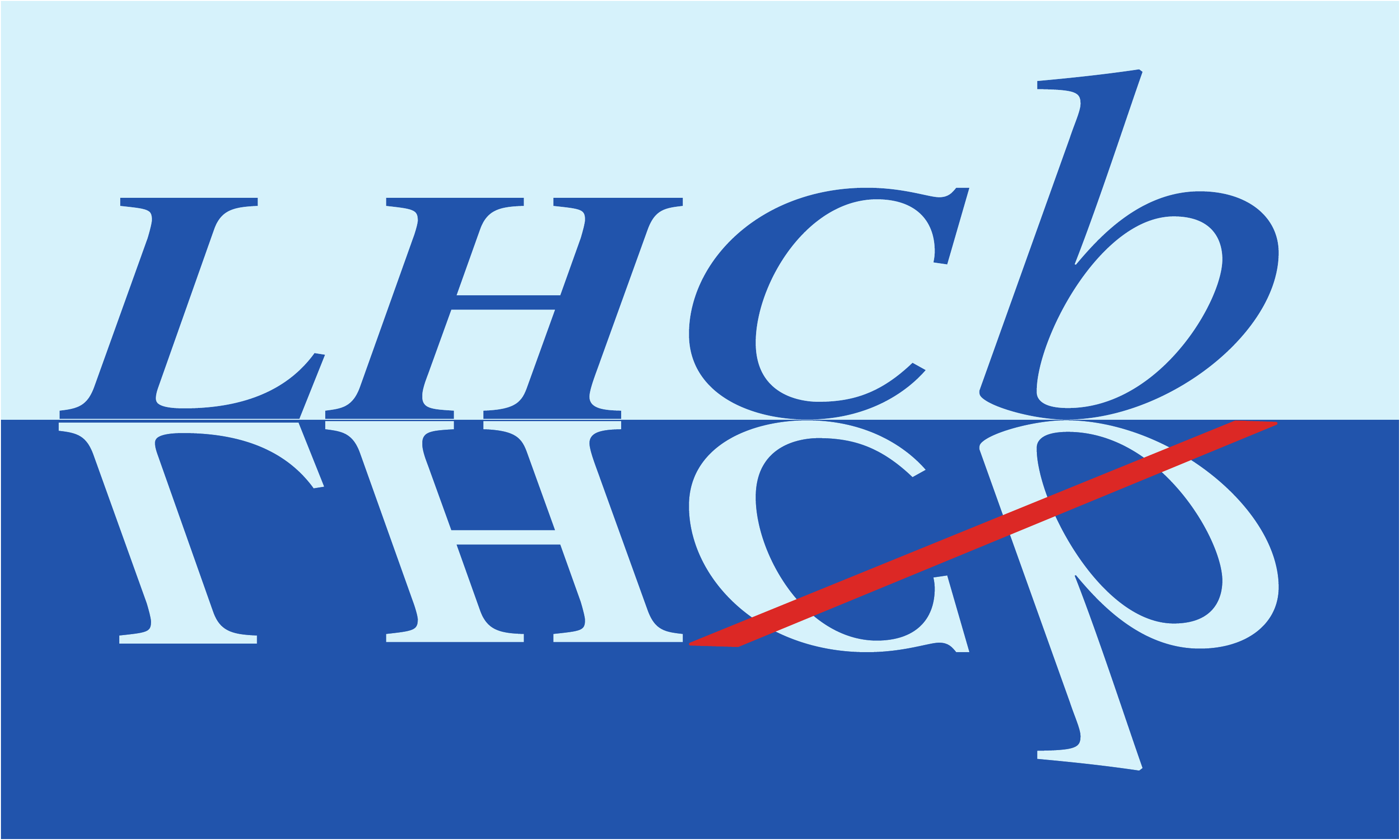}} & &}%
{\vspace*{-1.2cm}\mbox{\!\!\!\includegraphics[width=.12\textwidth]{figs/lhcb-logo.eps}} & &}%
\\
 & & CERN-EP-2024-116 \\  
 & & LHCb-PAPER-2024-013 \\  
 & & \today \\ 
 & & \\
\end{tabular*}

\vspace*{3.0cm}

{\normalfont\bfseries\boldmath\huge
\begin{center}
  \papertitle 
\end{center}
}

\vspace*{1.5cm}

\begin{center}
\paperauthors\footnote{Authors are listed at the end of this paper.}
\end{center}

\vspace{\fill}

\begin{abstract}
  \noindent
  Decays of $\mathit{\Xi}_{b}^-$ and $\mathit{\Omega}_{b}^-$ baryons to $\mathit{\Lambda}_{c}^+ h^- h^{\prime -}$ final states, with $h^- h^{\prime -}$ being $\pi^-\pi^-$, $K^-\pi^-$ and $K^-K^-$ meson pairs, are searched for using data collected with the LHCb detector.
  The data sample studied corresponds to an integrated luminosity of $8.7\,\mathrm{fb}^{-1}$ of $pp$ collisions collected at centre-of-mass energies $\sqrt{s} = 7$, $8$ and $13\,\mathrm{Te\kern -0.1em V}$.
  The products of the relative branching fractions and fragmentation fractions for each signal mode, relative to the $B^- \to \mathit{\Lambda}_{c}^+ \overline{p} \pi^-$ mode, are measured, with $\mathit{\Xi}_{b}^- \to\mathit{\Lambda}_{c}^+ K^- \pi^-$, $\mathit{\Xi}_{b}^- \to\mathit{\Lambda}_{c}^+ K^- K^-$ and $\mathit{\Omega}_{b}^- \to\mathit{\Lambda}_{c}^+ K^- K^-$ decays being observed at over $5\,\sigma$ significance.
  The $\mathit{\Xi}_{b}^- \to\mathit{\Lambda}_{c}^+ K^- \pi^-$ mode is also used to measure the $\mathit{\Xi}_{b}^-$ production asymmetry, which is found to be consistent with zero.
  In addition, the $B^- \to \mathit{\Lambda}_{c}^+ \overline{p} K^-$ decay is observed for the first time, and its branching fraction is measured relative to that of the $B^- \to \mathit{\Lambda}_{c}^+ \overline{p} \pi^-$ mode.
\end{abstract}

\vspace*{1.5cm}

\begin{center}
  Published in JHEP 08 (2024) 132
\end{center}

\vspace{\fill}

{\footnotesize 
\centerline{\copyright~\papercopyright. \href{\paperlicenceurl}{\paperlicence}.}}
\vspace*{2mm}

\end{titlepage}


\newpage
\setcounter{page}{2}
\mbox{~}

\renewcommand{\thefootnote}{\arabic{footnote}}
\setcounter{footnote}{0}

\cleardoublepage


\pagestyle{plain} 
\setcounter{page}{1}
\pagenumbering{arabic}


\section{Introduction}
\label{sec:introduction}

The large production rate of $b$ hadrons in LHC collisions has enabled a vastly increased range of $b$-baryon decays to be studied.
The majority of measurements in this sector have so far been made with \Lb\ baryons, with only relatively few \Xibm\ and \Omegab\ decays observed~\cite{LHCb-PAPER-2016-008,LHCb-PAPER-2016-050,LHCb-PAPER-2016-053,LHCb-PAPER-2020-017,LHCb-PAPER-2021-012,LHCb-PAPER-2022-053,LHCb-PAPER-2023-008,LHCb-PAPER-2023-015,CMS:2024rbi}.
New and improved measurements of the production and decays of $b$~baryons will constrain QCD models and also reduce systematic uncertainties in future measurements of the properties of related decays.
In particular, knowledge of production asymmetries is necessary to make precise searches for \CP\ violation in $b$-baryon decays~\cite{LHCb-PAPER-2020-017}.
Existing measurements of the production rates and asymmetries of \Xibm\ baryons, obtained using the $\Xibm\to\jpsi \Xires^-$ decay~\cite{LHCb-PAPER-2018-047}, have large uncertainties so improvements are well motivated.
Additionally, the phase-space distribution of multibody $b$-baryon decays can be probed to investigate the spectroscopy of charm baryons, including studies of potentially exotic hadrons.

\begin{figure}[!b]
\centering
\includegraphics[trim=0 45 0 0 , clip, scale=0.25]{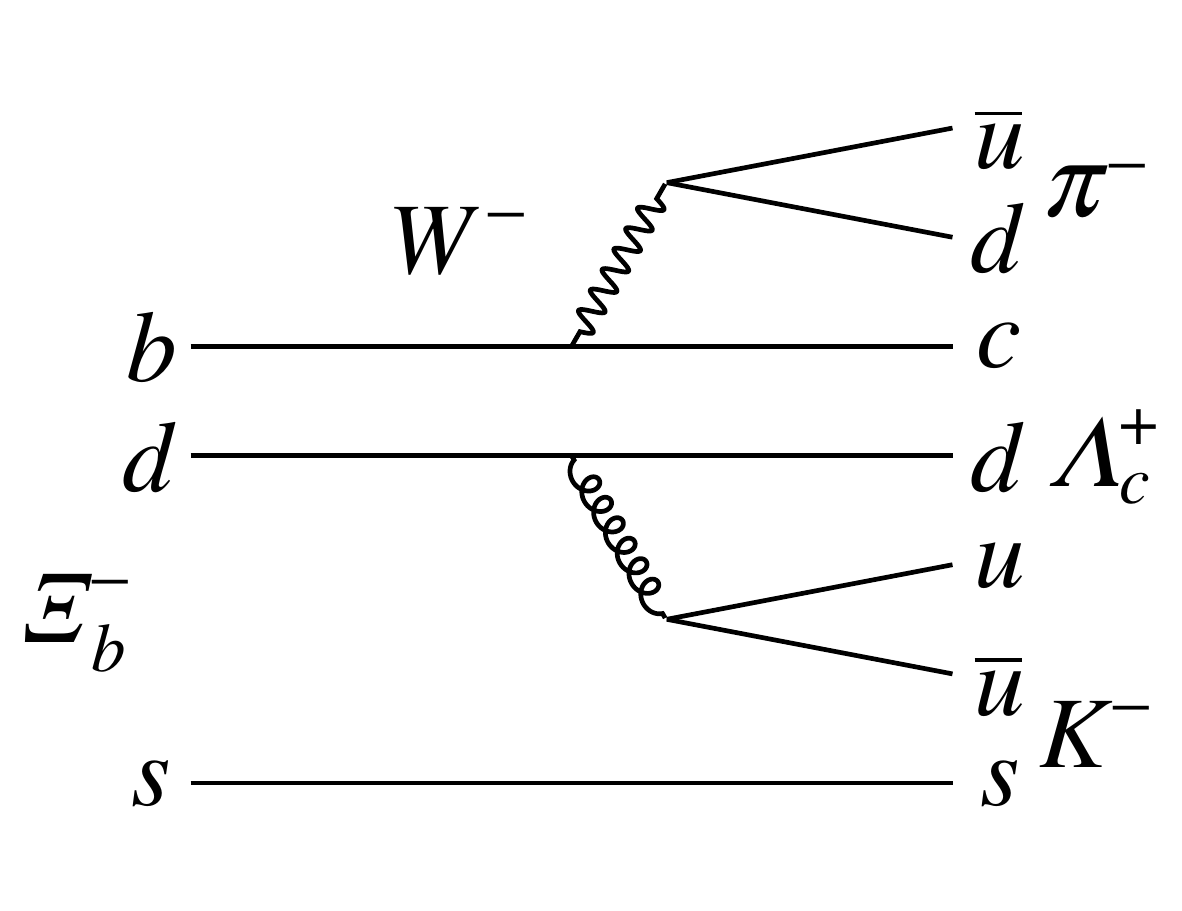}
\includegraphics[trim=0 45 0 0 , clip, scale=0.25]{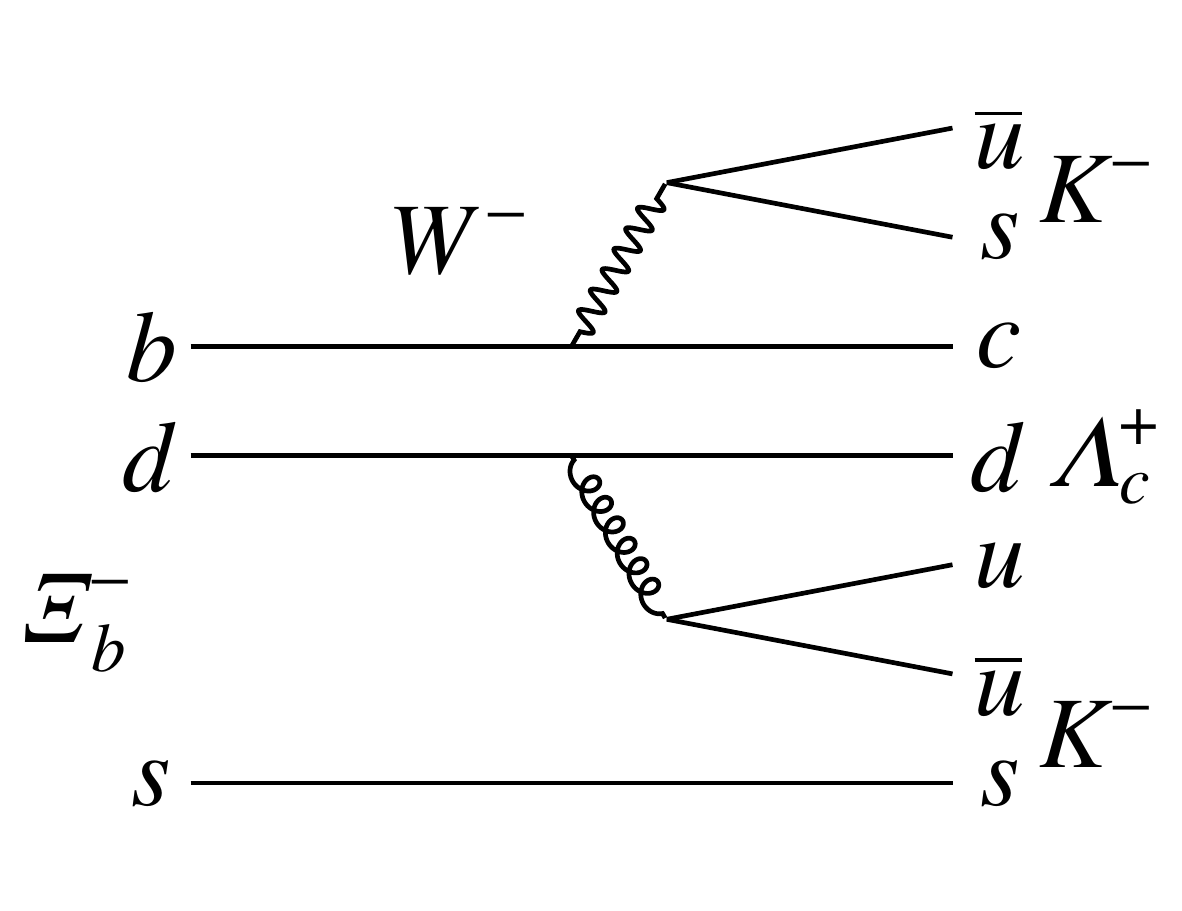}
\includegraphics[trim=0 45 0 0 , clip, scale=0.25]{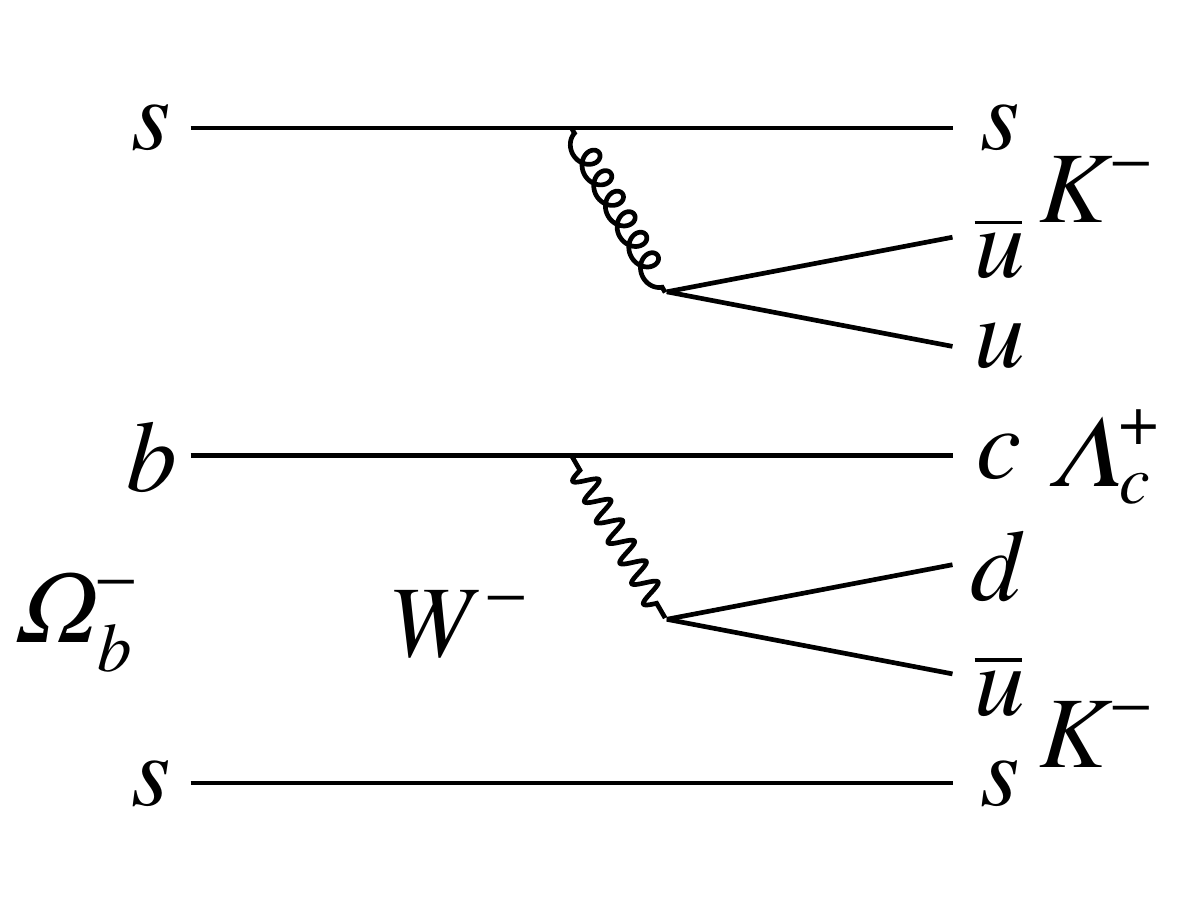}
\caption{\small 
  Examples of decay diagrams for the (left)~$\mathit{\Xi}_{b}^-\to\mathit{\Lambda}_{c}^+ K^-\pi^-$, (centre)~$\mathit{\Xi}_{b}^-\to\mathit{\Lambda}_{c}^+ K^- K^-$ and (right)~$\mathit{\Omega}_{b}^-\to\mathit{\Lambda}_{c}^+ K^- K^-$ channels.
}
\label{fig:dcydgms}
\end{figure}

At present there are no results on \Xibm\ and \Omegab\ decays to $\Lc h^- h^{\prime -}$ final states, where $h$ and $h^\prime$ are $K$ or $\pi$ mesons.
Figure~\ref{fig:dcydgms} shows possible decay diagrams for three such processes, specifically those where the branching fractions are expected to be largest. 
The $\Xibm\to\Lc\Km\pim$ decay shown in Fig.~\ref{fig:dcydgms}~(left) and the $\Omegab\to\Lc\Km\Km$ mode shown in Fig.~\ref{fig:dcydgms}~(right) are Cabibbo-favoured processes, while the $\Xibm\to\Lc\Km\Km$ decay shown in Fig.~\ref{fig:dcydgms}~(centre) is Cabibbo-suppressed. 
No tree-level diagram exists for the $\Xibm\to\Lc\pim\pim$, $\Omegab\to\Lc\Km\pim$ and $\Omegab\to\Lc\pim\pim$ decays, so these are expected to be significantly suppressed. 

In this paper, a search for $\Xibm$ and $\Omegab$ decays to the $\Lc h^- h^{\prime -}$ final states is described.
The analysis is based on the full \runone\ ($3\invfb$ of $pp$ collisions at $\sqrt{s}=7$ and $8\tev$) and \runtwo\ ($5.7\invfb$ at $\sqrt{s}=13\tev$) data sample collected with the LHCb detector.
The $\Bm\to\Lc\antiproton \pim$ decay, which has been previously observed~\cite{Belle:2004dmq,BaBar:2008get}, is used as a control and normalisation channel.
The previously unobserved $\Bm\to\Lc\antiproton \Km$ decay is also studied as part of the analysis.
In all cases, the $\Lc$ baryon is reconstructed in the $\proton\Km\pip$ final state.
The inclusion of charge-conjugate processes is implied throughout, except where asymmetries are discussed.
The analysis is performed blind, such that the signal mass regions for both the \Xibm\ ($5735$--$5865\mevcc$) and \Omegab\ ($5977$--$6107\mevcc$) baryons are not inspected until analysis procedures are established; the \Bm\ modes, however, are not blinded during the analysis.

Ratios of fragmentation fractions times branching fractions of two channels are determined from ratios of efficiency-corrected yields.
In the case that decays of the same $b$ hadron, $X_b$, to final states generically labelled $Y$ and $Z$ are considered, the ratio of branching fractions is determined as 
\begin{equation}\label{eq:BFratio:def}
  \frac{{\cal B}\left(X_b \to Y\right)}{{\cal B}\left(X_b \to Z\right)} = 
  \frac{N(Y)/\left<\epsilon(Y)\right>}{N(Z)/\left<\epsilon(Z)\right>} = 
  \frac{\sum_{i=0}^{n_Y} w^Y_i /\epsilon^Y_i}{\sum_{i=0}^{n_Z} w^Z_i /\epsilon^Z_i} = 
  \frac{N^{\rm corr}(Y)}{N^{\rm corr}(Z)} \,.
\end{equation}
The yield obtained from the fit to the $X_b$-candidate mass distribution, $N$, is equal to $\sum_{i=0}^{n} w_i$ where $w_i$ are signal weights~\cite{Pivk:2004ty} obtained from the same fit for all the $n$ candidates in the fit.  
Variation of the efficiency across the phase space of the decay is taken into account by use of candidate-by-candidate efficiencies, $\epsilon_i$, which depend on position in the decay phase-space and year of data taking. 
The efficiencies are determined from simulation with corrections, discussed below, applied in order to describe data more accurately. 
The efficiency-corrected yield is then $N^{\rm corr} = \sum_{i=0}^{n} w_i /\epsilon_i$, and an effective average efficiency can be defined as $\left<\epsilon\right> = N/N^{\rm corr}$.
In case the yield of a particular channel is not significant, the candidate-by-candidate efficiency correction of Eq.~\eqref{eq:BFratio:def} will not be reliable, and an average efficiency obtained directly from simulation with an assumption on the phase-space distribution is used instead.
In this case, the effect of the unknown phase-space distribution will be a source of systematic uncertainty.

When decays of different $b$ hadrons are considered, Eq.~\eqref{eq:BFratio:def} is modified so that the left-hand side is multiplied by the appropriate ratio of fragmentation fractions, but the procedure is otherwise identical.
The fragmentation fractions quantify the probability that a $b$ quark produced in the $pp$ collision hadronises into a particular $b$ hadron.
The fragmentation fractions can depend on $pp$ collision centre-of-mass energy, and therefore all ratios are determined separately for \runone\ and \runtwo.
Combined results are obtained where appropriate, taking correlations of systematic uncertainties into account.

For a channel $X_b \to Y$ with sufficient yield, the charge-conjugate processes can be separated to determine the $X_b$ production asymmetry, ${\cal A}_{\rm prod}$, from the measured efficiency-corrected yield asymmetry,
\begin{eqnarray}
  {\cal A}_{\rm meas}\left(X_b \to Y\right) & = & 
  \frac{\sum_{i=0}^{n_Y} \frac{w_i}{\epsilon_i} - \sum_{i=0}^{n_{\bar{Y}}} \frac{w_i}{\bar{\epsilon}_i}}
  {\sum_{i=0}^{n_Y} \frac{w_i}{\epsilon_i} + \sum_{i=0}^{n_{\bar{Y}}} \frac{w_i}{\bar{\epsilon}_i}} = 
  \frac{N^{\rm corr}(X_b \to Y) - N^{\rm corr}(\bar{X}_b \to \bar{Y})}
  {N^{\rm corr}(X_b \to Y) + N^{\rm corr}(\bar{X}_b \to \bar{Y})} \,,\label{eq:Aprod:def} \\
  & = & {\cal A}_{\CP}\left(X_b \to Y\right) + {\cal A}_{\rm prod}\left(X_b\right) \,. \nonumber
\end{eqnarray}
Here, the weights $w_i$ in all sums are obtained from a single fit to the sample including both charge conjugate final states, but sums over different candidates are performed for the two states.
The sense of the asymmetry is defined as being between the $X_b$ hadron that contains a $b$ quark and its $\bar{X}_b$ antiparticle that contains a $\bar{b}$ quark.
The efficiency maps used to obtain the $N^{\rm corr}$ values also differ for the two charge conjugate final states, and are denoted as $\epsilon$ and $\bar{\epsilon}$.
The use of separate efficiency maps corrects for detection asymmetry, although potential mismodelling of $\epsilon$ and $\bar{\epsilon}$ must be accounted for as a source of systematic uncertainty.
For decays where no \CP\ violation is expected, \ie\ where ${\cal A}_{\CP} = 0$ can be assumed, one has ${\cal A}_{\rm meas} = {\cal A}_{\rm prod}$.
This is the case for the channels studied here, which are expected to be dominated by tree-level diagrams as shown in Fig.~\ref{fig:dcydgms}.

Production asymmetries, and the relative fragmentation fractions of different hadrons, can depend on kinematics.
The above approach can be extended to study such effects, by making measurements in bins of, for example, $b$-hadron transverse momentum.
In this case the sum in Eq.~\eqref{eq:Aprod:def} includes only events in the relevant bin, and the efficiency maps are those appropriate for that bin.

\section{Detector and simulation}
\label{sec:Detector}

The \lhcb detector~\cite{LHCb-DP-2008-001,LHCb-DP-2014-002} is a single-arm forward spectrometer covering the \mbox{pseudorapidity} range $2<\eta <5$, designed for the study of particles containing \bquark or \cquark quarks.
The detector includes a high-precision tracking system consisting of a silicon-strip vertex detector surrounding the $pp$ interaction region~\cite{LHCb-DP-2014-001}, a large-area silicon-strip detector located upstream of a dipole magnet with a bending power of about $4\,{\mathrm{T\,m}}$, and three stations of silicon-strip detectors and straw drift tubes~\cite{LHCb-DP-2013-003,LHCb-DP-2017-001} placed downstream of the magnet.
The tracking system provides a measurement of the momentum, \ptot, of charged particles with relative uncertainty that varies from 0.5\% at low momentum to 1.0\% at $200\gevc$.
The minimum distance of a track to a primary $pp$ collision vertex (PV), the impact parameter (IP), is measured with a resolution of $(15+29/\pt)\mum$, where \pt is the component of the momentum transverse to the beam, in $\gevc$.
Different types of charged hadrons are distinguished using information from two ring-imaging Cherenkov detectors~\cite{LHCb-DP-2012-003}. 
Photons, electrons and hadrons are identified by a calorimeter system consisting of scintillating-pad and preshower detectors, an electromagnetic and a hadronic calorimeter.
Muons are identified by a system composed of alternating layers of iron and multiwire proportional chambers~\cite{LHCb-DP-2012-002}.

The online event selection is performed by a trigger~\cite{LHCb-DP-2012-004,LHCb-DP-2019-001}, which consists of a hardware stage, based on information from the calorimeter and muon systems, followed by a software stage, which applies a full event reconstruction.
At the hardware trigger stage, events are required to have a muon with high \pt or a hadron, photon or electron with high transverse energy in the calorimeters.
For hadrons, the transverse energy threshold is $3.5\gev$.
The software trigger requires a two-, three- or four-track secondary vertex with a significant displacement from any primary $pp$ interaction vertex.
At least one charged particle must have a transverse momentum $\pt > 1.6\gevc$ and be inconsistent with originating from a PV.
A multivariate algorithm~\cite{BBDT,LHCb-PROC-2015-018} is used for the identification of secondary vertices consistent with the decay of a \bquark hadron.
In the offline selection, trigger signals are associated with reconstructed particles.
Selection requirements can therefore be made on the trigger selection itself and on whether the decision was due to the signal candidate, other particles produced in the $pp$ collision, or a combination of both.

Simulation is used to model the effects of the detector acceptance and the imposed selection requirements, as well as to study potential backgrounds.
In the simulation, $pp$ collisions are generated using \pythia~\cite{Sjostrand:2007gs,*Sjostrand:2006za} with a specific \lhcb configuration~\cite{LHCb-PROC-2010-056}.
The signal processes are generated with the \texttt{FLATSQDALITZ} model, which generates events uniformly in the square Dalitz plot as described in Ref.~\cite{Back:2017zqt}.
Decays of unstable particles are described by \evtgen~\cite{Lange:2001uf}, in which final-state radiation is generated using \photos~\cite{davidson2015photos}.
The interaction of the generated particles with the detector, and its response, are implemented using the \geant\ toolkit~\cite{Allison:2006ve,*Agostinelli:2002hh} as described in Ref.~\cite{LHCb-PROC-2011-006}. 
The underlying $pp$ interaction is reused multiple times, with an independently generated signal decay for each~\cite{LHCb-DP-2018-004}.
The response of the ring-imaging Cherenkov detectors in simulation is known to not match the response in data, so the corresponding particle identification (PID) variables in the simulation samples are corrected based on calibration data samples of $\Dstarp\to\Dz\pip$, $\Dz \to \Km\pip$ decays for kaons and pions, and $\Lc \to \proton\Km\pip$ decays for protons~\cite{LHCb-DP-2018-001}.
The correction technique transforms the variables such that they match the data distributions, accounting for dependence on track kinematics and event multiplicity, while retaining correlations between different PID variables. 

\section{Selection}
\label{sec:selection}

Candidate signal and control channel decays are formed from five final-state tracks, where three of the tracks are consistent with originating from the \Lc\ decay and the other two are consistent with originating from the $b$-hadron decay.
The \Lc\ candidates are required to have reconstructed mass in the range $2260 < m(p\Km\pip) < 2310 \mevcc$.
The hardware trigger decision is required to have been caused by either calorimeter clusters associated with one or more of the final-state particles, or by particles produced in the $pp$ bunch crossing not involved in forming the $X_b$ candidate.
Candidates with clone tracks, reconstructed from the same detector hits, are removed.
The candidate is associated with the PV in the $pp$ bunch crossing with which it forms the smallest value of \chisqip, defined as the difference in \chisq of a given PV reconstructed with and without the considered particle.
Candidates are selected with an initial filtering, to reduce the sample size to a manageable level while maintaining high efficiency, followed by the use of multivariate algorithms to reject background processes in an optimal way.
Two dedicated boosted decision tree (BDT) algorithms, trained using the {\tt XGBoost} method~\cite{DBLP:journals/corr/ChenG16} separately for the \runone\ and \runtwo\ samples, are used to separate signal from combinatorial background.
Particle identification variables are used to suppress backgrounds involving misidentified final-state particles, including cross-feed between the different signal final states.
The same BDT requirements are used to select $\Xibm \to \Lc h^- h^{\prime -}$, $\Omegab \to \Lc h^- h^{\prime -}$ and $\Bm\to\Lc\antiproton h^-$ decays; only PID requirements distinguish the different final states.

The first BDT algorithm is designed to distinguish true $\Lc \to p\Km\pip$ decays originating from $b$-hadron decays from random combinations of tracks that produce fake \Lc\ candidates.
It uses as inputs the \Lc\ candidate's \chisqip, together with the significance of the separation from the \Lc\ decay vertex of each of the PV and the $X_b$ decay vertex, and the angle between the \Lc-candidate momentum vector and the vector from the PV to the \Lc\ decay vertex.
Variables that characterise the goodness of fit of each of the three tracks and their assumed common vertex are also included, together with the highest \pt\ of the three tracks and PID variables for each of them.
This BDT classifier is trained with a signal sample of the three \Xibm\ decays combined taken from simulation and a background sample taken from sideband regions of the $\Lc$-candidate mass distribution in data, defined as $2190<m(p\Km\pip)<2241\mevcc$ and $2331<m(p\Km\pip)<2380\mevcc$. 
A loose requirement is applied on the first BDT output to remove a significant fraction of the fake \Lc\ backgrounds.

The second BDT classifier suppresses combinatorial background composed of a true \Lc\ hadron with two random tracks. 
The input variables are the $X_b$-candidate's \chisqip\ and \pt, together with the significance of the separation between the $X_b$ decay vertex and the PV and the \chisq\ of the $X_b$ vertex.
The highest \pt\ among the $X_b$ decay products and the sum of their \chisqip\ values are also used, together with a variable that characterises the isolation of the candidate relative to other particles produced in the $pp$ collision in terms of a $\pt$ asymmetry~\cite{LHCb-PAPER-2012-001}. 
Finally, the first BDT output is taken as an input to the second.
This second BDT classifier is trained using a signal sample taken from simulation and a background sample taken from sideband regions of the $X_b$-candidate mass distribution in data, defined as $5400<m(X_b)<5742\mevcc$ and $5852<m(X_b)<7500\mevcc$.

The PID variables for \Lc\ decay products are included in the first BDT algorithm, and are not used further in the selection.
For the tracks originating directly from the $X_b$ decay, kaon and pion candidates are distinguished with requirements in the two-dimensional plane of the PID variables that characterise how well they fit each of those particle hypotheses.
These requirements are optimised as described below, and are required to be mutually exclusive so that no candidate can appear in more than one final state.
To select antiprotons in the $\Bm\to\Lc\antiproton h^-$ decays, a requirement on the consistency with the proton hypothesis is imposed.  
In order to ensure that reasonable PID information is available, fiducial requirements are imposed on the kinematics of all final-state tracks: $2.5< \ptot_{K,\pi} <100 \gevc$, $9.5< \ptot_p <100 \gevc$ and $2 < \eta_{p,K,\pi} < 5$. 
In addition, all tracks are required to be inconsistent with being muons, based on information from the muon system.  

The requirement on the second BDT output is optimised using the following figure of merit (FOM)~\cite{Punzi}, 
\begin{equation}
{\rm FOM}_{\rm BDT} = \frac{\epsilon_{\rm sig}}{(\frac{5}{2} + \sqrt{B})}\,,
\label{eq:fom_eq}
\end{equation}
where $B$ is the expected number of combinatorial background events in the signal region (\mbox{$5770 < m(X_b) < 5830\mevcc$}), obtained by extrapolating yields in data in the upper mass sideband (\mbox{$5830 < m(X_b) < 6800\mevcc$}), and $\epsilon_{\rm sig}$ is the signal efficiency determined from simulation.
The requirements on the kaon and pion identification variables are optimised using 
\begin{equation}
{\rm FOM}_{\rm PID} = \frac{\epsilon_{\rm sig}}{\epsilon_{\rm sig} + \sum_i r_{i}  \epsilon_{{\rm misID}\,i}}\,,
\label{eq:fom_eq2}
\end{equation}
where $\epsilon_{{\rm misID}\,i}$ is the misidentification rate for each potential cross-feed background $i$, taken from simulation with PID corrections applied, and $r_{i} = {\cal B}({\rm misID}\,i)/{\cal B}({\rm sig})$ is an estimate of the branching fraction of the cross-feed background, ${\cal B}({\rm misID}\,i)$, relative to that of the signal mode, ${\cal B}({\rm sig})$.
The optimisation of the requirements on the second BDT output and the PID variables is iterated to obtain the best overall performance.
The optimised requirements on the second BDT output reject around 99\% of background while retaining 40--50\% of signal depending on the final state and data-taking period.

The $\Lc\Km\pim$ and $\Lc\antiproton\Km$ samples contain misreconstructed candidates where the \Km\ from the \Lc\ decay is swapped with that directly from the $X_b$ decay. 
These are vetoed by removing candidates that satisfy $2260 < m(p_{\Lc}\Km\pi^+_{\Lc}) <2310 \mevcc$, where the subscript $\Lc$ indicates that the particle is one of the $\Lc$ decay products while the absence of a subscript indicates that the particle originates directly from the $X_b$ decay. 
Backgrounds also occur due to $\Lb \to \Lc\pim$ or $\Lb \to \Lc\Km$ decays combined with an extra track.
This source includes true \mbox{$\Xibm \to \Lb \pim$} decays~\cite{LHCb-PAPER-2023-015}, which are considered a background to this analysis, as well as cases where the extra track does not originate from the same \mbox{$b$-hadron} decay.
Vetoes that remove candidates with $5500 < m(\Lc\pim)<5700 \mevcc$ and $5500 < m(\Lc\Km)<5700 \mevcc$ are applied in the relevant final states to remove these backgrounds.
The loss in signal efficiency due to these veto requirements is considered as a source of systematic uncertainty.

Potential backgrounds from $X_b$ decays to the same final-state particles without intermediate \Lc\ baryons are investigated by inspecting the $X_b$-candidate mass distribution for candidates with \Lc-candidate masses both below and above the region populated by signal.
No significant contribution is visible and therefore no additional selection requirements are imposed to remove such backgrounds.

Backgrounds are expected from $b$-hadron decays to final states of $\Lc h^- h^{\prime -}$ plus one or more low-momentum particles, subsequently referred to as partially reconstructed backgrounds.
These include $b$-hadron decays to \mbox{$\Sigmac^{+(++)} h^- h^{\prime -}$} final states with \mbox{$\Sigmac^{+(++)} \to \Lc \pi^{0(+)}$}.
Similar arguments hold for partially reconstructed backgrounds to the $\Bm \to \Lc\antiproton h^-$ channels.
Due to the missing particles, these backgrounds tend to peak at $X_b$-candidate mass values below the signal peaks.
In order to minimise dependence on precise modelling of these backgrounds, the mass ranges retained for the fits are $5550 < m(X_b) < 6800\mevcc$ for the signal channels and $5200 < m(X_b) < 6000\mevcc$ for the \Bm\ channels.

A small fraction ($<1\%$) of selected events contain more than one candidate.  
All are retained, and the presence of multiple candidates is considered as source of systematic uncertainty. 

\section{Determination of signal yields}
\label{sec:massfits}

Fits to the $X_b$-candidate mass distributions are performed to determine the yields of the channels under study.
Simultaneous unbinned extended maximum-likelihood fits are implemented to allow parameters to be shared between the samples. 
One simultaneous fit is performed for the three signal final states, and another for the two \Bm\ modes, with separate fits for the \runone\ and \runtwo\ samples. 
The fits include components to describe the signals, and contributions due to cross-feed, partially reconstructed decays and combinatorial background.

The signal components are described using the sum of two Crystal Ball (CB) functions~\cite{Skwarnicki:1986xj}, with a common central value and width, and independent tails to each side of the peak. 
The tail parameters and the relative fraction of the two CB functions are determined from fits to simulation samples and are fixed in the fit to data. 

Cross-feed between the signal modes occurs due to misidentification of one of the particles originating directly from the $X_b$ decay.
The $X_b$-candidate mass distributions of these backgrounds depend on the momentum and misidentification probability of that particle.  
Therefore, these shapes depend (albeit fairly weakly) on the phase-space distribution of the misidentified decay.
For the $\Bm \to \Lc\antiproton\pim$ channel there are previous investigations of its resonant structure~\cite{Belle:2004dmq,BaBar:2008get}; however these were made with samples much smaller than available in this study.
Since the analysis of the \Bm\ modes is not blind, the phase-space distributions observed in data are used to weight the simulation and obtain a best estimate of the shape of the cross-feed background components.
On the other hand, since the analysis is performed blind for the $\Xibm$ and $\Omegab$ baryon decays, their phase-space distributions are taken from simulation without any phase-space weighting.
This procedure is considered as a source of systematic uncertainty.
The distribution for each sample under each $\kaon \leftrightarrow \pion$ misidentification hypothesis is fitted with a double CB function.
Misidentification of protons as pions or kaons, and vice versa, does not lead to background contributions in the fitted mass ranges.  

Residual partially reconstructed backgrounds due to the decays $\Xibm\to\Sigmacp h^- h^{\prime -}$ and \mbox{$\Xibz\to\Sigmacpp h^- h^{\prime -}$} are described as a single component in the fit.
The shape is investigated with a simplified simulation~\cite{Cowan:2016tnm}, and is modelled with a superposition of Gaussian kernels~\cite{Cranmer:2000du}.
Partially reconstructed backgrounds to the \Bm\ modes are found to be negligible in the chosen mass fit range.
The combinatorial background is modelled with an exponential shape. 
The slope of the exponent in each final state is a free parameter in the fit. 

The yields of all signal, partially reconstructed, and combinatorial components are free parameters in the simultaneous fit to the three signal final states.
The yields of cross-feed backgrounds are constrained relative to the signal yield of the same decay when it is correctly reconstructed, based on knowledge of the relative misidentification rates and correct identification efficiencies.
The shapes of all components are fixed, except that the slope parameters of the exponential function used to describe combinatorial background in each final state are free to vary independently of each other.
In addition, a peak position offset and a width scale factor, which account for possible discrepancies between data and simulation and that are shared between all components described by a double CB function, are free parameters of the fit.
In total there are seventeen free parameters in each simultaneous fit to the signal final states: six signal yields, three combinatorial background yields and three slope parameters, three partially reconstructed background yields, and the peak shift and width scale parameters.
There are, in addition, ten Gaussian constraints that control the yields of the cross-feed backgrounds.

A similar configuration is used for the simultaneous fit to the two \Bm\ modes.
In this case there are eight free parameters: two signal yields, two combinatorial background yields and two slope parameters, as well as the peak shift and width scale parameters.
In addition, there are two Gaussian constraints that control the yields of the cross-feed backgrounds.

The performance of each fit is evaluated using ensembles of pseudoexperiments.
For the fit to the signal final states, this is done prior to unblinding with guessed signal yields, and subsequently repeated with the yields observed in data.
For the \Bm\ modes the yields observed in data are used in the generation of the pseudoexperiments.
In all cases, the fits are found to return the signal yields with reasonable coverage and without large biases; small possible biases are accounted for as a source of systematic uncertainty.

\begin{figure}[!tb]
\centering
\includegraphics[scale=0.39]{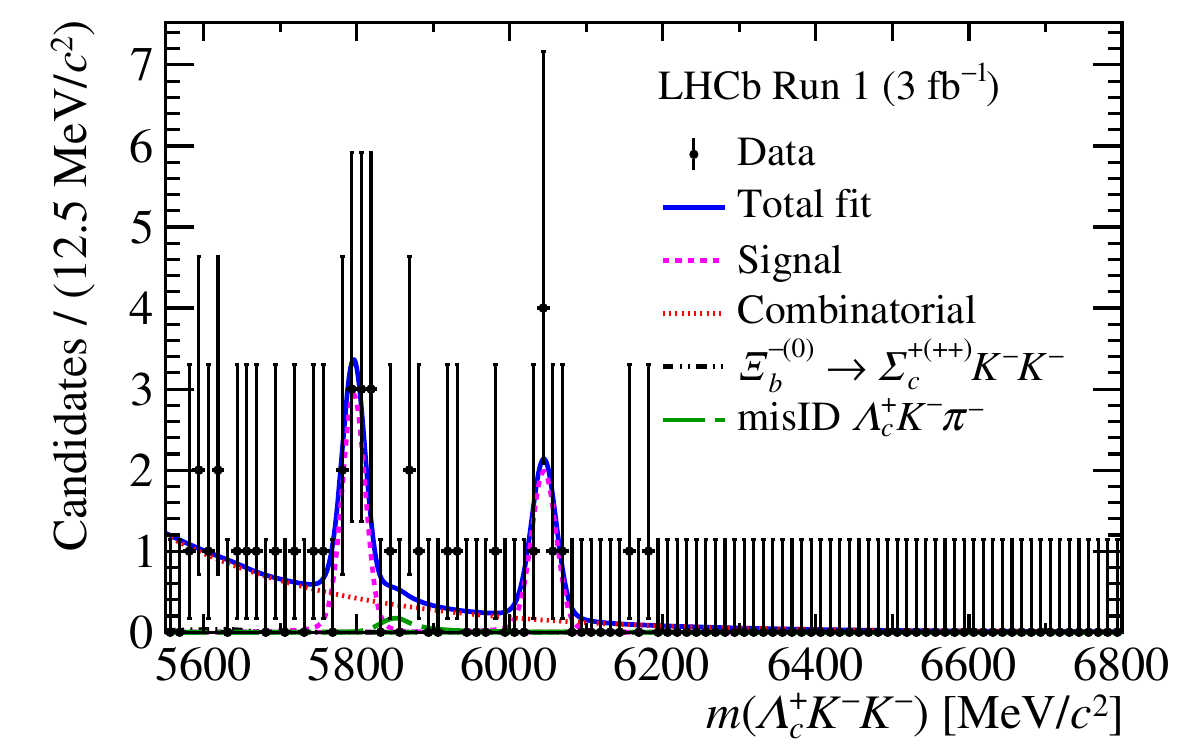}
\includegraphics[scale=0.39]{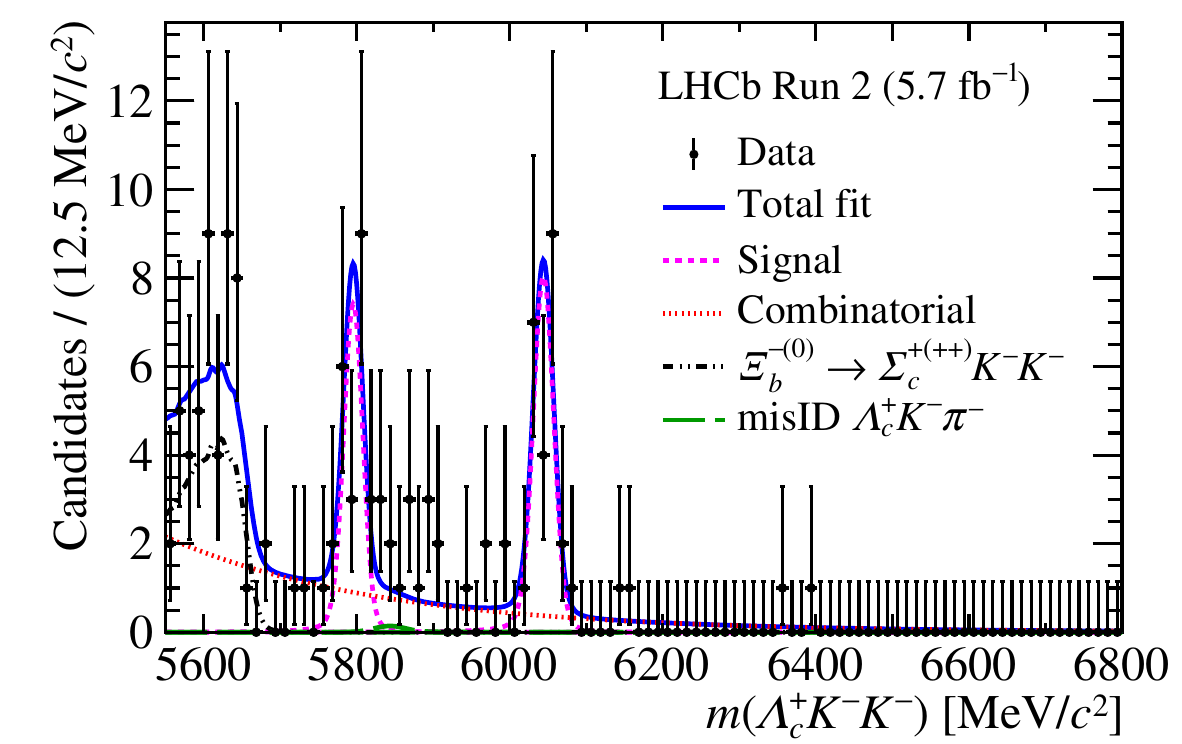} 
\includegraphics[scale=0.39]{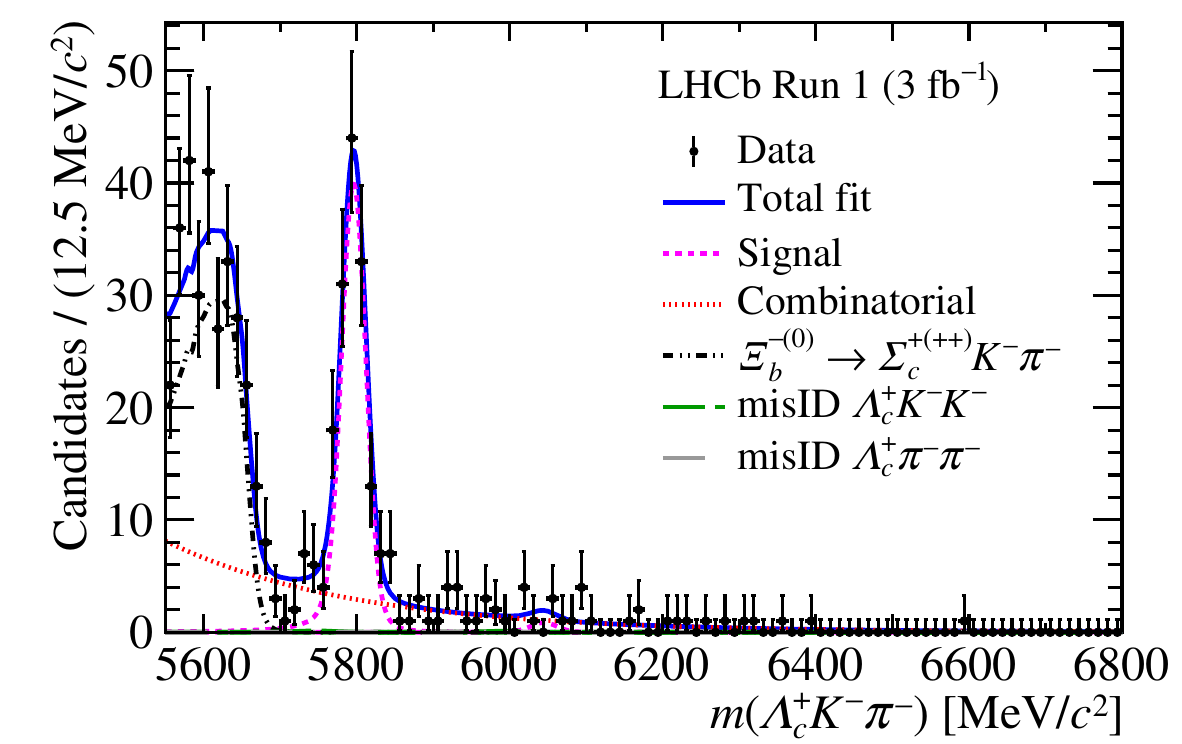}
\includegraphics[scale=0.39]{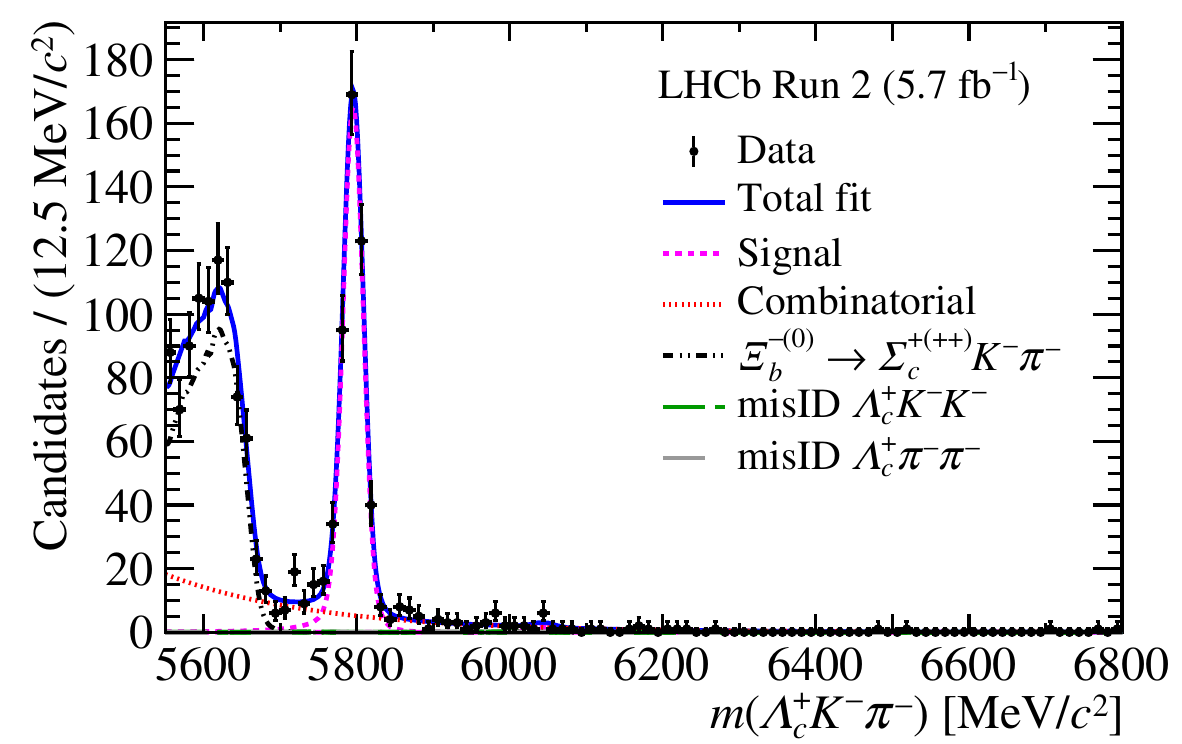}
\includegraphics[scale=0.39]{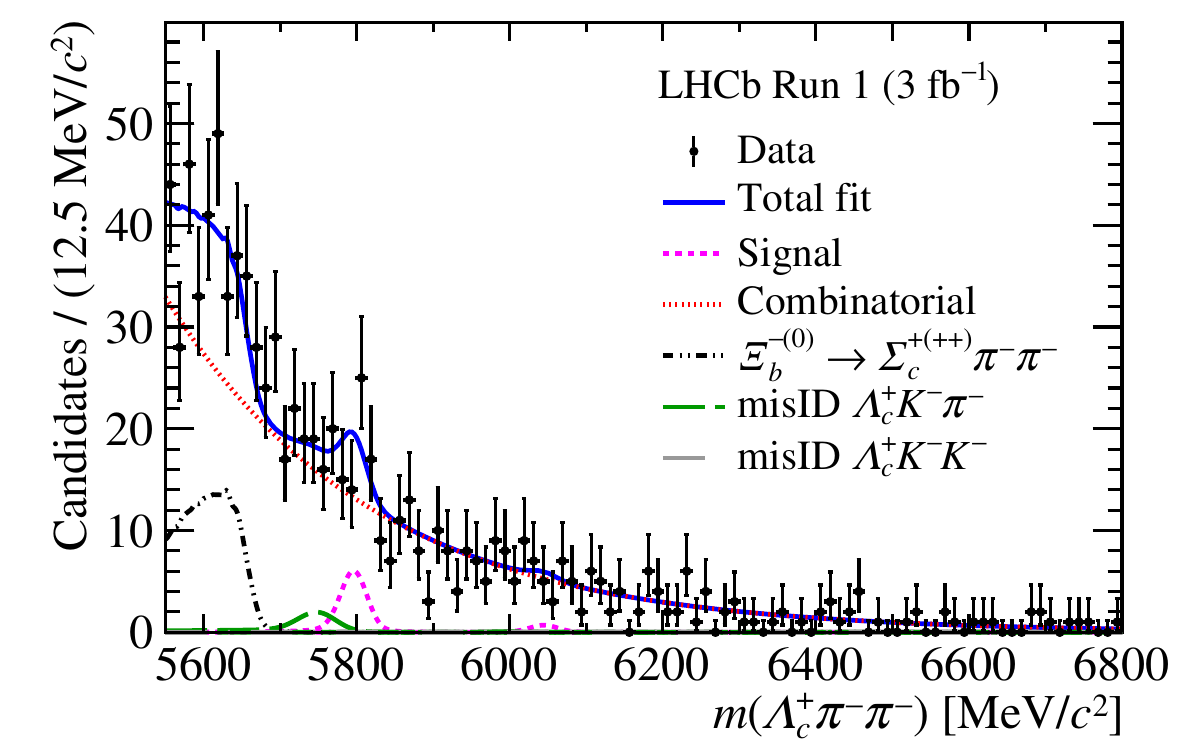}
\includegraphics[scale=0.39]{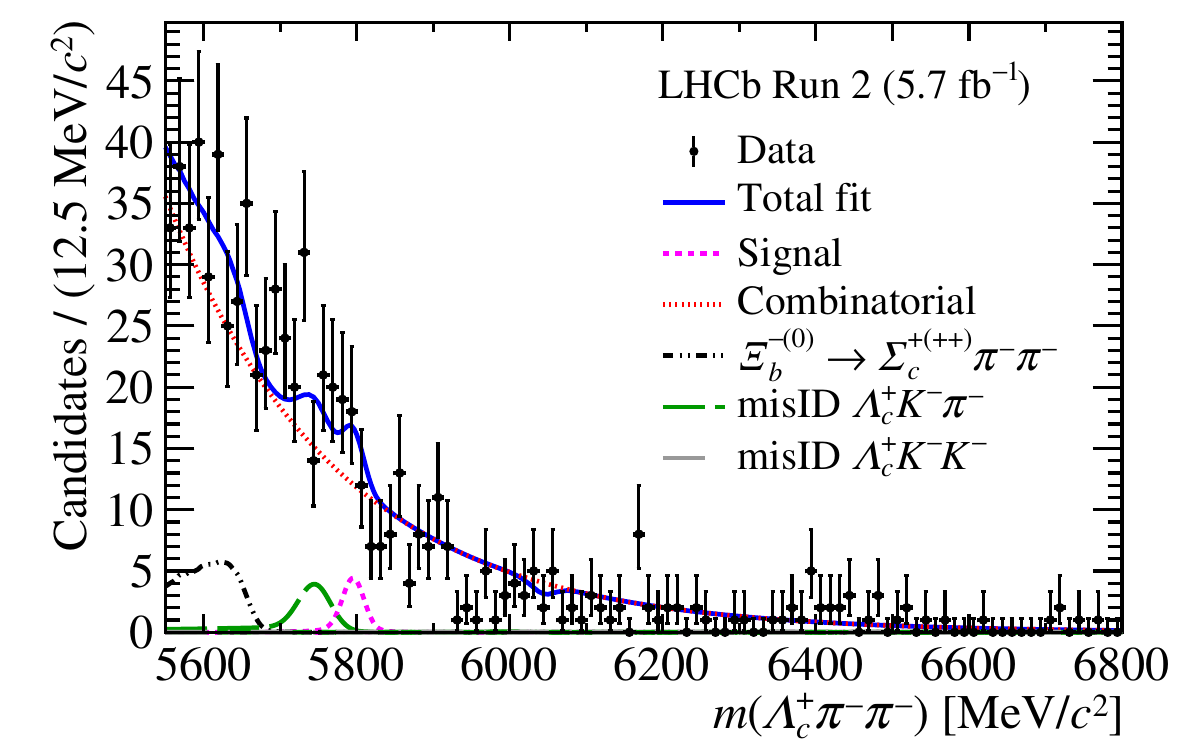}
\caption{\small 
    Mass distributions of the signal channels with results of the fit superimposed: (top)~$\mathit{\Lambda}_{c}^+ K^- K^-$, (middle)~$\mathit{\Lambda}_{c}^+ K^-\pi^-$ and (bottom)~$\mathit{\Lambda}_{c}^+\pi^-\pi^-$ final states for (left)~Run~1 and (right)~Run~2.
    The different components in the fit are shown as indicated in the legends, where the signal components include independent contributions from $\mathit{\Xi}_{b}^-$ and $\mathit{\Omega}_{b}^-$ decays.
}
\label{fig:SM_simMfit_unblinded}
\end{figure}

\begin{figure}[!tb]
\centering
\includegraphics[scale=0.39]{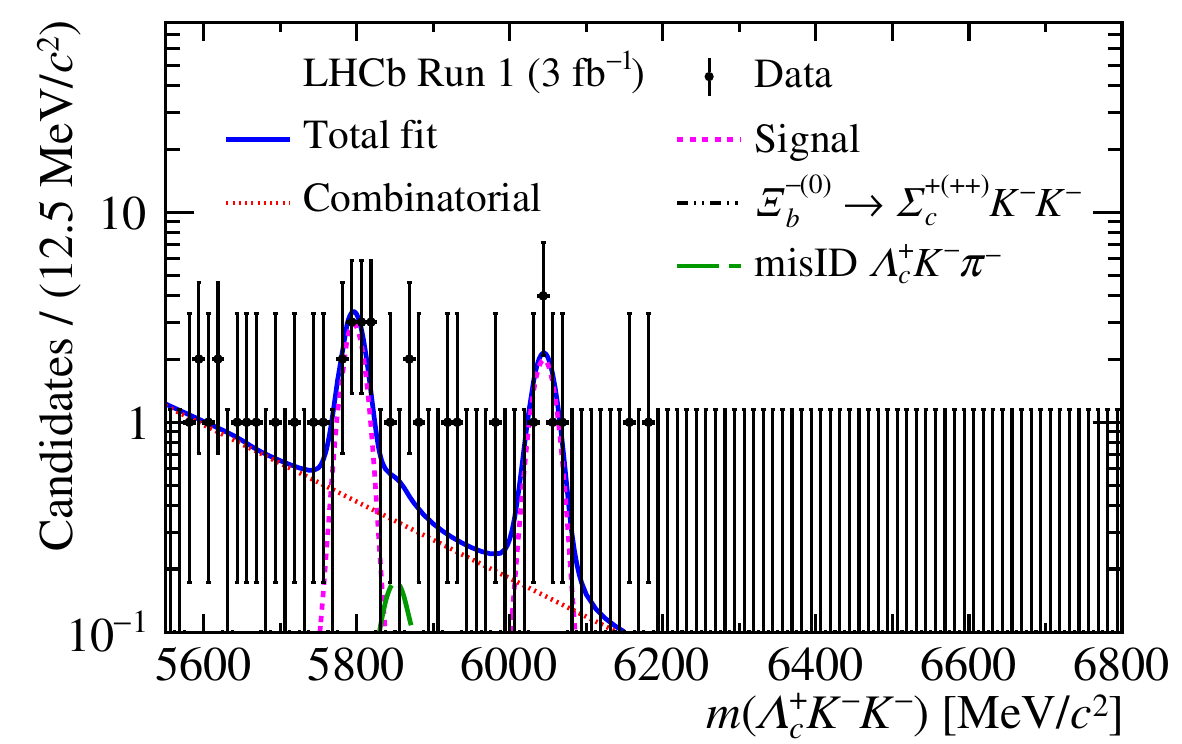}
\includegraphics[scale=0.39]{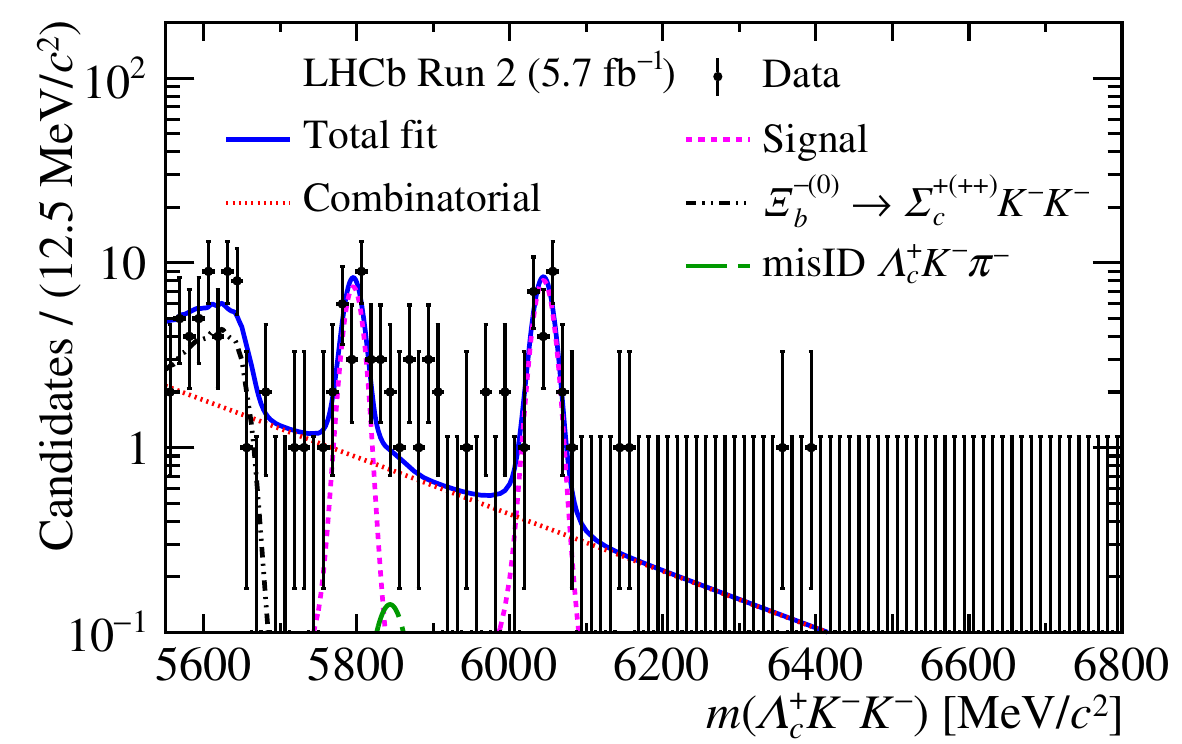}
\includegraphics[scale=0.39]{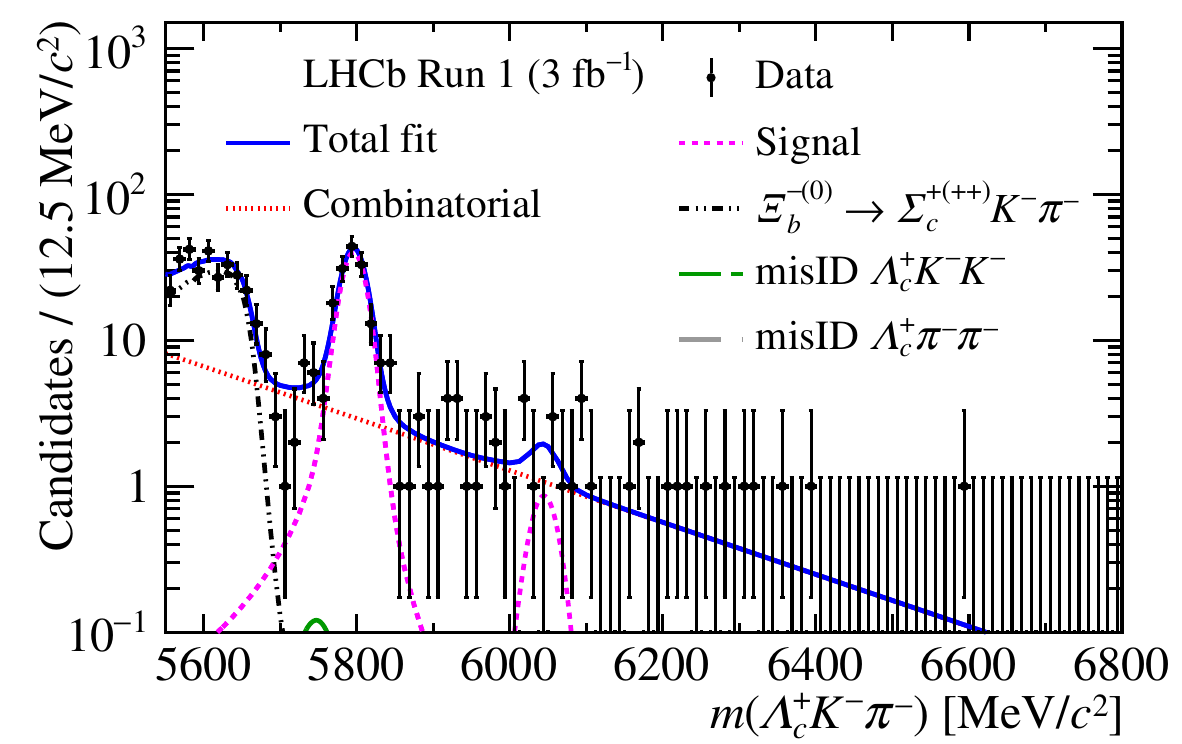}
\includegraphics[scale=0.39]{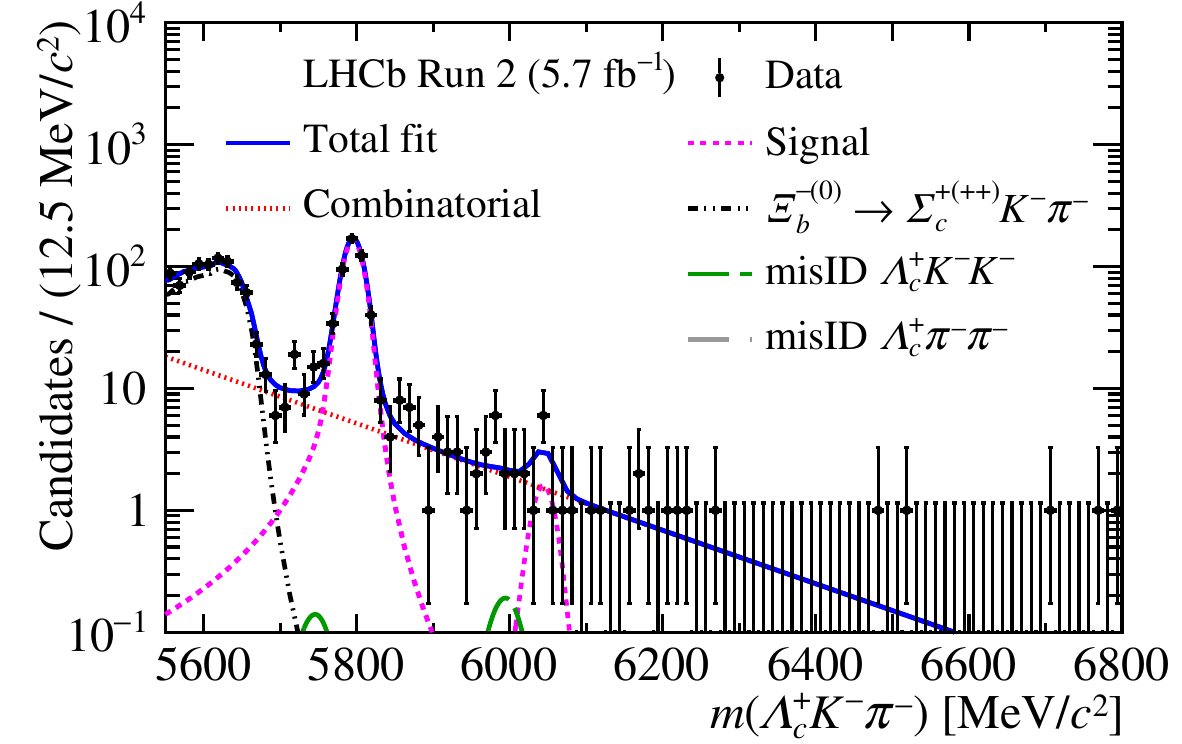}
\includegraphics[scale=0.39]{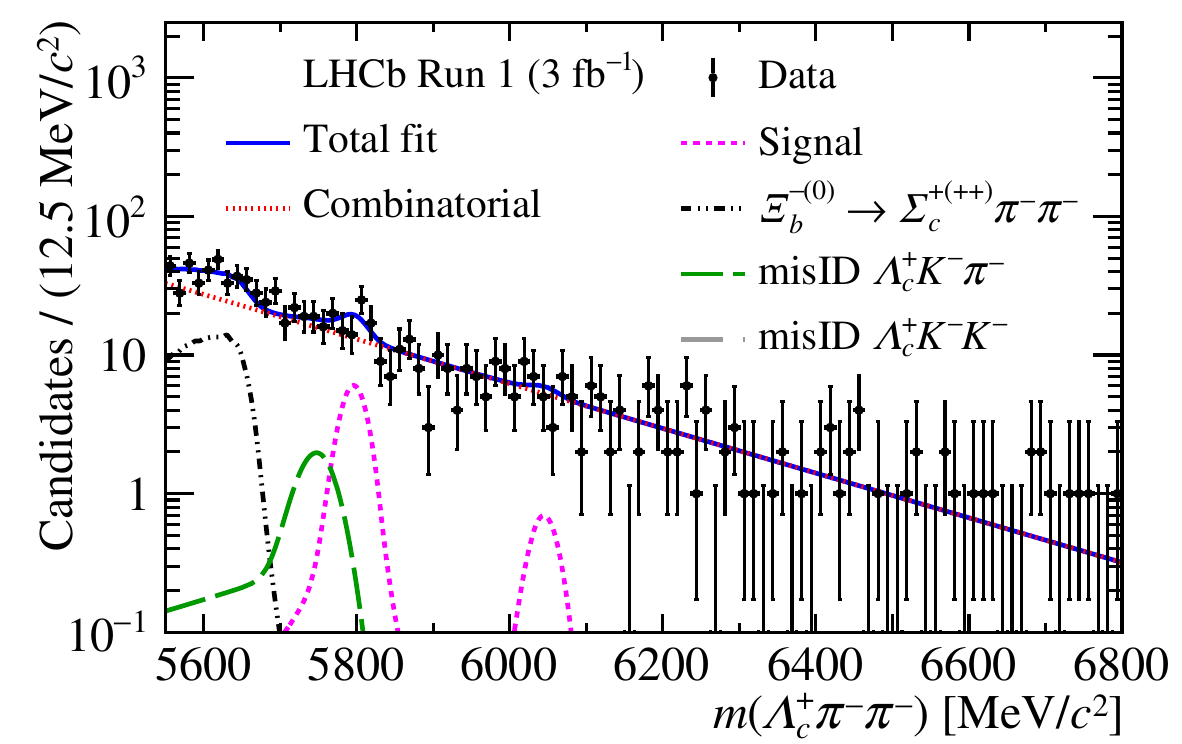}
\includegraphics[scale=0.39]{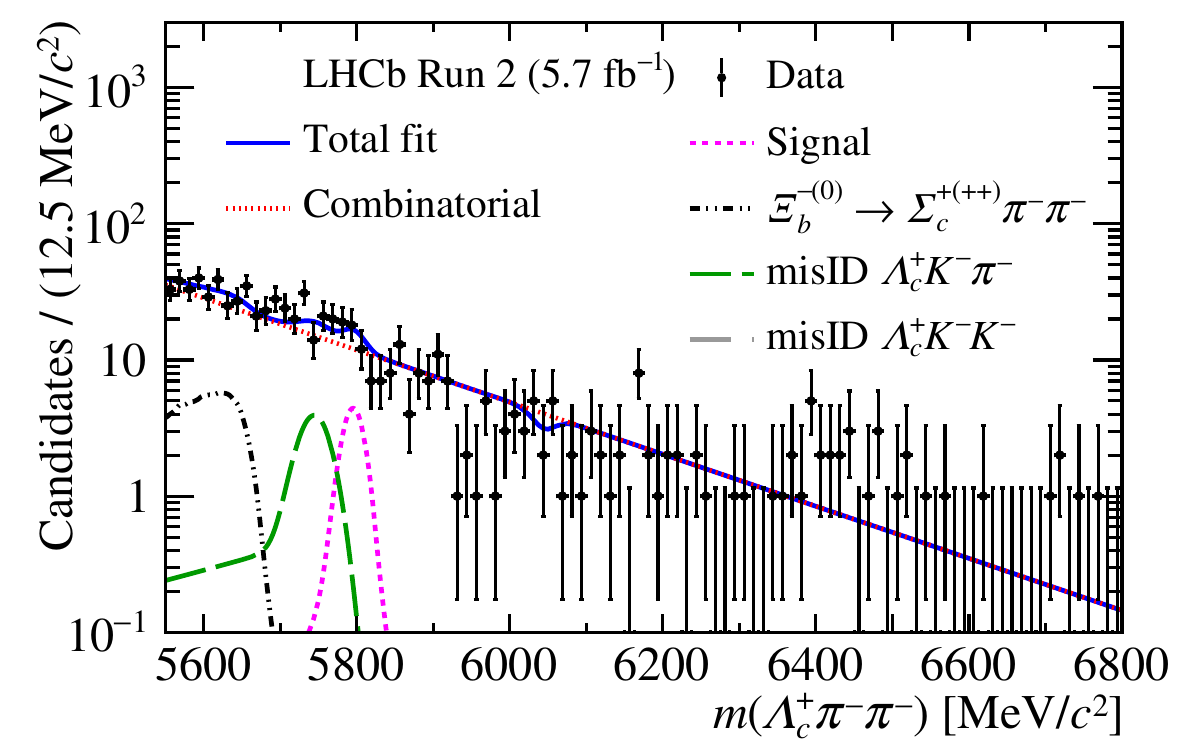}
\caption{\small 
    Mass distributions of the signal channels with results of the fit superimposed: (top)~$\mathit{\Lambda}_{c}^+ K^- K^-$, (middle)~$\mathit{\Lambda}_{c}^+ K^-\pi^-$ and (bottom)~$\mathit{\Lambda}_{c}^+\pi^-\pi^-$ final states for (left)~Run~1 and (right)~Run~2 (log scale).
    The different components in the fit are shown as indicated in the legends, where the signal components include independent contributions from $\mathit{\Xi}_{b}^-$ and $\mathit{\Omega}_{b}^-$ decays.
}
\label{fig:SM_simMfitlog_unblinded}
\end{figure}

\begin{figure}[!tb]
\centering
\includegraphics[scale=0.39]{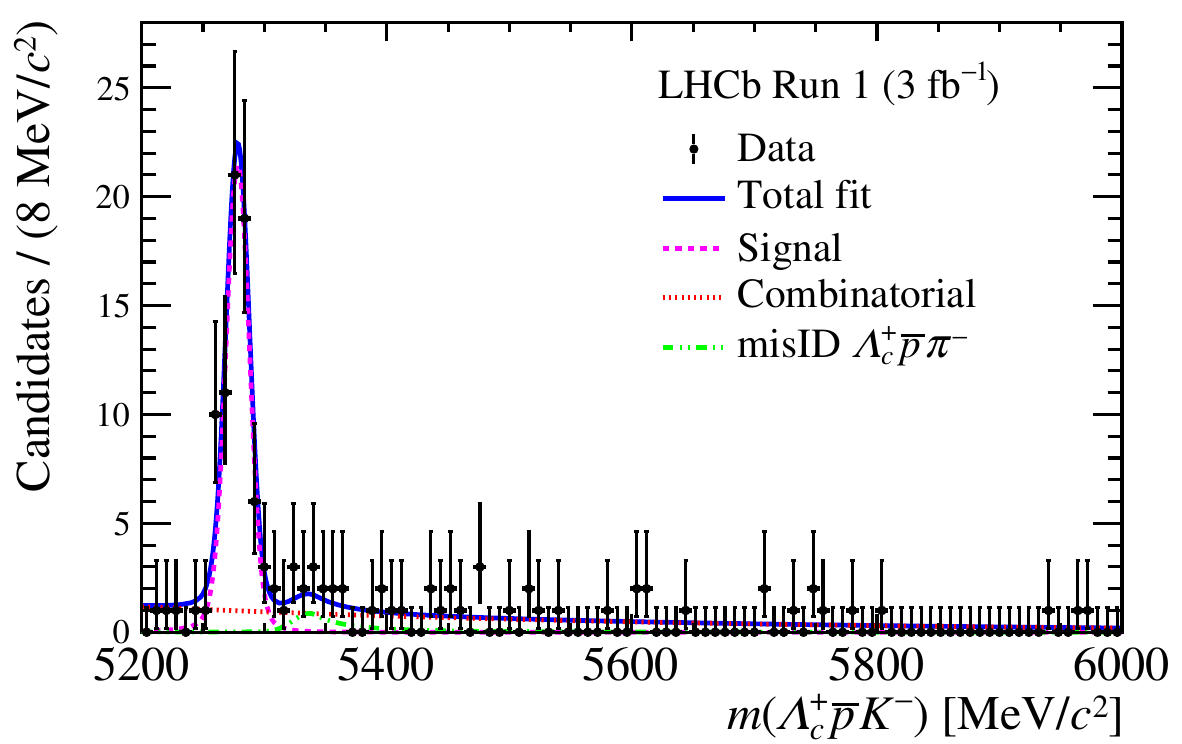}
\includegraphics[scale=0.39]{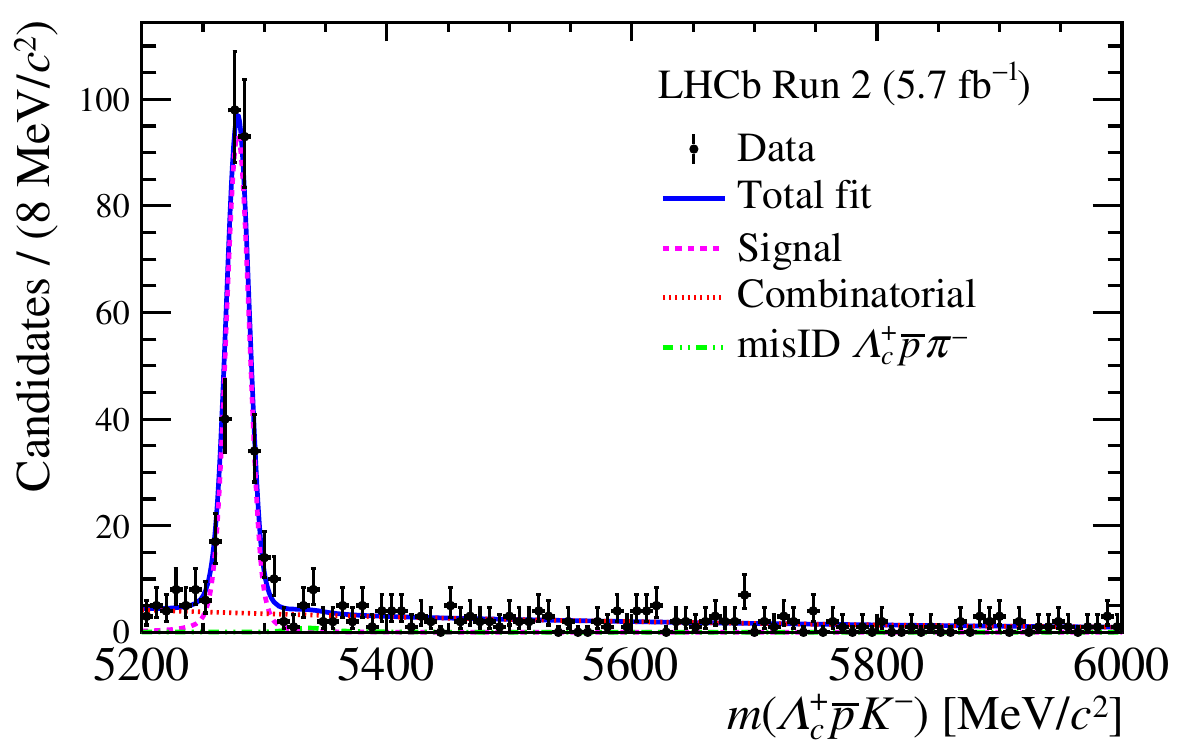}
\includegraphics[scale=0.39]{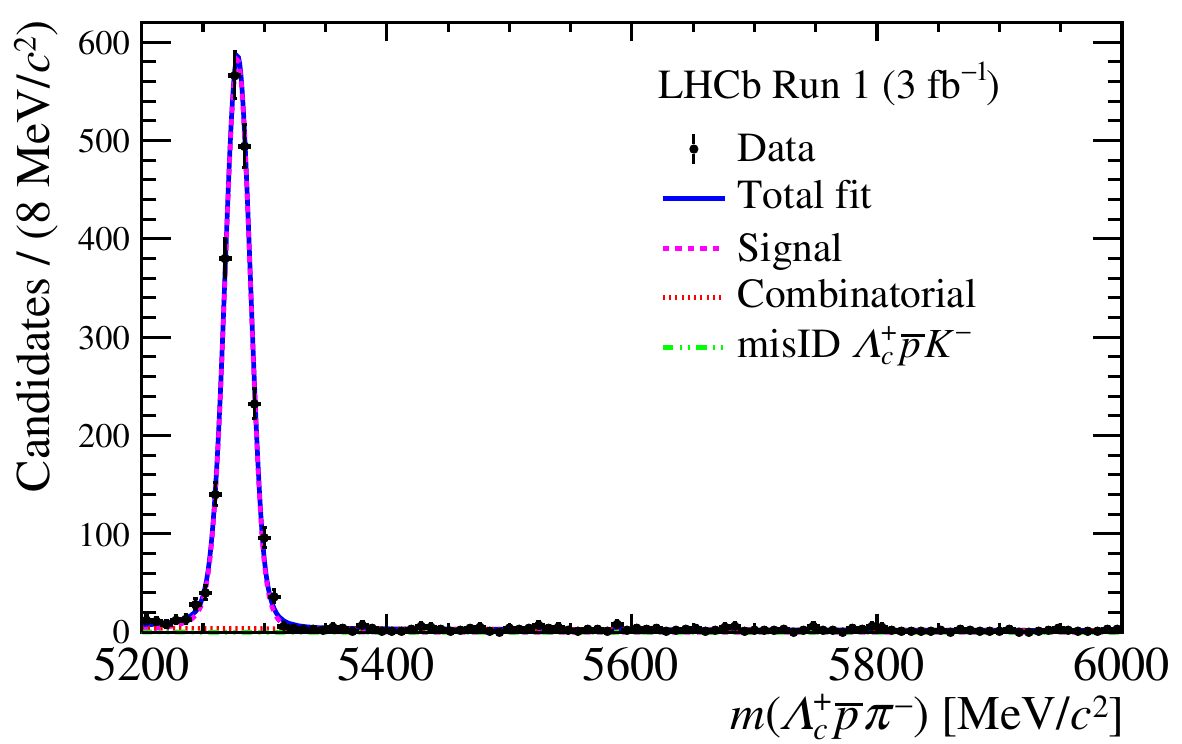}
\includegraphics[scale=0.39]{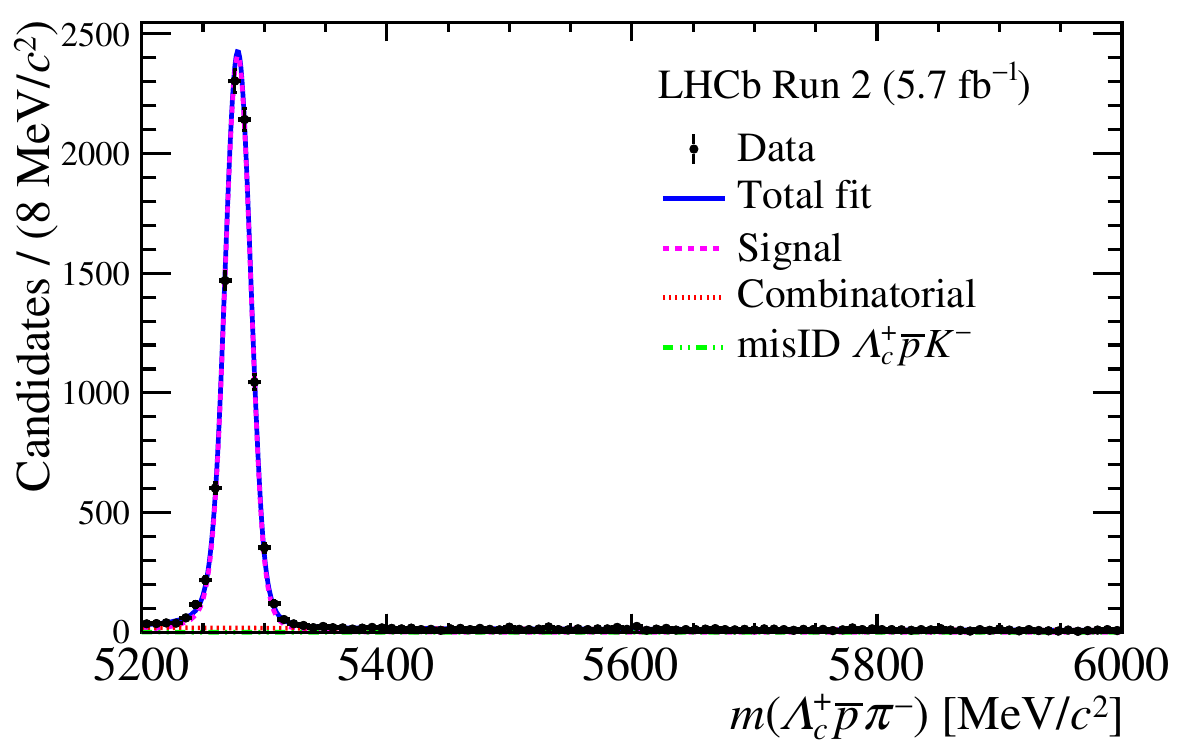}
\caption{\small 
    Mass distributions of the $B^-$ modes with results of the fit superimposed: (top)~$\mathit{\Lambda}_{c}^+\overline{p} K^-$ and (bottom)~$\mathit{\Lambda}_{c}^+\overline{p}\pi^-$ final states for (left)~Run~1 and (right)~Run~2.
    The different components in the fit are shown as indicated in the legends.
}
\label{fig:CM_simMfit}
\end{figure}

\begin{figure}[!tb]
\centering
\includegraphics[scale=0.39]{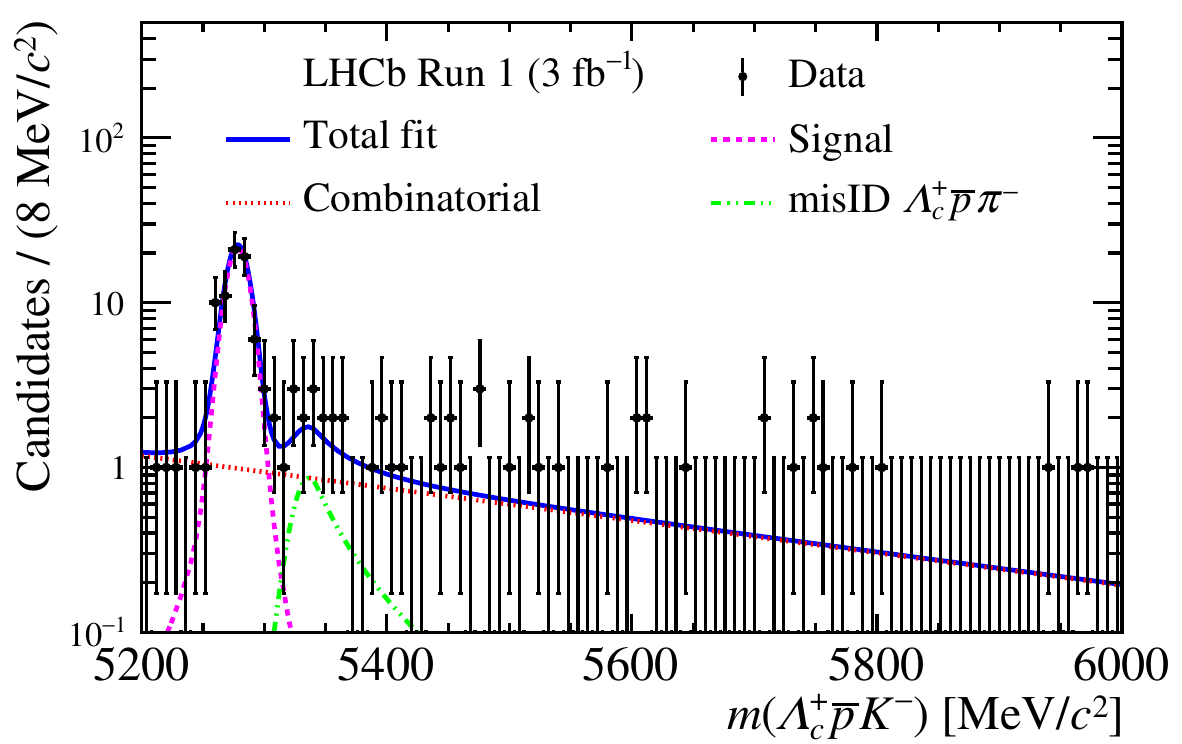}
\includegraphics[scale=0.39]{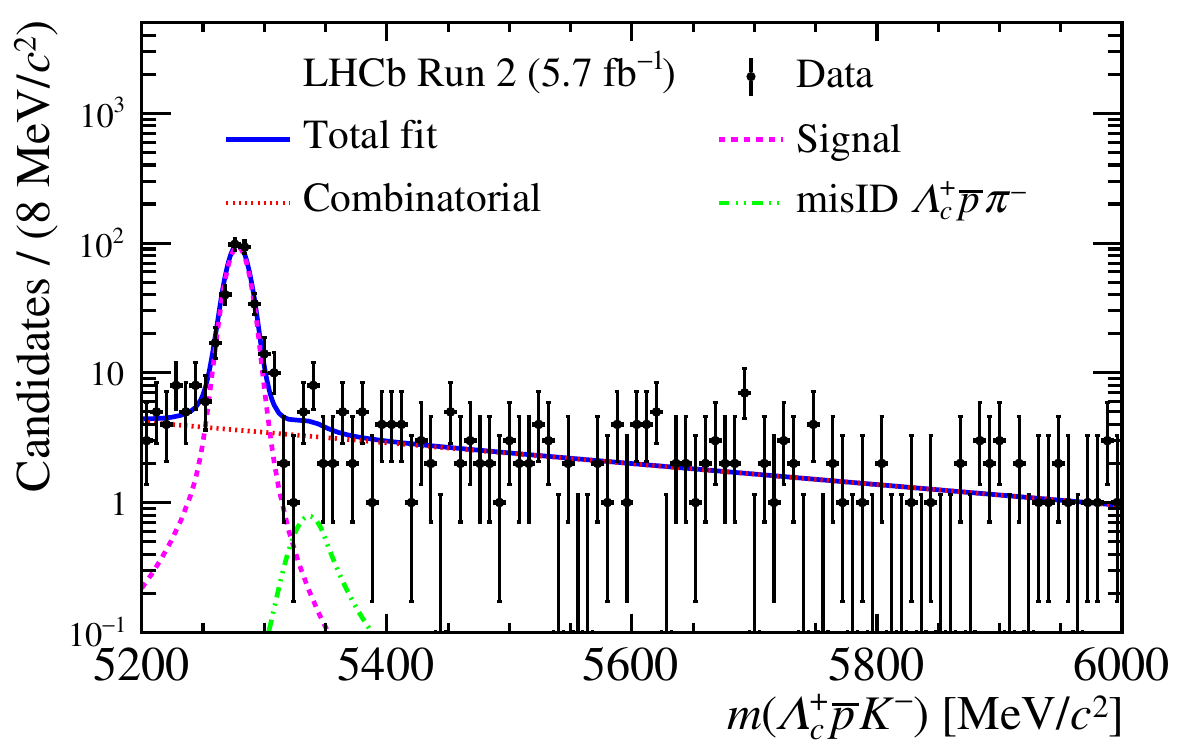}
\includegraphics[scale=0.39]{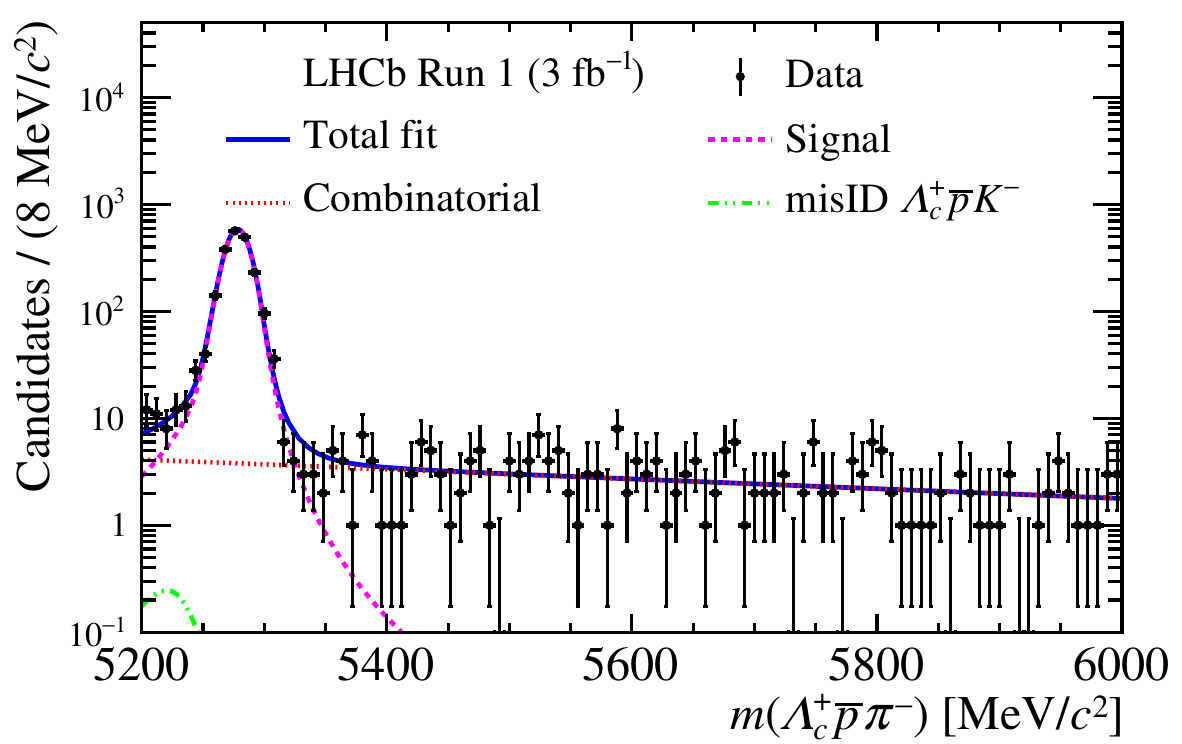}
\includegraphics[scale=0.39]{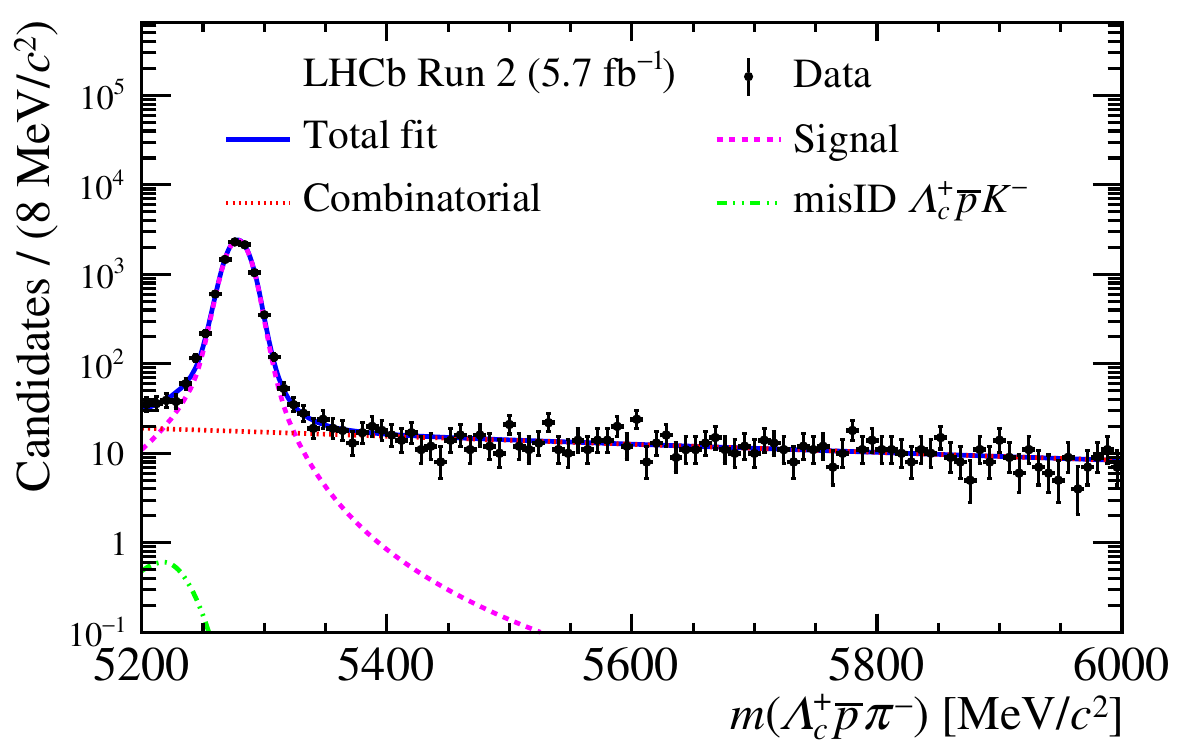}
\caption{\small 
    Mass distributions of the $B^-$ modes with results of the fit superimposed: (top)~$\mathit{\Lambda}_{c}^+\overline{p} K^-$ and (bottom)~$\mathit{\Lambda}_{c}^+\overline{p}\pi^-$ final states for (left)~Run~1 and (right)~Run~2 (log scale).
    The different components in the fit are shown as indicated in the legends.
}
\label{fig:CM_simMfitlog}
\end{figure}

The $X_b$-candidate mass distributions for the three signal final states for both \runone\ and \runtwo\ are shown in Figs.~\ref{fig:SM_simMfit_unblinded} and~\ref{fig:SM_simMfitlog_unblinded} with linear and logarithmic $y$-axis scales, respectively, with results of the fits superimposed.
The mass distributions and results of the fits to the two \Bm\ modes are similarly shown in Figs.~\ref{fig:CM_simMfit} and~\ref{fig:CM_simMfitlog}.
Pronounced peaks corresponding to $\Xibm\to\Lc\Km\pim$ and $\Bm\to\Lc\antiproton\Km$ decays, as well as the previously established $\Bm\to\Lc\antiproton\pim$ decay, are observed.
Smaller, but still clear, peaks corresponding to $\Xibm\to\Lc\Km\Km$ and $\Omegab\to\Lc\Km\Km$ decays are also visible.
The fitted signal yields are given in Table~\ref{tab:yields}.
The uncertainties on all fitted yields are found to be symmetric to within 20\% and therefore symmetrised uncertainties are quoted in Table~\ref{tab:yields}; the full likelihood curves are, however, used to set upper limits on channels with yields that are not significantly different from zero, as discussed below. 
The significance of each of the signals is discussed in Sec.~\ref{sec:results}.

\begin{table}[!tb]
\centering
\caption{
Fitted yields for $\mathit{\Xi}_{b}^-$ and $\mathit{\Omega}_{b}^-$ decays to the $\mathit{\Lambda}_{c}^+ h^- h^{\prime -}$ final states, and for $B^- \to \mathit{\Lambda}_{c}^+ \overline{p} h^-$ decays.
}
\label{tab:yields}
\renewcommand{\arraystretch}{1.1}
\begin{tabular}{lr@{$\,\pm\,$}lr@{$\,\pm\,$}l}
\hline
& \multicolumn{2}{c}{\runone} & \multicolumn{2}{c}{\runtwo} \\
\hline
$N(\Xibm\to\Lc\Km\Km$) & $9.1 $ & $ 3.6$  & $19.4 $ & $ 5.2$ \\
$N(\Xibm\to\Lc\Km\pim)$ & $137 $ & $14$ & $471$ & $24$ \\
$N(\Xibm\to\Lc\pim\pim)$   & $23$ & $11$ & $14 $ & $9$ \\
$N(\Omegab\to\Lc\Km\Km$) & $6.7 $ & $2.8 $ & $22.6 $ & $5.2$ \\
$N(\Omegab\to\Lc\Km\pim$) & $3.2 $ & $3.4$ & $ 4.9$ & $3.5$ \\
$N(\Omegab\to\Lc\pim\pim$) & $2.8$ & $6.5$ & $-3.0$ & $4.2$ \\
\hline
$N(\Bm\to\Lc\antiproton\Km)$           & $66$ & $ 9$ & $288$ & $18$ \\
$N(\Bm\to\Lc\antiproton\pim)$          & $2020$ & $ 50$ & $8410$ & $90$ \\
\hline
\end{tabular}
\end{table}

\section{Efficiency}
\label{sec:efficiency}

The total efficiency includes effects from the geometrical acceptance of the LHCb detector, the reconstruction efficiency, and the probabilities to pass both trigger stages and offline selection requirements.
All these effects are estimated using simulation, with \mbox{data-driven} corrections applied where appropriate.
In particular, the efficiencies of the PID requirements can be obtained directly from the simulation samples, since the corresponding variables are corrected using calibration data.

\begin{figure}[!tb]
    \centering
    \includegraphics[width=0.49\textwidth]{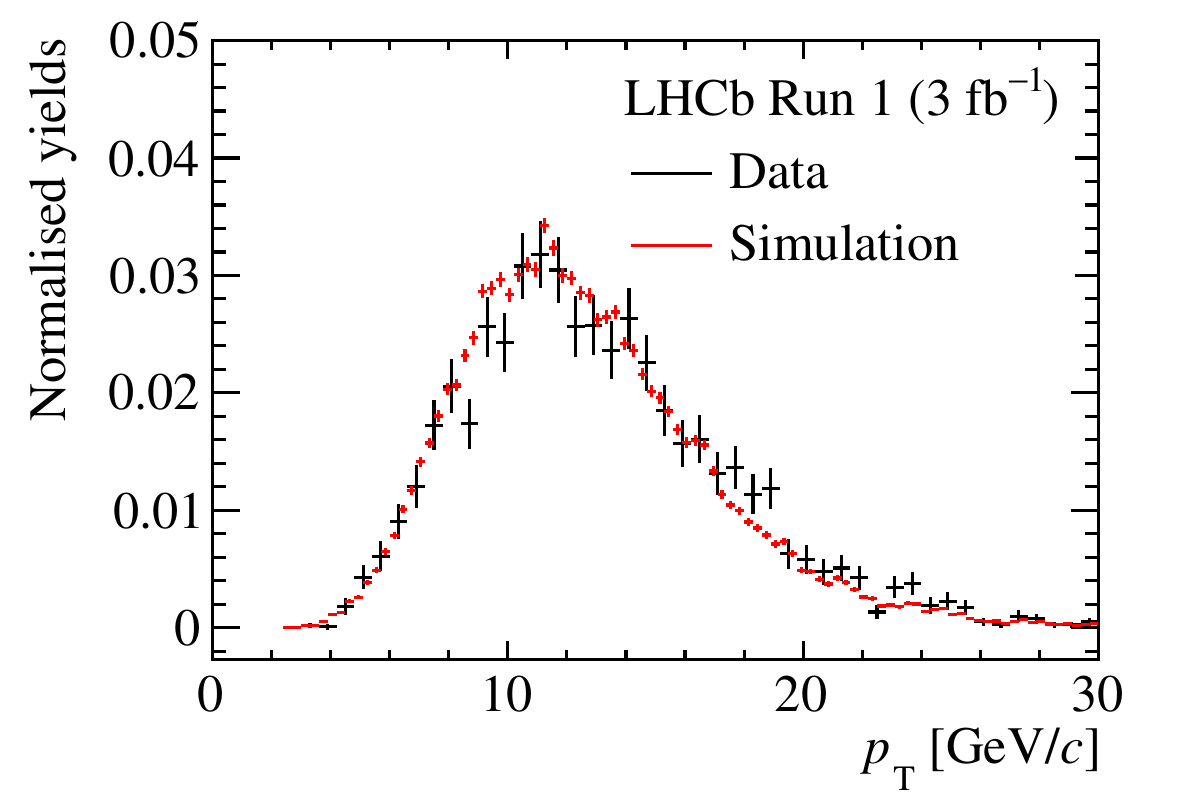}
    \includegraphics[width=0.49\textwidth]{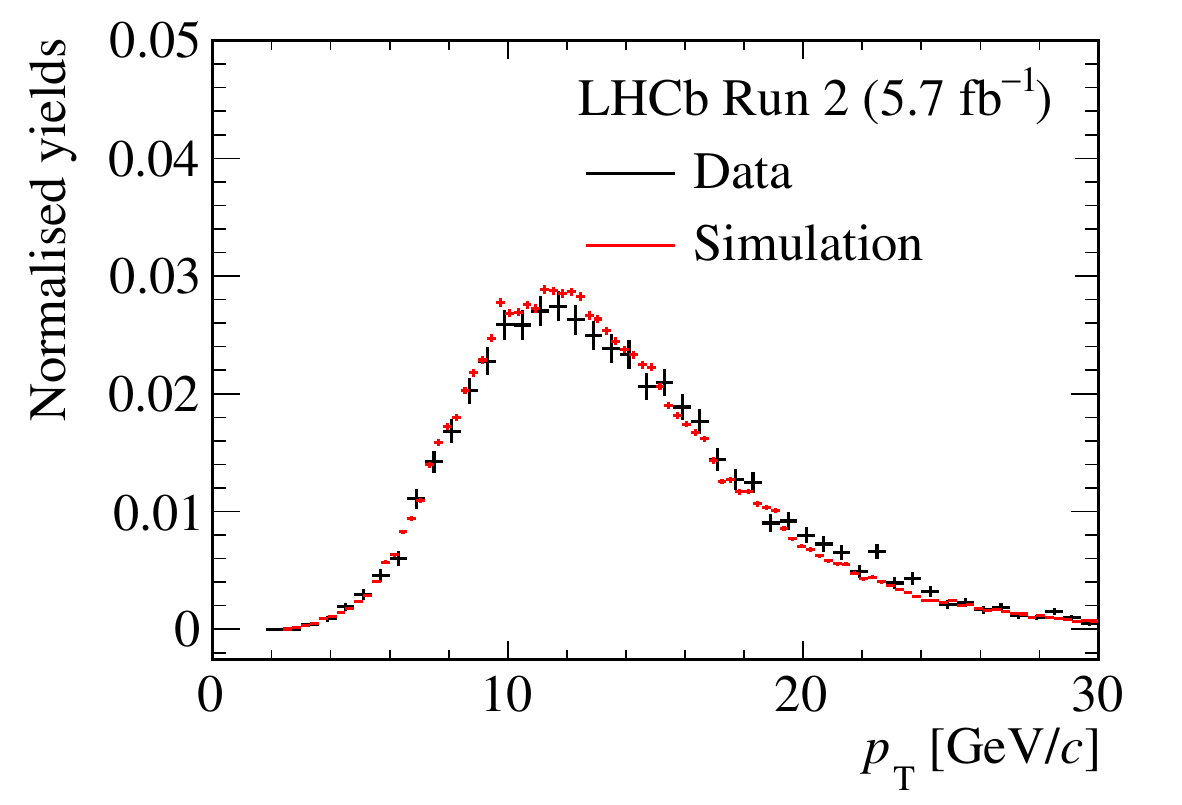}
    \includegraphics[width=0.49\textwidth]{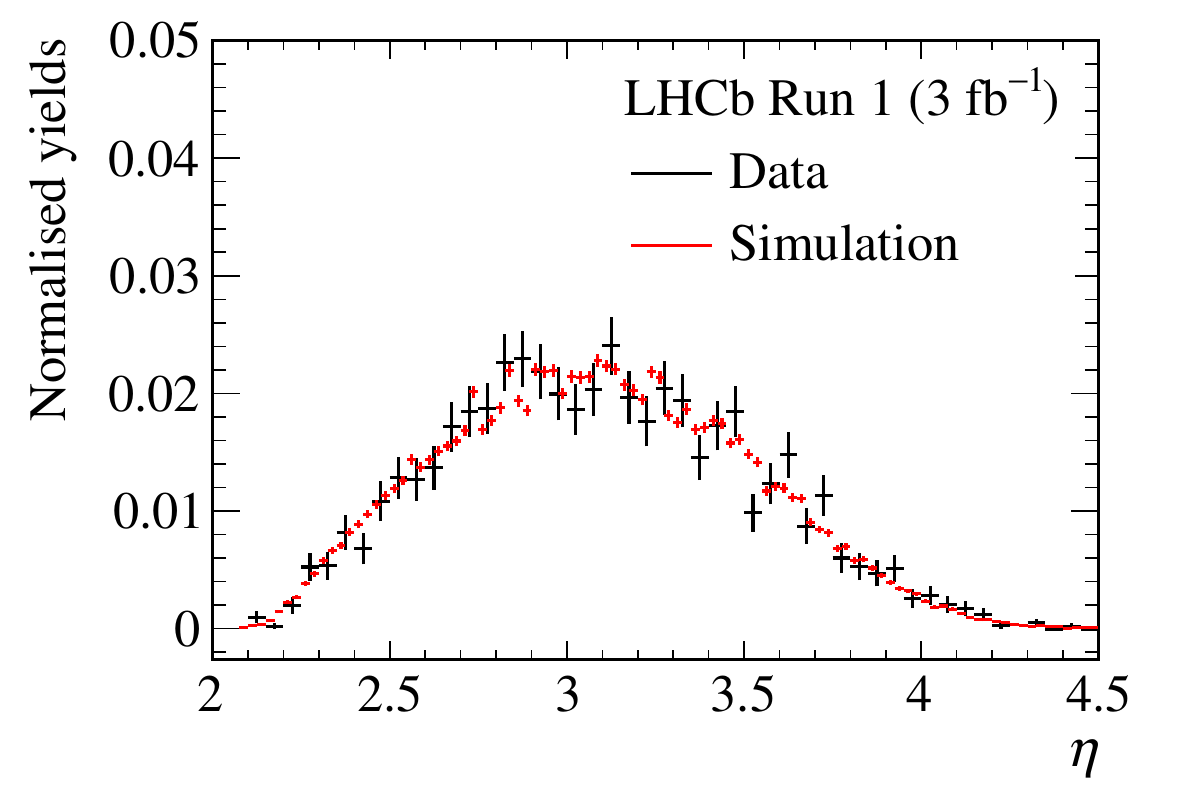}
    \includegraphics[width=0.49\textwidth]{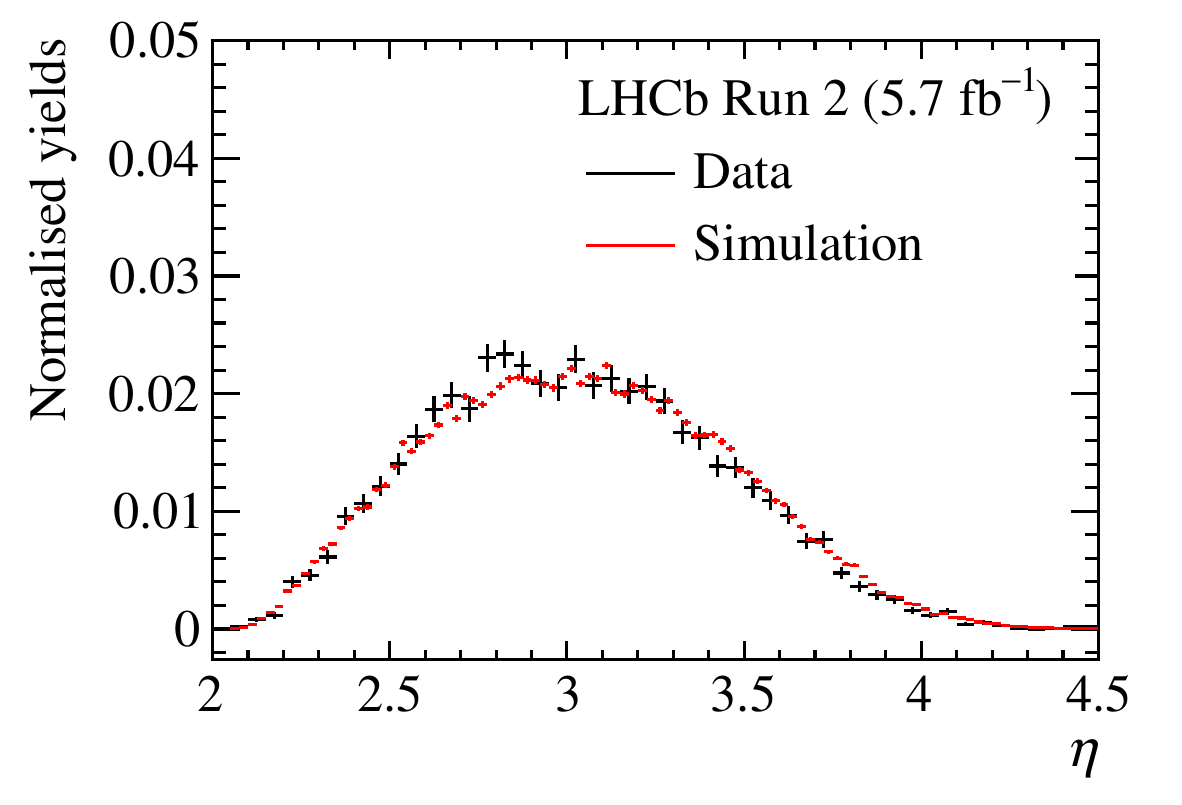}
\caption{
Distributions of (top)~$p_{\mathrm{T}}$ and (bottom)~$\eta$ for $B^- \to \mathit{\Lambda}_{c}^+\overline{p}\pi^-$ decays in (left)~Run~1 and (right)~Run~2, obtained from background-subtracted data and compared to the uncorrected distributions from simulation.
}
    \label{fig:sweighted-kinematics1}
\end{figure}

\begin{figure}[!tb]
    \centering
    \includegraphics[width=0.49\textwidth]{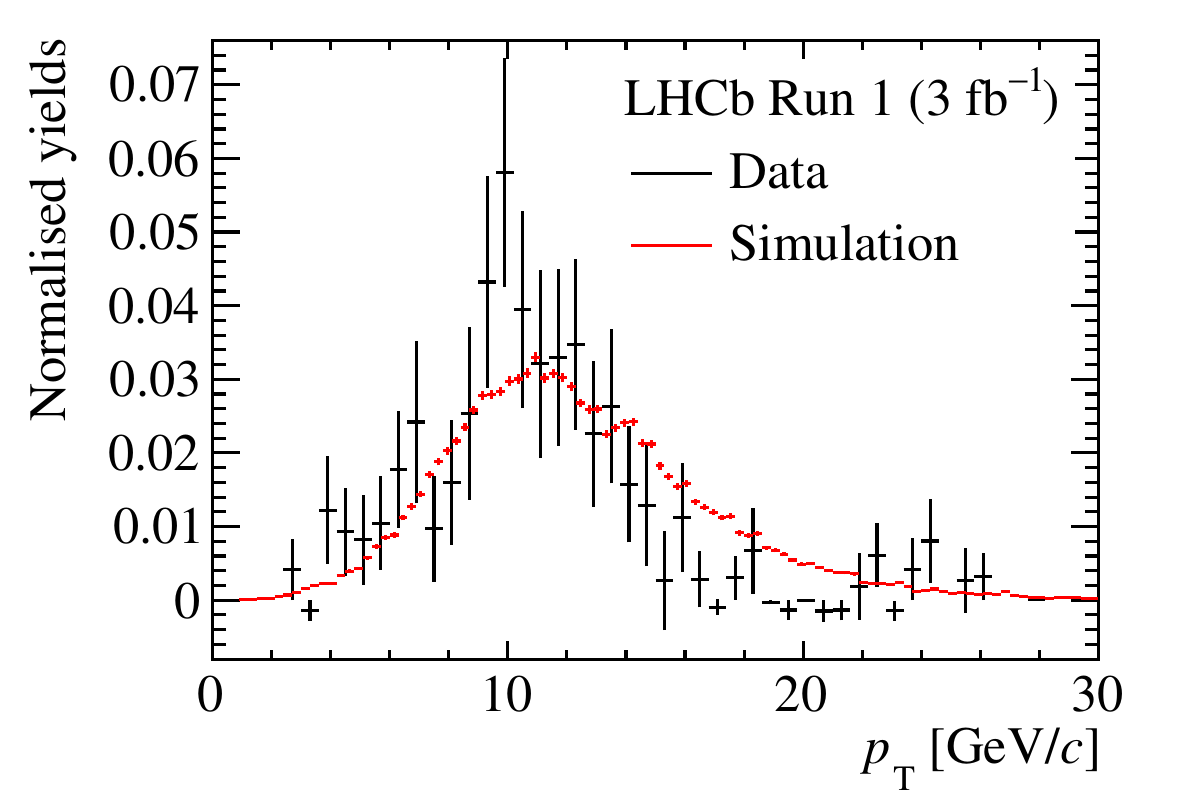}
    \includegraphics[width=0.49\textwidth]{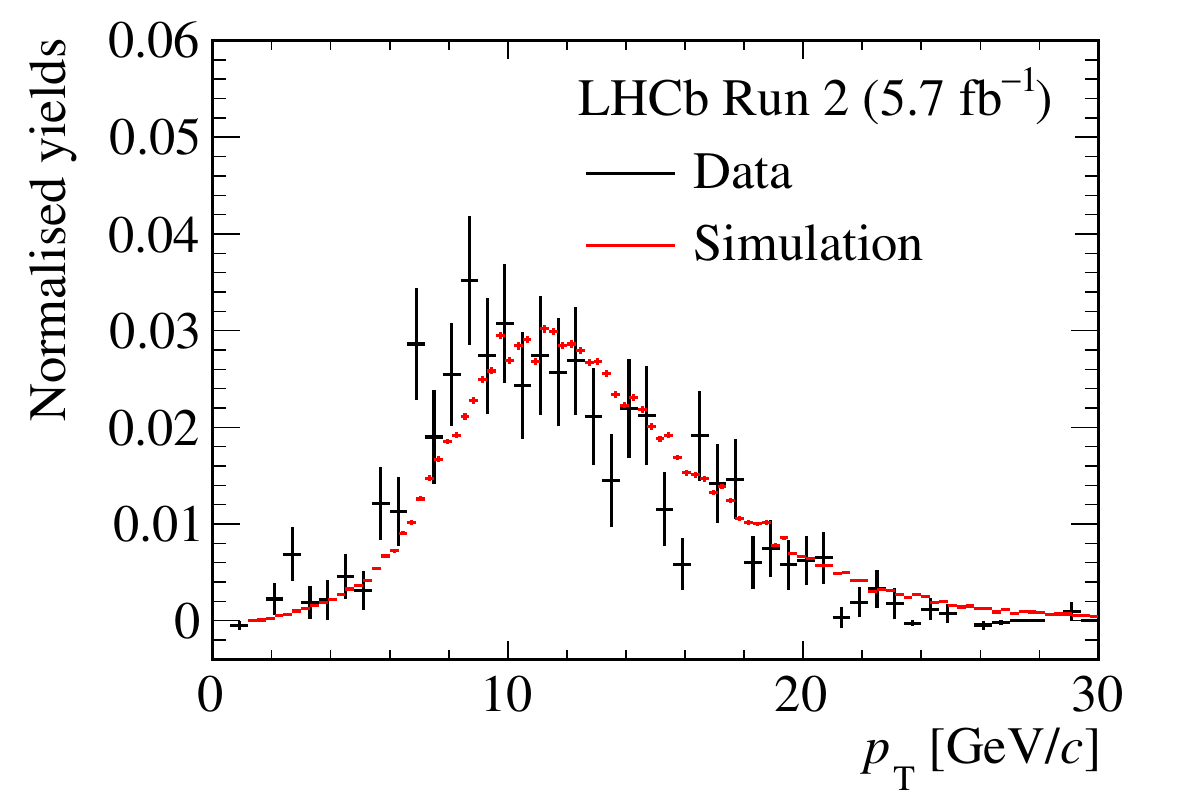}
    \includegraphics[width=0.49\textwidth]{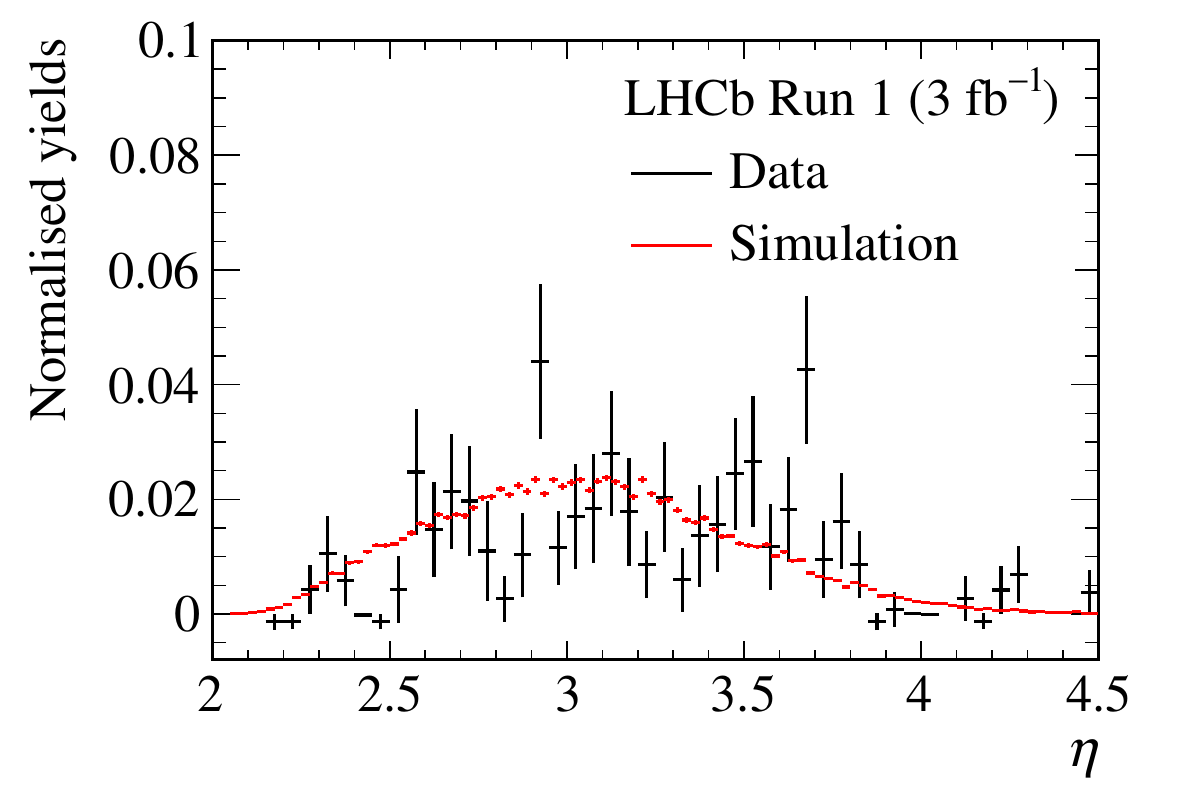}
    \includegraphics[width=0.49\textwidth]{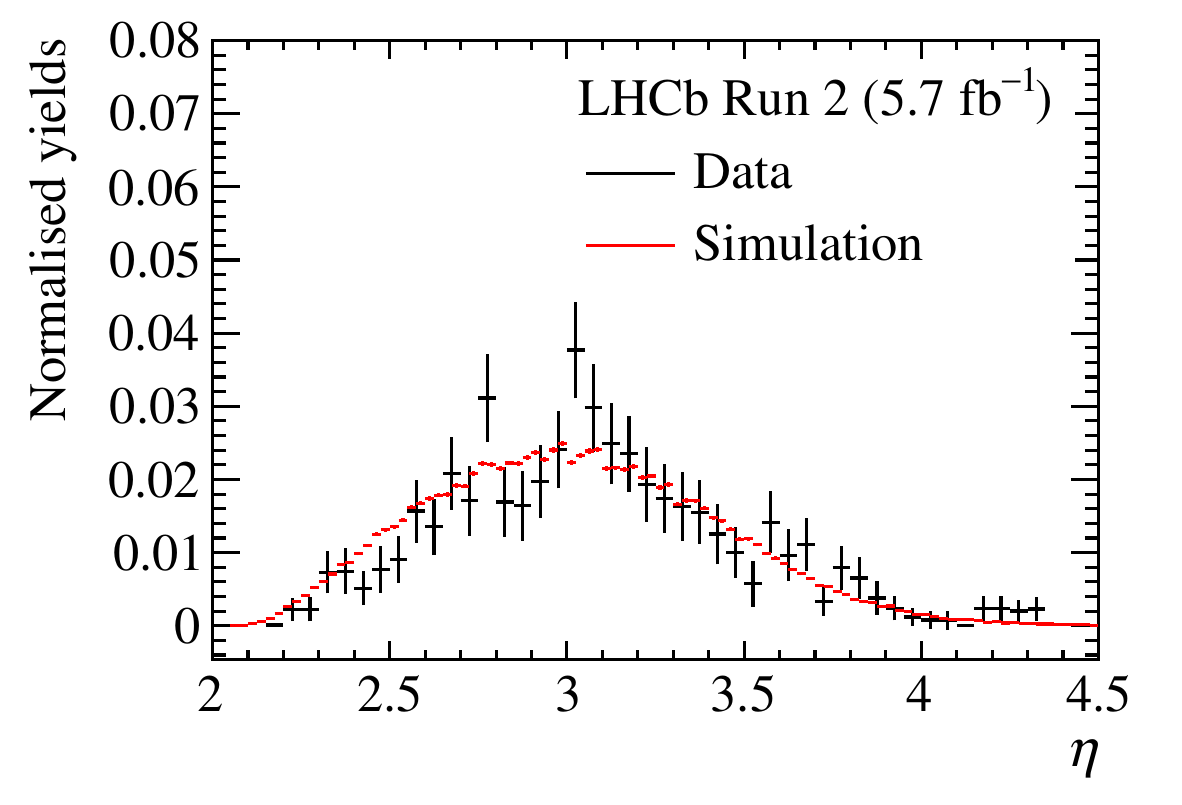}
\caption{
Distributions of (top)~$p_{\mathrm{T}}$ and (bottom)~$\eta$ for $\mathit{\Xi}_{b}^- \to \mathit{\Lambda}_{c}^+ K^-\pi^-$ decays in (left)~Run~1 and (right)~Run~2, obtained from background-subtracted data and compared to the uncorrected distributions from simulation.
}
    \label{fig:sweighted-kinematics2}
\end{figure}

Separate efficiency maps for each channel are obtained for each year of data taking, using square Dalitz-plot coordinates~\cite{Back:2017zqt} to represent the phase space of the three-body decay.
The use of candidate-by-candidate efficiencies, depending on position in the phase space, ensures that the distribution of decays in data is accounted for when evaluating results using Eq.~\eqref{eq:BFratio:def}.
The simulated samples are additionally weighted to correct for two sources of data-simulation disagreement.
The first of these is due to the $\Omegab$ lifetime, where improved measurements~\cite{CDF:2014mon,LHCb-PAPER-2014-010,LHCb-PAPER-2016-008} are available compared to the value used when generating the simulation.
A weighting factor is applied to the $\Omegab$ signal samples based on the difference in the decay-time distribution associated to updating the lifetime from the generated value.
The second is due to the $b$-hadron production kinematics, which for $b$-baryons are not well tuned in the simulation due to the absence of previous measurements.
The reconstructed \pt\ and pseudorapidity ($\eta$) distributions are obtained from background-subtracted data in the $\Bm\to\Lc\antiproton\pim$ and $\Xibm\to\Lc\Km\pim$ channels, as shown in Figs.~\ref{fig:sweighted-kinematics1} and~\ref{fig:sweighted-kinematics2}.
The distributions obtained from $\Xibm\to\Lc\Km\pim$ decays are used to weight both the $\Xibm$ and $\Omegab$ signal samples.
Since the efficiencies depend quite strongly on the \mbox{$X_b$-candidate} flight distance and transverse momentum, these corrections impact the results and are a source of systematic uncertainty.

\section{Systematic uncertainties}
\label{sec:systematics}

As seen in Eqs.~\eqref{eq:BFratio:def} and~\eqref{eq:Aprod:def}, each calculation of a ratio of fragmentation fractions times branching fractions or of a production asymmetry depends essentially on two inputs: the signal weights obtained from the $X_b$-candidate mass fits and the efficiency maps as a function of phase-space position.  
Sources of systematic uncertainty are hence considered as being related to one of these inputs, although in some cases the choice of which to relate to is arbitrary. 
Since the $\Bm\to\Lc\antiproton\pim$ control channel has the same topology as the signal channels, and the selection of different channels differs only by PID requirements, several potentially large systematic uncertainties are reduced in the ratios of fragmentation fractions times branching fractions. 

Systematic uncertainties on the yields arise from the model chosen to fit the $X_b$-candidate mass distributions, 
from fixed parameters in the fit model, and from the dependence of the shape of the misidentified $\Xibm\to\Lc\Km\pim$ component on its {\it a~priori} unknown phase-space distribution.
These uncertainties are evaluated by assessing the impact on the results of using alternative models for each of the fit components, varying fixed parameters within their uncertainties, and obtaining the misidentified $\Xibm\to\Lc\Km\pim$ shape from simulation weighted to match the phase-space distribution observed in data, respectively.
The alternative models considered are a combination of Johnson functions and Gaussian shapes for the signal components, second-order polynomials for the combinatorial background, shapes obtained with kernel density estimates for the cross-feed components and modified kernel density estimates for the partially reconstructed backgrounds.
Systematic uncertainties due to potential biases in the fit are determined from large ensembles of pseudoexperiments generated according to the yields obtained from, and the probability density functions used in, the fit to the $X_b$-candidate mass distributions.
Additionally, effects due to the presence of multiple candidates and from the impact of vetoing regions of the phase space are considered as being related to the yields.  
The former is evaluated from the impact on the results of allowing at most one candidate per $pp$ bunch crossing, and the latter from changing the width of the veto window.

Systematic uncertainties on the efficiency maps occur due to the finite size of the simulation samples and potential inaccuracies in the simulation description of the detector and its response.
The former is evaluated by varying the efficiency maps within their uncertainties, while for the latter several possible data-simulation inconsistencies are considered.
These include the amount of detector material~\cite{LHCb-DP-2013-002,LHCb-PAPER-2021-016}, the efficiency of the tracking algorithms~\cite{LHCb-DP-2013-002} and the response of the hardware trigger~\cite{LHCb-DP-2020-001}.
The impact of possible effects due to inaccuracies in the simulation description of the detector response are evaluated by weighting the simulated samples to correct the distributions of key variables. 
Uncertainties also arise due to the procedure to correct PID variables in simulation, and the granularity over phase space of the efficiency maps.
The effect of correcting the $\Lc\to\proton\Km\pip$ decay model in simulation to match the latest measurement~\cite{LHCb-PAPER-2022-002} is also considered as a systematic uncertainty.
Uncertainty in the $b$-baryon production kinematics is accounted for by considering the impact on the results of changing the weighting procedure to account for only the \pt\ distribution, rather than both the \pt\ and $\eta$ distributions.
Finally, for channels in which the signal yield is not significant, so that the candidate-by-candidate efficiency correction procedure is not reliable, an uncertainty is assigned due to the unknown phase-space distribution.

Systematic uncertainties on the production asymmetries are evaluated in the same way as for the ratios of fragmentation fractions and branching fractions, which is possible as the measurements are made in very similar ways, as seen in Eqs.~\eqref{eq:BFratio:def} and~\eqref{eq:Aprod:def}.  
This includes systematic uncertainties associated to detection asymmetries, which arise because of the combined effect of different interaction probabilities of particles and antiparticles and the potential mismatch in the material description between data and simulation. 

For the majority of measurements, the systematic uncertainties are significantly below the statistical uncertainties, as seen in Table~\ref{tab:Ncorr_CM_stat_syst}.
Some measurements involving unobserved modes have either or both fit model and veto-related systematic uncertainties that are comparable to the statistical uncertainty, which is understood due to the size of the backgrounds in the corresponding final states.
Additionally, as the statistical precision of the \runtwo\ measurement of $\frac{f_{\Xibm}}{f_{\Bm}} \cdot \frac{{\cal B}\left(\Xibm\to\Lc\Km\pim\right)}{{\cal B}\left(\Bm\to\Lc\antiproton\pim\right)}$ is around $5\%$, it is unsurprising that its systematic uncertainties are not negligible.
In particular, effects due to uncertainty in the $b$-hadron production kinematics are significant due to the rapid variation in efficiency with transverse momentum.
These effects, however, tend to cancel in ratios where decays of the same $b$ hadron are considered in both numerator and denominator.

In order to combine results from \runone\ and \runtwo, the effects of the systematic uncertainties are considered as being either 100\% correlated or completely uncorrelated between the two data-taking periods.
The dominant systematic uncertainties discussed above are all correlated between \runone\ and \runtwo.

\section{Results}
\label{sec:results}

Results for the ratios of fragmentation fractions and branching fractions are obtained for all channels studied relative to the $\Bm \to \Lc \antiproton \pim$ normalisation channel.
In addition, since significant signals are observed for $\Xibm\to\Lc\Km\pim$ and $\Omegab\to\Lc\Km\Km$ decays, the relative branching fractions of the other $\Xibm$ and $\Omegab$ decays, respectively, to those are also determined.
These results are given separately for \runone\ and \runtwo\ in Table~\ref{tab:Ncorr_CM_stat_syst}.

\begin{table}[!b]
\centering
\caption{
Ratios of branching fractions, multiplied where appropriate by ratios of fragmentation fractions, separately for Run~1 and Run~2. 
In each result the first uncertainty is statistical and the second is systematic.
}
\label{tab:Ncorr_CM_stat_syst}
\begin{tabular}{lr@{$\,\pm\,$}c@{$\,\pm\,$}cr@{$\,\pm\,$}c@{$\,\pm\,$}c}
\hline
 & \multicolumn{3}{c}{\runone} & \multicolumn{3}{c}{\runtwo}  \\
\hline
$\frac{\mathcal{B}(\Bm \rightarrow \Lc \antiproton \Km)}{\mathcal{B}(\Bm \rightarrow \Lc \antiproton \pim)}$ & $0.0404$ & $0.0056$ & $0.0021$ & $0.0395$ & $0.0025$ & $0.0013$ \\
$\frac{f_{\Xibm}}{f_{\Bm}} \cdot \frac{{\cal B}\left(\Xibm\to\Lc\Km\Km\right)}{{\cal B}\left(\Bm\to\Lc\antiproton\pim\right)}$ & $0.0085$ & $0.0037$ & $0.0024$ & $0.0032$ & $0.0009$ & $0.0003$ \\
$\frac{f_{\Xibm}}{f_{\Bm}} \cdot \frac{{\cal B}\left(\Xibm\to\Lc\Km\pim\right)}{{\cal B}\left(\Bm\to\Lc\antiproton\pim\right)}$ & $0.113$ & $0.012$ & $0.029$ & $0.076$ & $0.004$ & $0.006$ \\
$\frac{f_{\Xibm}}{f_{\Bm}} \cdot \frac{{\cal B}\left(\Xibm\to\Lc\pim\pim\right)}{{\cal B}\left(\Bm\to\Lc\antiproton\pim\right)}$ & $0.0110$ & $0.0054$ & $0.0057$ & $0.0015$ & $0.0010$ & $0.0014$ \\
$\frac{f_{\Omegab}}{f_{\Bm}} \cdot \frac{{\cal B}\left(\Omegab\to\Lc\Km\Km\right)}{{\cal B}\left(\Bm\to\Lc\antiproton\pim\right)}$ & $0.0053$ & $0.0023$ & $0.0012$ & $0.0037$ & $0.0009$ & $0.0004$ \\ 
$\frac{f_{\Omegab}}{f_{\Bm}} \cdot \frac{{\cal B}\left(\Omegab\to\Lc\Km\pim\right)}{{\cal B}\left(\Bm\to\Lc\antiproton\pim\right)}$ & $0.0020$ & $0.0021$ & $0.00013$ & $0.0006$ & $0.0005$ & $0.0004$ \\
$\frac{f_{\Omegab}}{f_{\Bm}} \cdot \frac{{\cal B}\left(\Omegab\to\Lc\pim\pim\right)}{{\cal B}\left(\Bm\to\Lc\antiproton\pim\right)}$ & $0.0013$ & $0.0029$ & $0.0027$ & $-0.0003$ & $0.0004$ & $0.0005$ \\
$\frac{{\cal B}\left(\Xibm\to\Lc\Km\Km\right)}{{\cal B}\left(\Xibm\to\Lc\Km\pim\right)}$ & $0.075$ & $0.034$ & $0.011$ & $0.041$ & $0.012$ & $0.005$ \\ 
$\frac{{\cal B}\left(\Xibm\to\Lc\pim\pim\right)}{{\cal B}\left(\Xibm\to\Lc\Km\pim\right)}$ & $0.097$ & $0.049$ & $0.046$ & $0.019$ & $0.013$ & $0.017$ \\
$\frac{{\cal B}\left(\Omegab\to\Lc\Km\pim\right)}{{\cal B}\left(\Omegab\to\Lc\Km\Km\right)}$ & $0.38$ & $0.43$ & $0.24$ & $0.17$ & $0.13$ & $0.09$ \\
$\frac{{\cal B}\left(\Omegab\to\Lc\pim\pim\right)}{{\cal B}\left(\Omegab\to\Lc\Km\Km\right)}$ & $0.24$ & $0.57$ & $0.53$ & $-0.08$ & $0.12$ & $0.15$ \\
\hline
\end{tabular}
\end{table}

Results for ratios of branching fractions of the same $b$ hadron do not depend on $pp$ collision centre-of-mass energy, and therefore the \runone\ and \runtwo\ results can be combined.
This is done taking correlations of systematic uncertainties into account, using the method described in Ref.~\cite{LHCb-PAPER-2021-043}, to obtain
\begingroup\allowdisplaybreaks
\begin{eqnarray*}
\frac{\mathcal{B}\left(B^- \rightarrow \Lc \antiproton K^{-}\right)}{\mathcal{B}\left(B^- \rightarrow \Lc \antiproton \pi^{-}\right)}& = &\phantom{-}0.0397\pm0.0023\stat\pm0.0012\syst \, ,\\
\frac{{\cal B}\left(\Xibm\to\Lc\Km\Km\right)}{{\cal B}\left(\Xibm\to\Lc\Km\pim\right)}& = &\phantom{-}0.045\pm0.011\stat\pm0.005\syst \, ,\\
\frac{{\cal B}\left(\Xibm\to\Lc\pim\pim\right)}{{\cal B}\left(\Xibm\to\Lc\Km\pim\right)}& = &\phantom{-}0.025\pm0.013\stat\pm0.019\syst \, ,\\
\frac{{\cal B}\left(\Omegab\to\Lc\Km\pim\right)}{{\cal B}\left(\Omegab\to\Lc\Km\Km\right)}& = & \phantom{-}0.19\pm0.12\stat\pm0.10\syst \, ,\\
\frac{{\cal B}\left(\Omegab\to\Lc\pim\pim\right)}{{\cal B}\left(\Omegab\to\Lc\Km\Km\right)}& = &-0.07\pm0.12\stat\pm0.16\syst \, ,
\end{eqnarray*}
\endgroup
where the first uncertainty is statistical and the second is systematic.

The significance of the signal in each mode is determined using Wilks' theorem, \ie\ as $\sqrt{2\Delta({\rm NLL})}$ where $2\Delta({\rm NLL})$ is double the difference in negative log-likelihood obtained in the best fit (with the signal yield free) and in another fit where the relevant signal yield is fixed to zero.
Significances are determined separately for \runone\ and \runtwo, using the likelihood functions obtained from the fits convolved with Gaussian functions to account for relevant sources of systematic uncertainty, and then combined taking correlations of systematic uncertainties into account.
As the $\Bm\to\Lc\antiproton\Km$ and $\Xibm\to\Lc\Km\pim$ signals are clearly far in excess of $5\,\sigma$ significance, they are not quantified further.
Similarly, as the statistical significances of the $\Xibm\to\Lc\pim\pim$, $\Omegab\to\Lc\Km\pim$ and $\Omegab\to\Lc\pim\pim$ signals are found to be below $3\,\sigma$, they are not quantified further.
For the $\Xibm\to\Lc\Km\Km$ and $\Omegab\to\Lc\Km\Km$ channels, the combined significances are found to be above $5\,\sigma$, as shown in Table~\ref{tab:significances_syst}.

\begin{table}[!tb]
\centering
\caption{
    Significances ($\sigma$) of the signals for $\mathit{\Xi}_{b}^-\to\mathit{\Lambda}_{c}^+ K^- K^-$ and $\mathit{\Omega}_{b}^-\to\mathit{\Lambda}_{c}^+ K^- K^-$ decays.
}
\label{tab:significances_syst}
\renewcommand{\arraystretch}{1.1}
\begin{tabular}{cccc}
\hline
Decay & Run 1 &  Run 2 & Combined  \\
\hline
\multicolumn{4}{c}{Statistical uncertainties only} \\
\hline
$\Xibm\to\Lc\Km\Km$                 & 3.6 & 5.5 & 6.3\\
$\Omegab\to\Lc\Km\Km$               & 4.2 & 7.9 & 8.8\\
\hline
\multicolumn{4}{c}{Including systematic uncertainties} \\
\hline
$\Xibm\to\Lc\Km\Km$                 & 3.5 & 5.2 & 5.9\\
$\Omegab\to\Lc\Km\Km$               & 3.6 & 6.9 & 7.5\\
\hline
\end{tabular}
\end{table}

For the modes where no significant signal is observed, upper limits on the relative ratios of fragmentation fractions times branching fractions are reported. 
The limits are obtained for the combined \runone\ and \runtwo\ ratios, so in cases where the decays in the numerator and denominator involve different $b$ hadrons, the ratio of fragmentation fractions in the combination should be considered as an effective average of the values for \runone\ and \runtwo\ (dominated by the latter due to the larger sample size).
A combined likelihood for \runone\ and \runtwo\ is obtained, taking correlations of systematic uncertainties into account, and upper limits at 90\%~(95\%) confidence level (CL) are determined by integrating the likelihood in the physical region of non-negative relative fragmentation fractions times branching fractions.
The combined limits are 
\begingroup\allowdisplaybreaks
\begin{eqnarray*}
\frac{f_{\Xibm}}{f_{\Bm}} \cdot \frac{{\cal B}\left(\Xibm\to\Lc\pim\pim\right)}{{\cal B}\left(\Bm\to\Lc\antiproton\pim\right)}& < &0.0049~(0.0057)\ \text{at 90\%~(95\%) CL}\, ,\\
\frac{f_{\Omegab}}{f_{\Bm}} \cdot \frac{{\cal B}\left(\Omegab\to\Lc\Km\pim\right)}{{\cal B}\left(\Bm\to\Lc\antiproton\pim\right)}& < &0.0019~(0.0022)\ \text{at 90\%~(95\%) CL}\, ,\\
\frac{f_{\Omegab}}{f_{\Bm}} \cdot \frac{{\cal B}\left(\Omegab\to\Lc\pim\pim\right)}{{\cal B}\left(\Bm\to\Lc\antiproton\pim\right)}& < &0.0012~(0.0015)\ \text{at 90\%~(95\%) CL}\, ,\\
\frac{{\cal B}\left(\Xibm\to\Lc\pim\pim\right)}{{\cal B}\left(\Xibm\to\Lc\Km\pim\right)}& < &0.065~(0.074)\ \text{at 90\%~(95\%) CL}\, ,\\
\frac{{\cal B}\left(\Omegab\to\Lc\Km\pim\right)}{{\cal B}\left(\Omegab\to\Lc\Km\Km\right)}& < &0.56~(0.64)\ \text{at 90\%~(95\%) CL}\, ,\\
\frac{{\cal B}\left(\Omegab\to\Lc\pim\pim\right)}{{\cal B}\left(\Omegab\to\Lc\Km\Km\right)}& < &0.37~(0.45)\ \text{at 90\%~(95\%) CL}\, .\\
\end{eqnarray*}
\endgroup

\begin{figure}[!tb]
    \centering
    \includegraphics[scale=0.36]{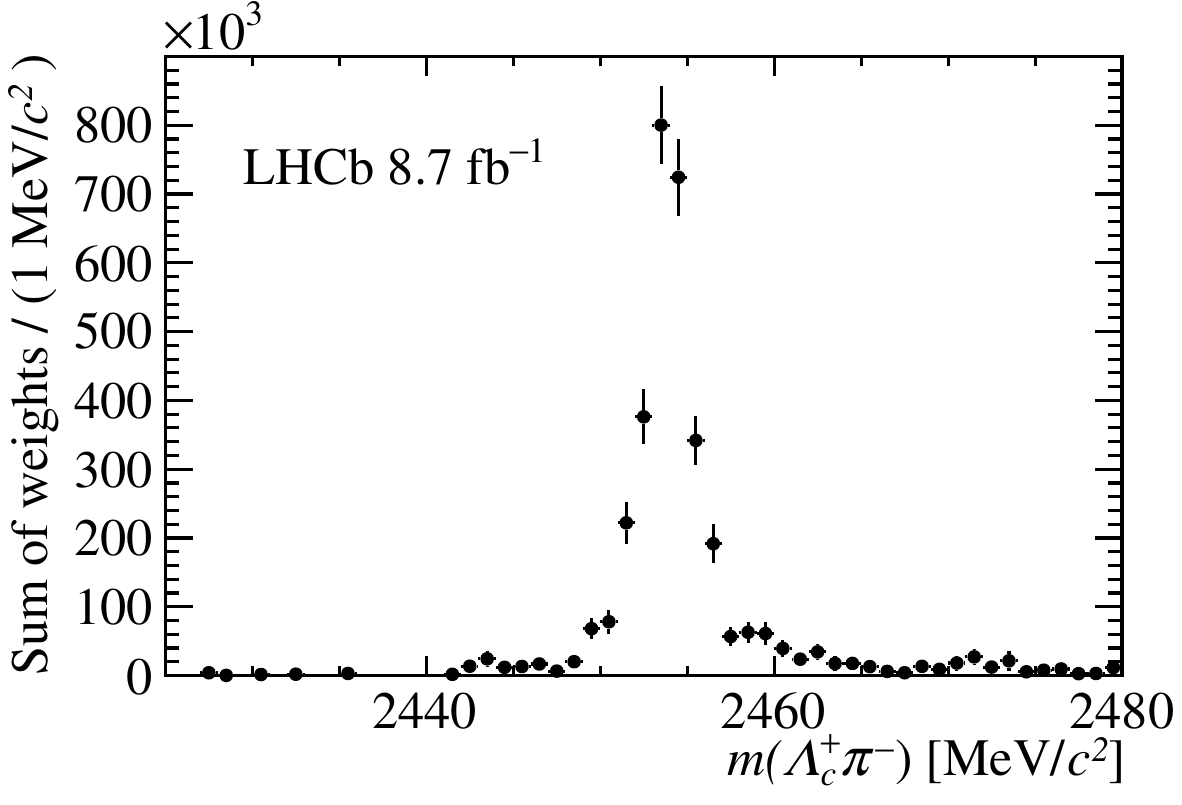}
    \includegraphics[scale=0.36]{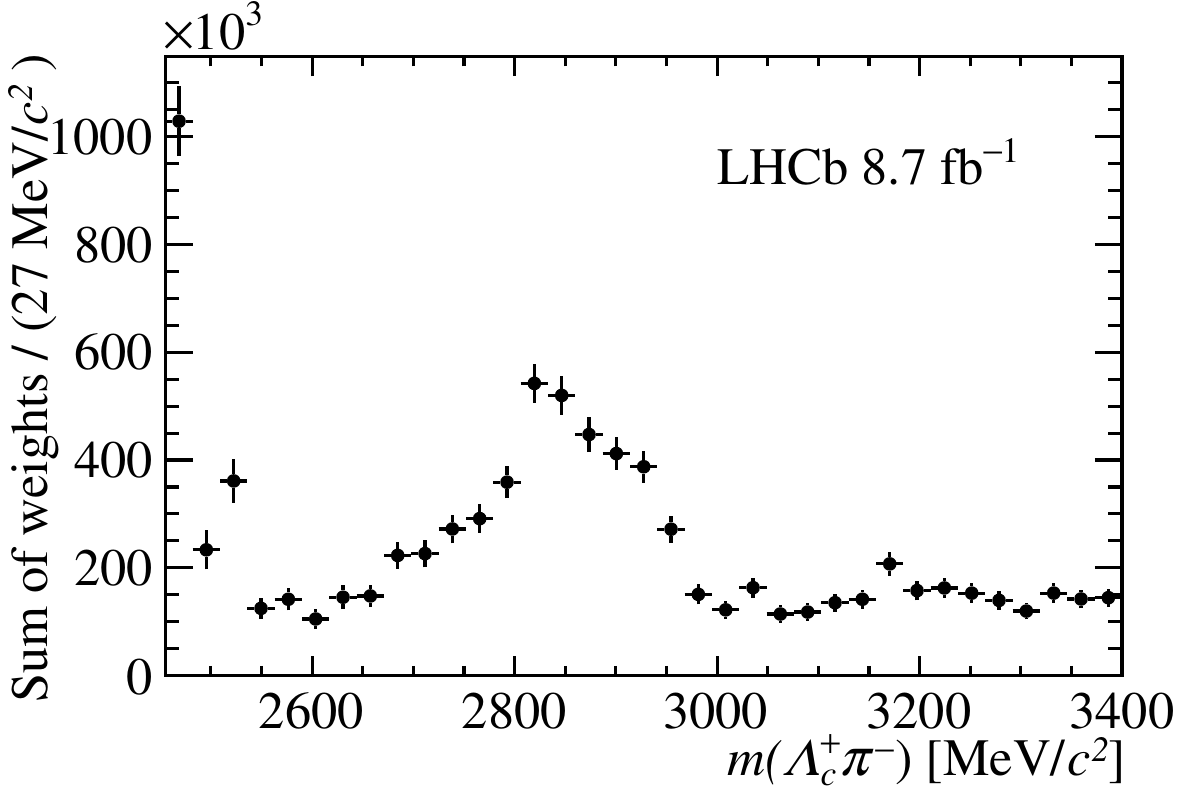} \\
    \includegraphics[scale=0.36]{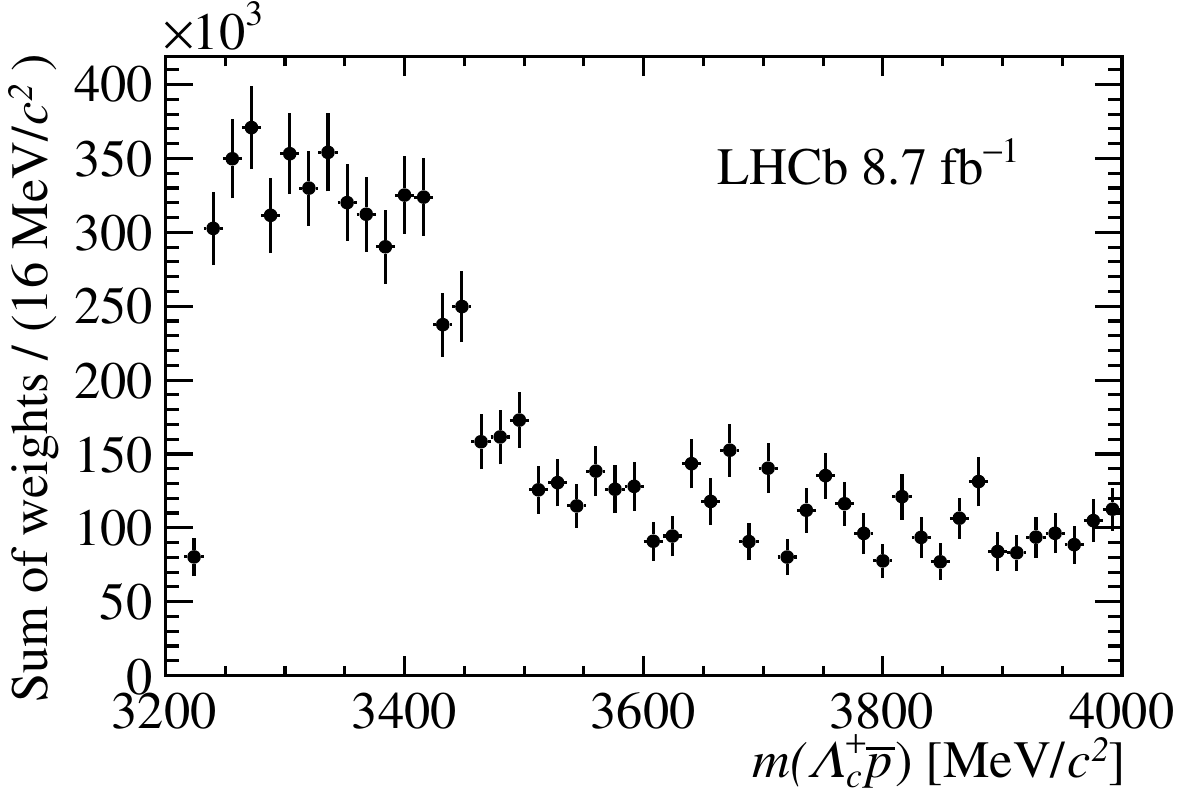}
\caption{
    Background-subtracted and efficiency-corrected distributions of $B^-\to\mathit{\Lambda}_{c}^+\overline{p}\pi^-$ decays:
    (top~left)~$m(\mathit{\Lambda}_{c}^+\pi^-)$ at low mass, showing the $\mathit{\Sigma}_{c}(2455)^0$ resonance peak;
    (top~right)~$m(\mathit{\Lambda}_{c}^+\pi^-)$ at higher mass, showing additional $\mathit{\Sigma}_{c}$ resonance structures;
    (bottom)~$m(\mathit{\Lambda}_{c}^+\overline{p})$ at low mass, where $m(\mathit{\Lambda}_{c}^+\pi^-)>2460\,\mathrm{Me\kern -0.1em V\!/}c^2$ has been required to remove contributions from the $\mathit{\Sigma}_{c}(2455)^0$ resonance.
} 
    \label{fig:Lcppi-DP-projs}
\end{figure}

\begin{figure}[!tb]
    \centering
    \includegraphics[scale=0.36]{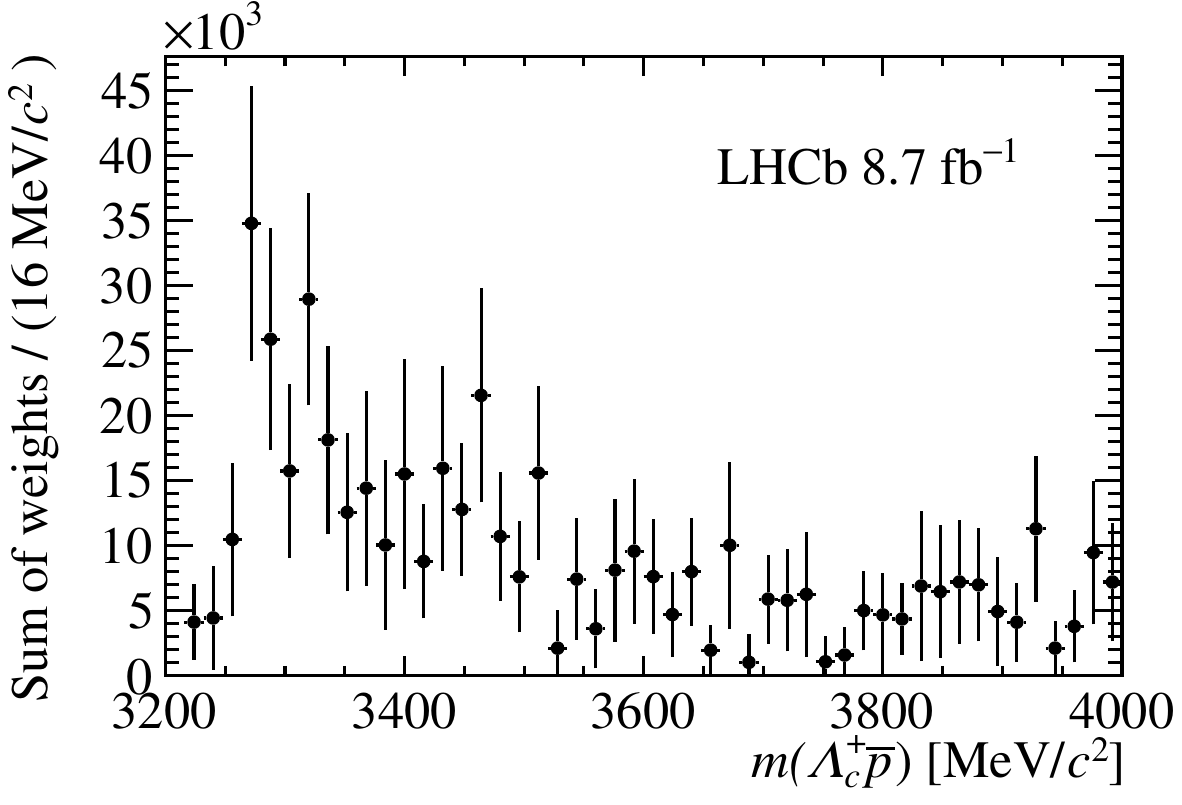}
    \includegraphics[scale=0.36]{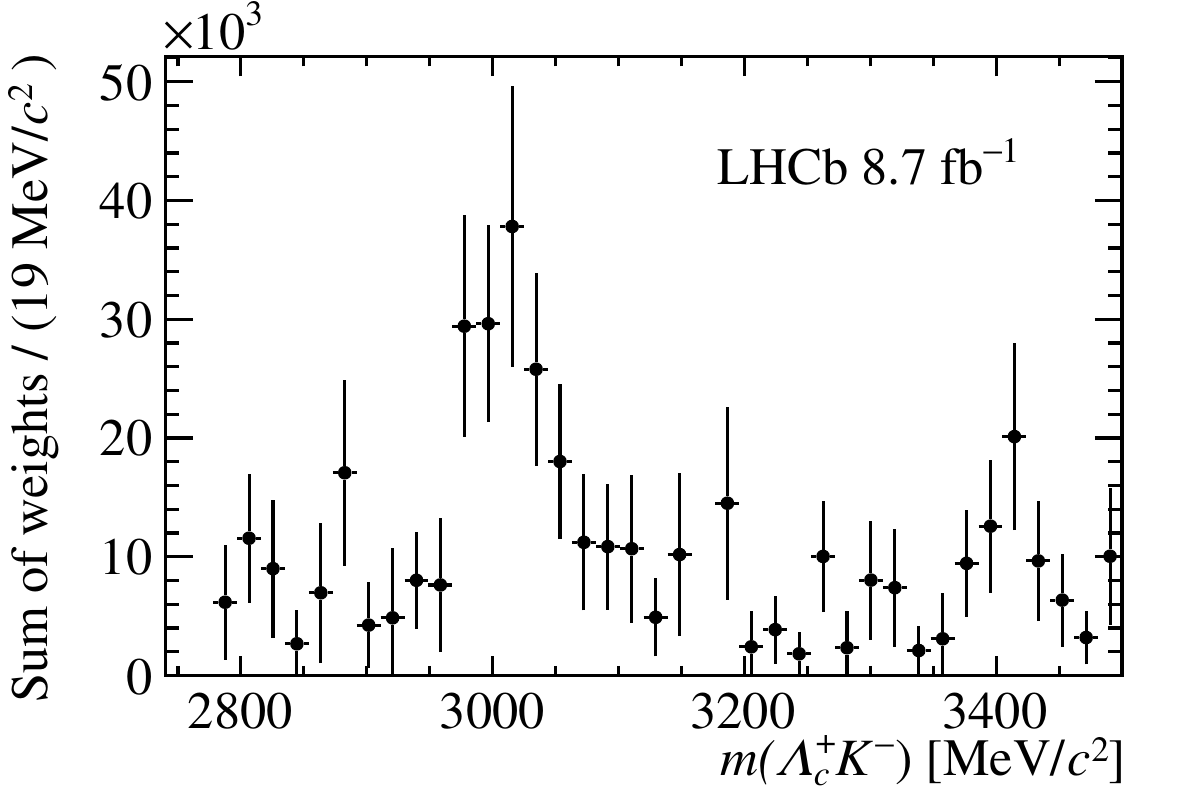}
\caption{
    Background-subtracted and efficiency-corrected distributions of $B^-\to\mathit{\Lambda}_{c}^+\overline{p} K^-$ decays:
    (left)~$m(\mathit{\Lambda}_{c}^+\overline{p})$ at low mass;
    (right)~$m(\mathit{\Lambda}_{c}^+ K^-)$ at low mass, where $m(\mathit{\Lambda}_{c}^+\overline{p})>3500\,\mathrm{Me\kern -0.1em V\!/}c^2$ has been required to remove contributions from the $\mathit{\Lambda}_{c}^+\overline{p}$ threshold enhancement.
}
    \label{fig:LcpK-DP-projs}
\end{figure}

\begin{figure}[!tb]
    \centering
    \includegraphics[scale=0.36]{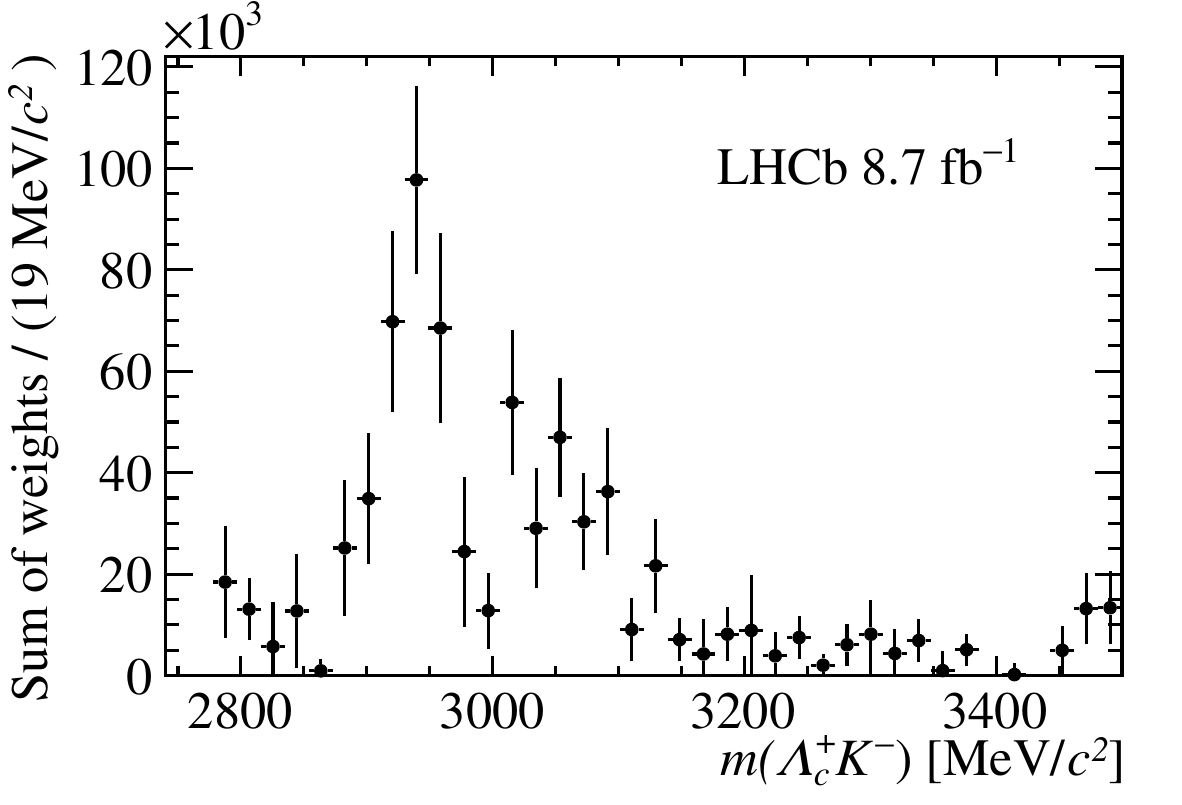}
    \includegraphics[scale=0.36]{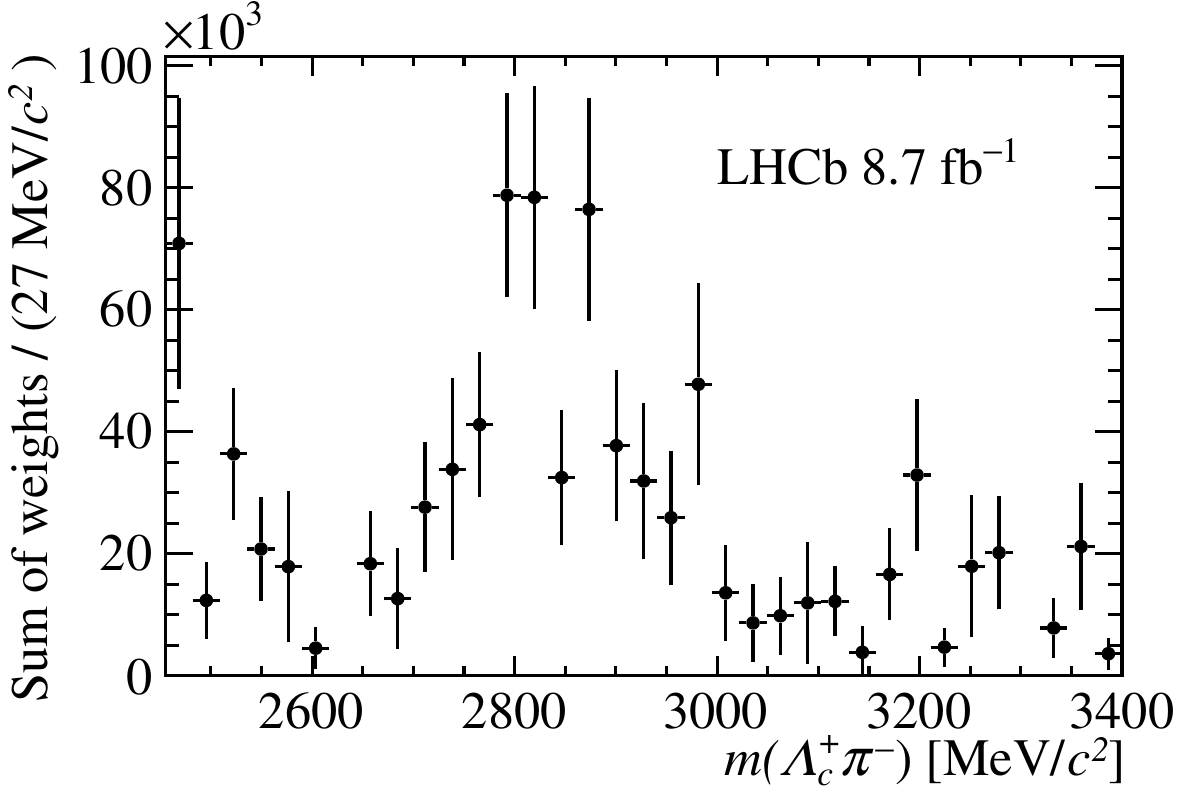}
\caption{
    Background-subtracted and efficiency-corrected distributions of $\Xibm\to\mathit{\Lambda}_{c}^+ K^-\pi^-$ decays:
    (left)~$m(\mathit{\Lambda}_{c}^+ K^-)$ at low mass; (right)~$m(\mathit{\Lambda}_{c}^+\pi^-)$ at low mass.
}
    \label{fig:LcKpi-DP-projs}
\end{figure}

In channels with large enough yields, namely $\Bm\to\Lc\antiproton\pim$, $\Bm\to\Lc\antiproton\Km$ and $\Xibm\to\Lc\Km\pim$, the candidate-by-candidate efficiency correction procedure allows background-subtracted and efficiency-corrected phase-space distributions to be obtained.
Two-body mass projections of these distributions are shown in Figs.~\ref{fig:Lcppi-DP-projs},~\ref{fig:LcpK-DP-projs} and~\ref{fig:LcKpi-DP-projs}.
Clear resonant structures are seen in $\Lc\pim$ and $\Lc\Km$ mass distributions, while enhancements are observed near threshold in $\Lc\antiproton$ mass.
The $\Lc\pim$ and $\Lc\antiproton$ structures observed in \mbox{$\Bm\to\Lc\antiproton\pim$} decays appear consistent with those seen in previous studies of this decay~\cite{Belle:2004dmq,BaBar:2008get}.
A $\Lc\antiproton$ threshold enhancement is also observed in $\Bm\to\Lc\antiproton\Km$ decays, where $\Lc\Km$ resonances are also seen that may be the same as those observed in other recent studies~\cite{LHCb-PAPER-2020-004,LHCb-PAPER-2022-028}.
Both $\Lc\pim$ and $\Lc\Km$ resonances are also seen in $\Xibm\to\Lc\Km\pim$ decays.
Detailed investigations of these structures is beyond the scope of this work.

\begin{figure}[!tb]
\centering
\includegraphics[width=0.48\textwidth]{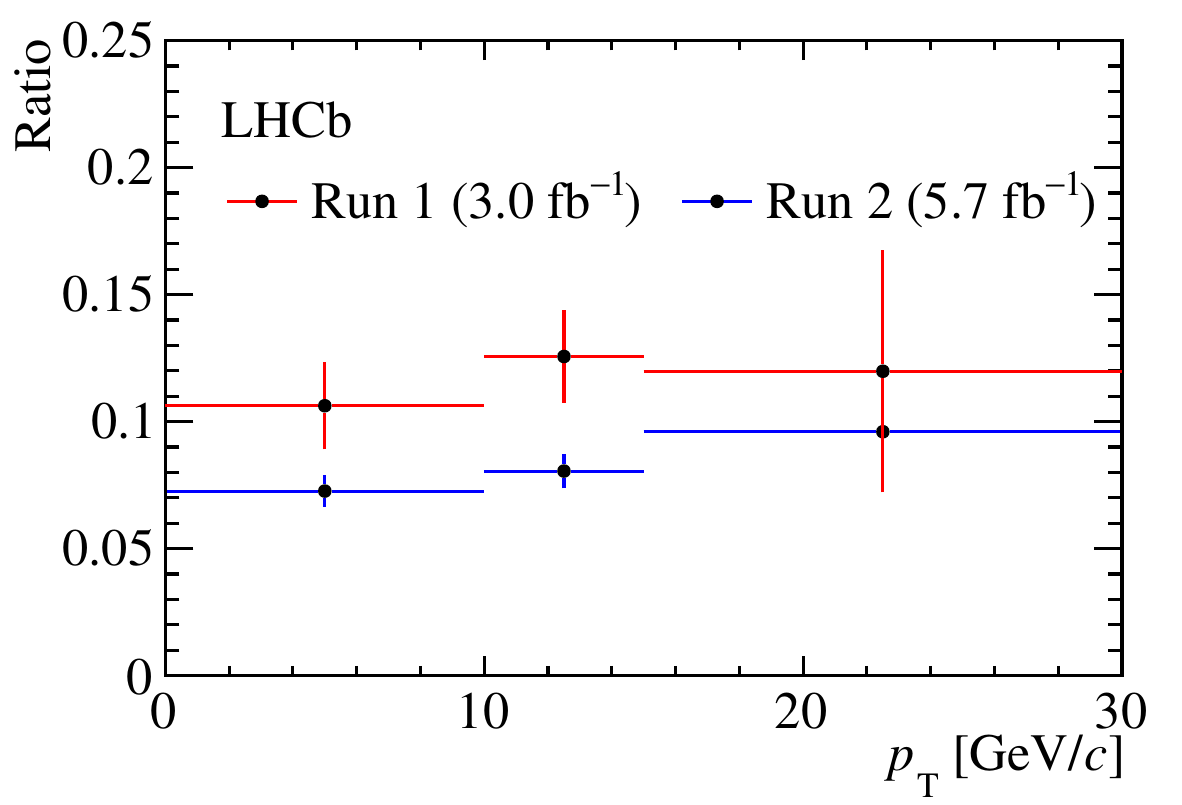}
\includegraphics[width=0.48\textwidth]{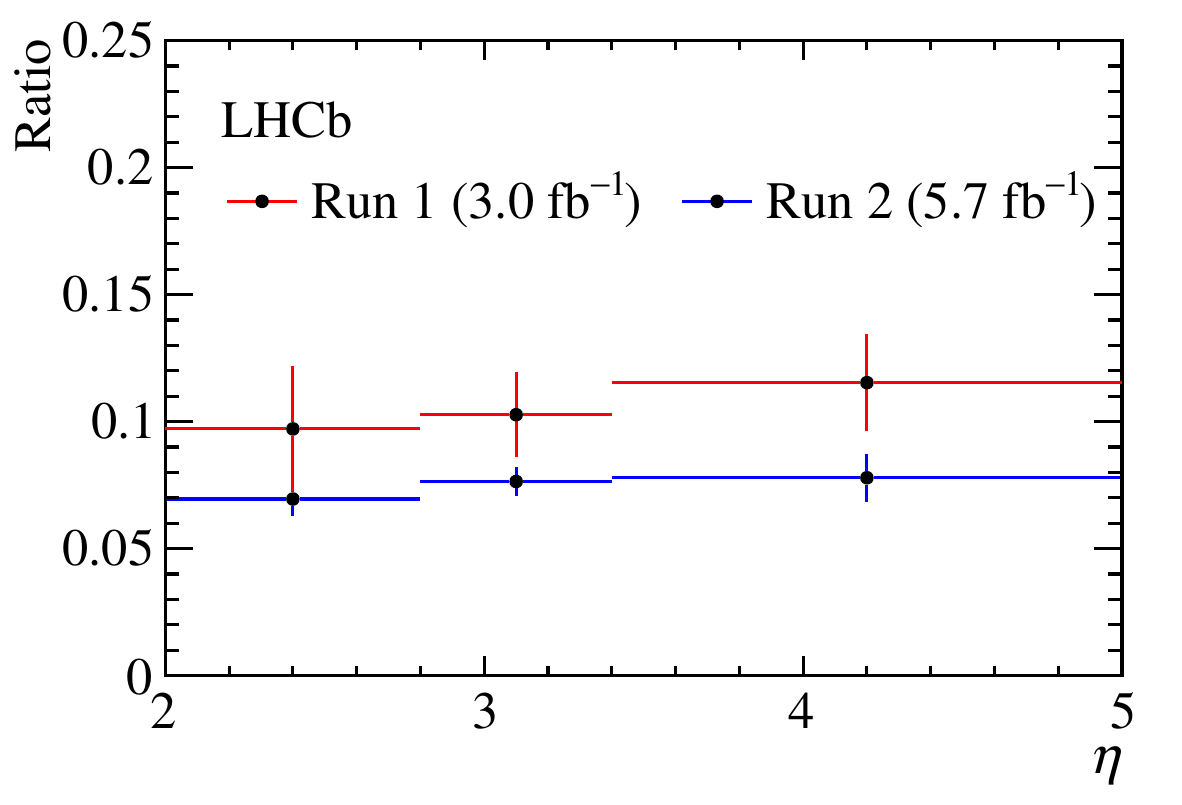}
\caption{\small 
  Variation of the ratio of fragmentation fractions times branching fractions of \mbox{$\mathit{\Xi}_{b}^- \to \mathit{\Lambda}_{c}^+ K^-\pi^-$} and $B^- \to \mathit{\Lambda}_{c}^+\overline{p}\pi^-$ decays with (left)~$b$-hadron transverse momentum and (right)~pseudorapidity, separately for Run~1 and Run~2, with statistical uncertainties only.
}
\label{fig:signal_BF_kinematics}
\end{figure}

\begin{figure}[!tb]
\centering
\includegraphics[width=0.48\textwidth]{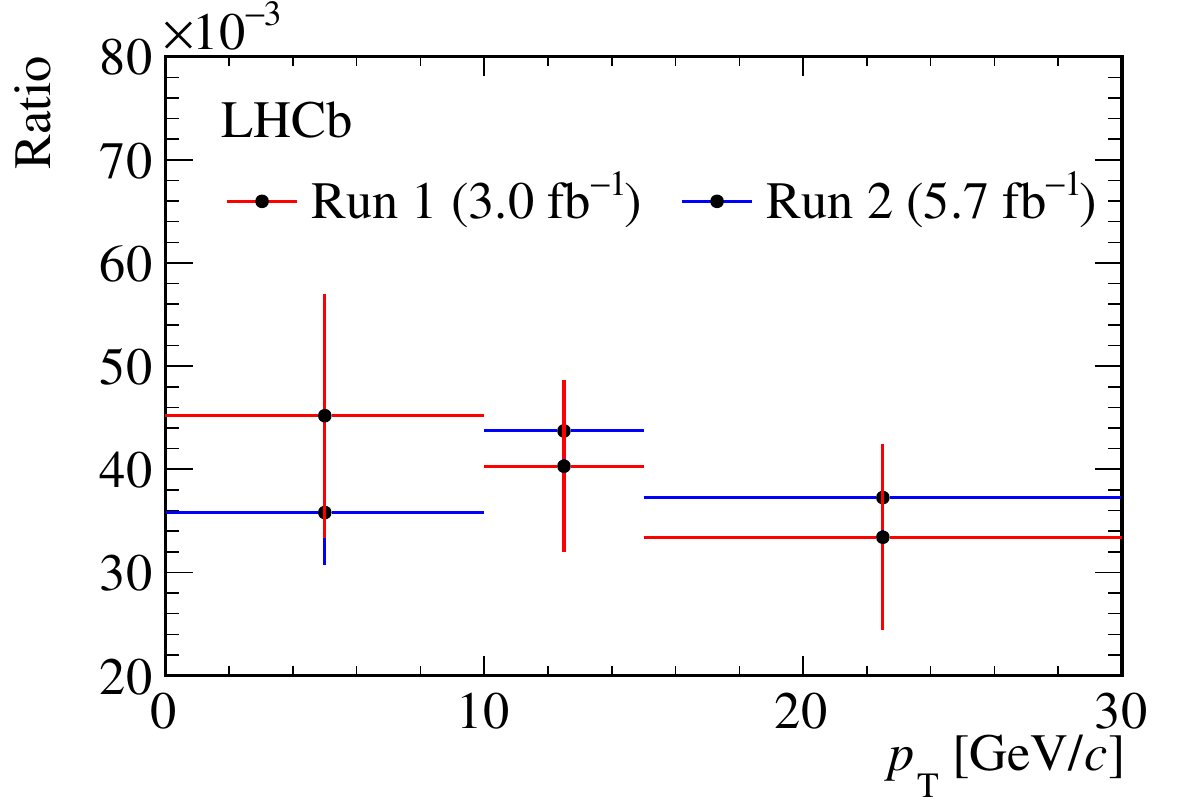}
\includegraphics[width=0.48\textwidth]{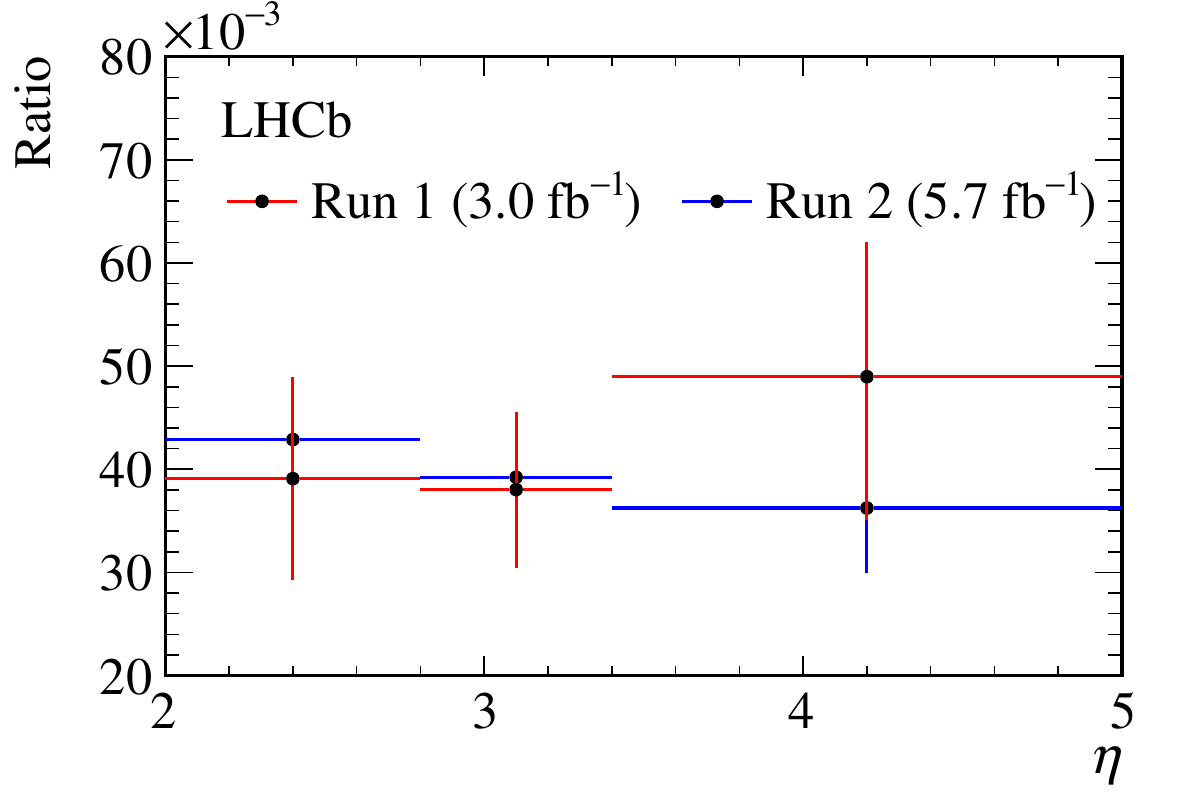}
\caption{\small 
  Variation of the ratio of branching fractions of $B^-\to\mathit{\Lambda}_{c}^+\overline{p} K^-$ and $B^-\to\mathit{\Lambda}_{c}^+\overline{p}\pi^-$ decays with (left)~$b$-hadron transverse momentum and (right)~pseudorapidity, separately for Run~1 and Run~2, with statistical uncertainties only.
}
\label{fig:CM_BF_crosscheck}
\end{figure}

The reasonably large yield observed in $\Xibm\to\Lc\Km\pim$ decays also allows the corresponding ratio of fragmentation fractions times branching fractions, relative to the $\Bm\to\Lc\antiproton\pim$ control channel, to be determined in intervals of the $b$-hadron kinematic variables.  
Since the ratio of branching fractions is constant, this quantity probes dependence of the ratio of fragmentation fractions on the production kinematics.
Significant dependence on \pt\ of the \Lb\ and \Bs\ fragmentation fractions relative to those of \Bd\ and \Bu\ mesons, has previously been observed~\cite{LHCb-PAPER-2018-050,LHCb-PAPER-2019-020,LHCb-PAPER-2020-046}.
Knowledge of whether similar effects are present in the ratio of \Xibm\ to \Bm\ fragmentation fractions would provide insight into production mechanisms. 
Results are shown in Fig.~\ref{fig:signal_BF_kinematics}, with statistical uncertainties only (the majority of systematic effects are correlated bin-to-bin and therefore not relevant to the dependence).
The limited statistical precision precludes any strong conclusion, except that the \runone\ result is higher than that of \runtwo, as also seen in Table~\ref{tab:Ncorr_CM_stat_syst}.  
As a cross-check, the relative branching fraction of $\Bm\to\Lc\antiproton\Km$ and $\Bm\to\Lc\antiproton\pim$ decays is shown in the same kinematic bins in Fig.~\ref{fig:CM_BF_crosscheck}.
No dependence is seen, as expected.

\begin{figure}[!tb]
\centering
\includegraphics[width=0.49\textwidth]{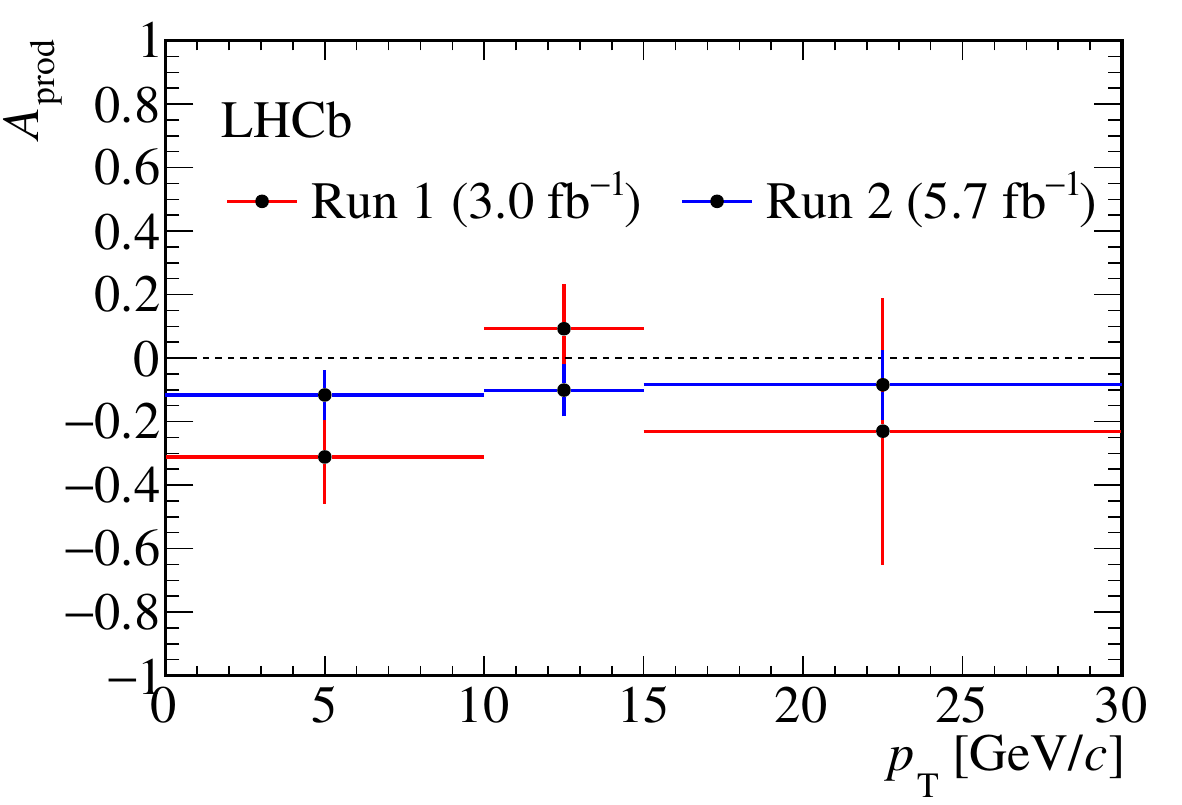}
\includegraphics[width=0.49\textwidth]{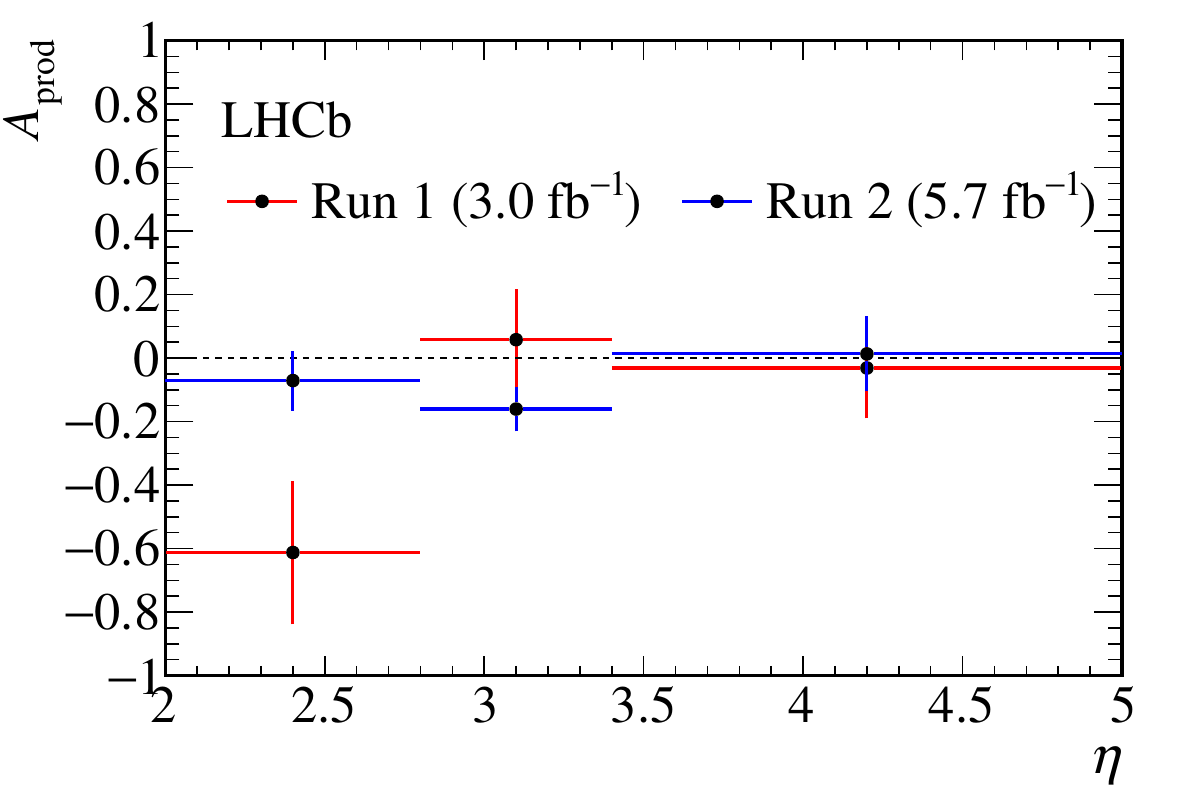}
\caption{\small 
  Variation of $\mathit{\Xi}_{b}^-$ production asymmetry with (left)~$\mathit{\Xi}_{b}^-$ transverse momentum and (right)~pseudorapidity determined with $\mathit{\Xi}_{b}^-\to\mathit{\Lambda}_{c}^+ K^-\pi^-$ decays.
}
\label{fig:signal_Araw}
\end{figure}

\begin{figure}[!tb]
\centering
\includegraphics[width=0.49\textwidth]{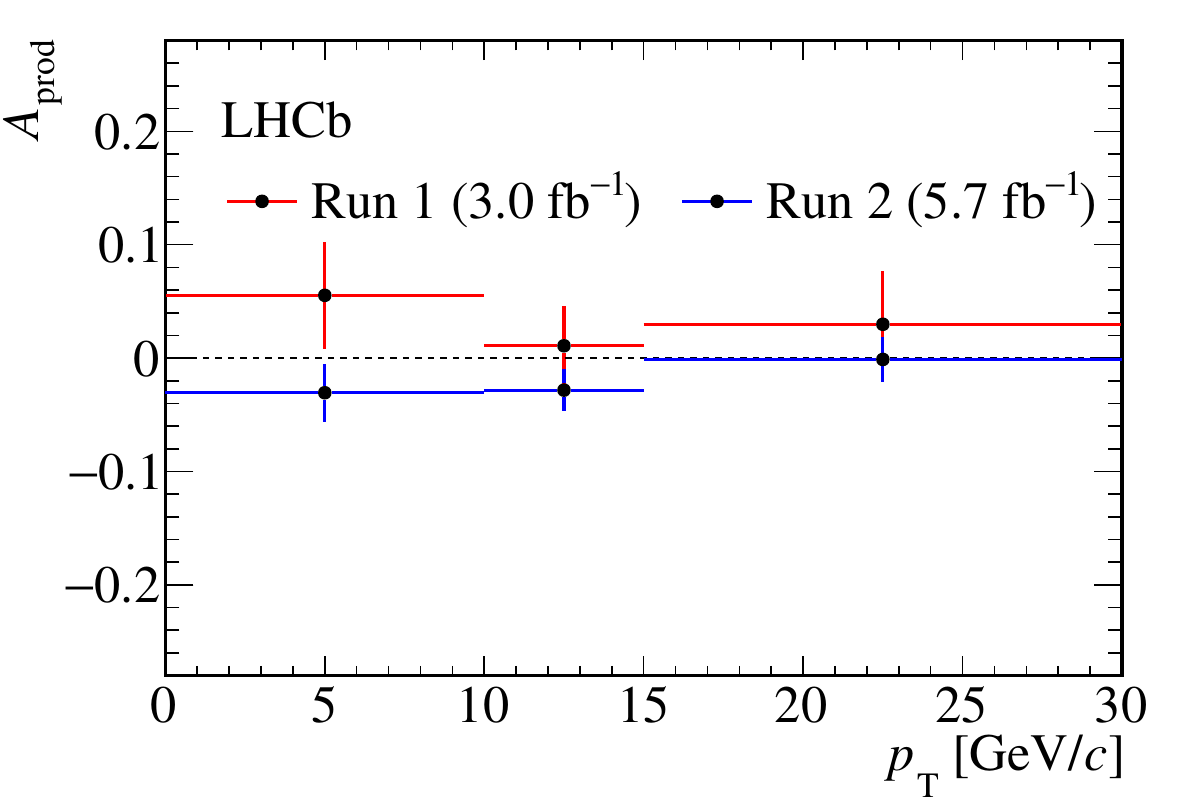}
\includegraphics[width=0.49\textwidth]{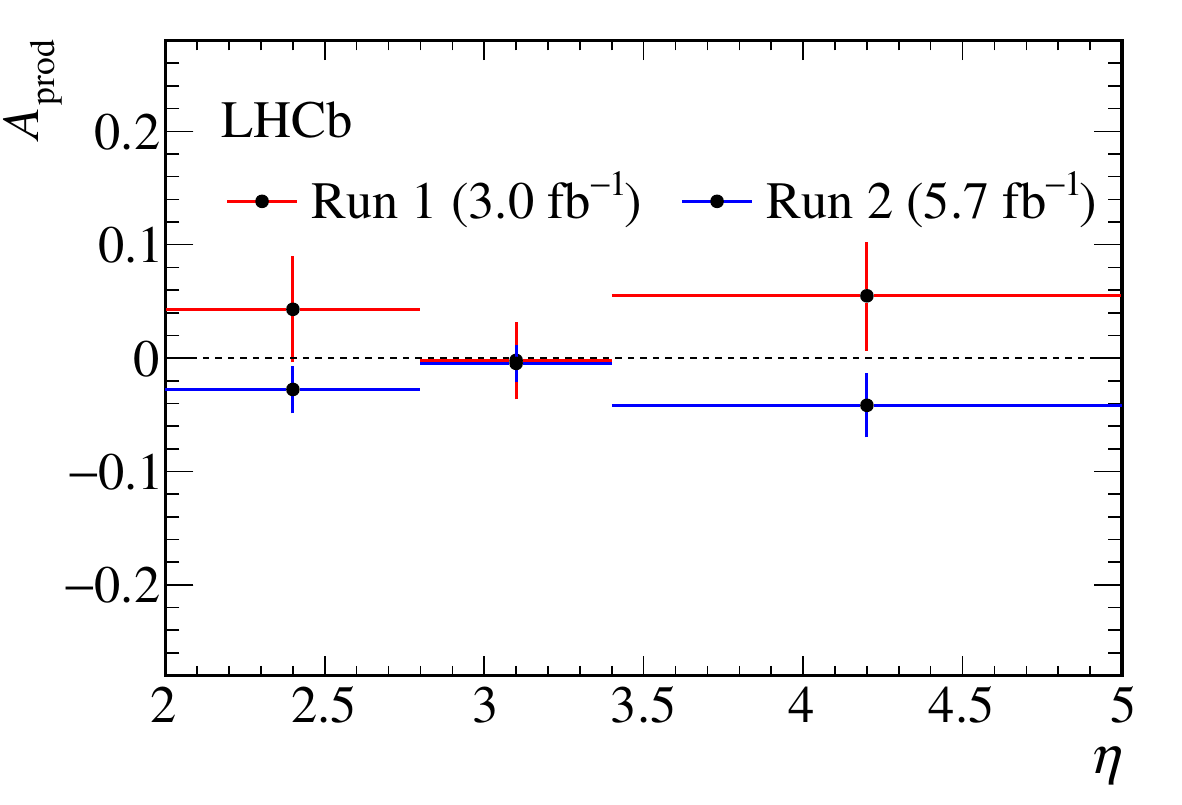}
\caption{\small 
  Variation of $B^-$ production asymmetry with (left)~$B^-$ transverse momentum and (right)~pseudorapidity determined with the $B^- \rightarrow \mathit{\Lambda}_{c}^+ \overline{p} \pi^-$ control mode.
}
\label{fig:CM_Araw}
\end{figure}

Similarly, the \Xibm\ production asymmetry is studied in intervals of $b$-hadron kinematics using the $\Xibm\to\Lc\Km\pim$ channel, with the $\Bm\to\Lc\antiproton\pim$ mode used as a control channel.
The results are shown, with statistical uncertainties only, in Figs.~\ref{fig:signal_Araw} and~\ref{fig:CM_Araw} respectively.
No strong dependence of the $\Xibm$ production asymmetry with either \pt\ or $\eta$ is observed, and all results are consistent with zero asymmetry within uncertainties.  
This is as expected since although a significant $\Lb$ production asymmetry, with a potential dependence on kinematics, has been observed~\cite{LHCb-PAPER-2021-016}, the effect is at the level of 2\% or smaller and hence a comparably sized asymmetry would not be detectable in the current analysis.
The results for the \Xibm\ production asymmetry in $pp$ collisions, integrated over the LHCb acceptance, are 
\begin{eqnarray*}
    {\cal A}_{\rm prod}(\Xibm; \text{\runone}) & = & -0.10\pm0.10\stat\pm0.03\syst\,, \\
    {\cal A}_{\rm prod}(\Xibm; \text{\runtwo}) & = & -0.10\pm0.05\stat\pm0.02\syst\,,
\end{eqnarray*}
which are also consistent with zero and with previous measurements of the \Xibm\ production asymmetry~\cite{LHCb-PAPER-2018-047}.
The control channel results are seen to be consistent with zero and with previous measurements of the \Bm\ production asymmetry in $pp$ collisions in the LHCb acceptance~\cite{LHCb-PAPER-2016-054,LHCb-PAPER-2016-062}. 

\section{Conclusion}
\label{sec:conclusion}

The decays $\Xibm\to\Lc h^- h^{\prime -}$ and $\Omegab\to\Lc h^- h^{\prime -}$, with $h^- h'^-$ being $\pim\pim$, $\Km\pim$ and $\Km\Km$ are studied using a data sample of $pp$ collisions corresponding to an integrated luminosity of $8.7~{\rm fb}^{-1}$ collected by LHCb during the \runone\ and \runtwo\ operation periods of the LHC. 
The $\Xibm \to\Lc\Km\pim$ mode is observed with large significance, and the \mbox{$\Xibm\to\Lc\Km\Km$} and \mbox{$\Omegab\to\Lc\Km\Km$} modes are also observed with over $5\,\sigma$ significance.
Ratios of fragmentation fractions times branching fractions are determined for all channels relative to the control mode $\Bm \to \Lc \antiproton\pim$, and additionally ratios of branching fractions are determined relative to the most significant decay for each $b$ baryon.
The $\Xibm \to\Lc\Km\pim$ mode is also used to study the variation with kinematics of the $\Xibm/\Bm$ production ratio, and the \Xibm\ production asymmetry.
In both cases the variation is found to be consistent with zero; the integrated \Xibm\ production asymmetry within the LHCb acceptance is also found to be consistent with zero. 
Furthermore, the $\Bm \to \Lc \antiproton\Km$ decay is observed with large significance, and its branching fraction measured relative to that of the $\Bm \to \Lc \antiproton\pim$ control mode.

These results significantly increase the currently available experimental knowledge of $\Xibm$ and $\Omegab$ decays.
The decay modes observed can in future be used to improve knowledge of the processes behind $b$-baryon production in high-energy $pp$ collisions, including production asymmetries, and to study charm baryon spectroscopy.

\section*{Acknowledgements}
\noindent We express our gratitude to our colleagues in the CERN
accelerator departments for the excellent performance of the LHC. We
thank the technical and administrative staff at the LHCb
institutes.
We acknowledge support from CERN and from the national agencies:
CAPES, CNPq, FAPERJ and FINEP (Brazil); 
MOST and NSFC (China); 
CNRS/IN2P3 (France); 
BMBF, DFG and MPG (Germany); 
INFN (Italy); 
NWO (Netherlands); 
MNiSW and NCN (Poland); 
MCID/IFA (Romania); 
MICINN (Spain); 
SNSF and SER (Switzerland); 
NASU (Ukraine); 
STFC (United Kingdom); 
DOE NP and NSF (USA).
We acknowledge the computing resources that are provided by CERN, IN2P3
(France), KIT and DESY (Germany), INFN (Italy), SURF (Netherlands),
PIC (Spain), GridPP (United Kingdom), 
CSCS (Switzerland), IFIN-HH (Romania), CBPF (Brazil),
and Polish WLCG (Poland).
We are indebted to the communities behind the multiple open-source
software packages on which we depend.
Individual groups or members have received support from
ARC and ARDC (Australia);
Key Research Program of Frontier Sciences of CAS, CAS PIFI, CAS CCEPP, 
Fundamental Research Funds for the Central Universities, 
and Sci. \& Tech. Program of Guangzhou (China);
Minciencias (Colombia);
EPLANET, Marie Sk\l{}odowska-Curie Actions, ERC and NextGenerationEU (European Union);
A*MIDEX, ANR, IPhU and Labex P2IO, and R\'{e}gion Auvergne-Rh\^{o}ne-Alpes (France);
AvH Foundation (Germany);
ICSC (Italy); 
GVA, XuntaGal, GENCAT, Inditex, InTalent and Prog.~Atracci\'on Talento, CM (Spain);
SRC (Sweden);
the Leverhulme Trust, the Royal Society and UKRI (United Kingdom).
We also acknowledge support from the commissionerate of social welfare, Government of Maharashtra (India) through the Rajarshi Shahu Maharaj scholarship.

\addcontentsline{toc}{section}{References}
\bibliographystyle{LHCb}
\bibliography{main,standard,LHCb-PAPER,LHCb-CONF,LHCb-DP,LHCb-TDR}

\newpage
\centerline
{\large\bf LHCb collaboration}
\begin
{flushleft}
\small
R.~Aaij$^{36}$\lhcborcid{0000-0003-0533-1952},
A.S.W.~Abdelmotteleb$^{55}$\lhcborcid{0000-0001-7905-0542},
C.~Abellan~Beteta$^{49}$,
F.~Abudin{\'e}n$^{55}$\lhcborcid{0000-0002-6737-3528},
T.~Ackernley$^{59}$\lhcborcid{0000-0002-5951-3498},
A. A. ~Adefisoye$^{67}$\lhcborcid{0000-0003-2448-1550},
B.~Adeva$^{45}$\lhcborcid{0000-0001-9756-3712},
M.~Adinolfi$^{53}$\lhcborcid{0000-0002-1326-1264},
P.~Adlarson$^{79}$\lhcborcid{0000-0001-6280-3851},
C.~Agapopoulou$^{13}$\lhcborcid{0000-0002-2368-0147}
% ...

{\footnotesize \it

% ...
$^{13}$Universit{\'e} Paris-Saclay, CNRS/IN2P3, IJCLab, Orsay, France\\
% ...
$^{36}$Nikhef National Institute for Subatomic Physics, Amsterdam, Netherlands\\
% ...
$^{45}$Instituto Galego de F{\'\i}sica de Altas Enerx{\'\i}as (IGFAE), Universidade de Santiago de Compostela, Santiago de Compostela, Spain\\
% ...
$^{49}$Physik-Institut, Universit{\"a}t Z{\"u}rich, Z{\"u}rich, Switzerland\\
% ...
$^{53}$H.H. Wills Physics Laboratory, University of Bristol, Bristol, United Kingdom\\
% ...
$^{55}$Department of Physics, University of Warwick, Coventry, United Kingdom\\
% ...
$^{58}$School of Physics and Astronomy, University of Glasgow, Glasgow, United Kingdom\\
$^{59}$Oliver Lodge Laboratory, University of Liverpool, Liverpool, United Kingdom\\
% ...
$^{67}$Syracuse University, Syracuse, NY, United States\\
% ...
$^{79}$Department of Physics and Astronomy, Uppsala University, Uppsala, Sweden, associated to $^{58}$\\
% ...
}
\end{flushleft}

\end{document}